\documentclass[preprint,review,11pt,sort&compress]{elsarticle}
\usepackage[margin=1.10 in]{geometry}
\usepackage{multirow}
\usepackage{textcomp}
\usepackage[usenames, dvipsnames]{color}
\definecolor{UW}{RGB}{64, 38, 96}
\usepackage{caption}
\usepackage{subcaption}
\usepackage{makecell}
\usepackage{xcolor}
\usepackage{array}
\usepackage{tabularx}
\newcolumntype{Y}{>{\setlength{\baselineskip}{0.75\baselineskip}\centering\arraybackslash}X}
\newcolumntype{Z}{>{\setlength{\baselineskip}{0.75\baselineskip}\centering\arraybackslash\hsize=.69\hsize}X}

\usepackage{amssymb}

\usepackage{float}

\journal{Sustainable Materials and Technologies}

\begin{document}

\begin{titlepage}

\clearpage\thispagestyle{empty}



\noindent

\hrulefill

\begin{figure}[h!]

\centering

\includegraphics[width=1.5 in]{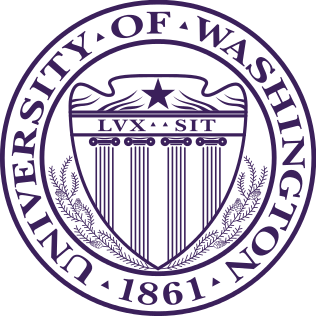}

\end{figure}

\begin{center}

{\color{UW}{

{\bf A\&A Program in Structures} \\ [0.1in]

William E. Boeing Department of Aeronautics and Astronautics \\ [0.1in]

University of Washington \\ [0.1in]

Seattle, Washington 98195, USA

}

}

\end{center} 

\hrulefill \\ \vskip 2mm

\vskip 0.5in

\begin{center}

{\large {\bf Effects of Out Time on the Mechanical and Fracture Properties of Chopped Fiber Composites Made From Aerospace Prepreg Scrap and Waste
}}\\[0.5in]

{\large {\sc Troy Nakagawa, Seunghyun Ko, Cory Slaughter, Talal Abdullah, Guy Houser, Marco Salviato}}\\[0.75in]

{\sf \bf INTERNAL REPORT No. E--9}\\[0.75in]

\end{center}

%


\end{titlepage}

\newpage

\begin{frontmatter}


\cortext[cor1]{Corresponding Author, \ead{salviato@aa.washington.edu}}

\title{Effects of Out Time on the Mechanical and Fracture Properties of Chopped Fiber Composites Made From Repurposed Aerospace\\ Prepreg Scrap and Waste}


\author[address]{Troy Nakagawa}
\author[address]{Seunghyun Ko}
\author[address]{Cory Slaughter}
\author[address2]{Talal Abdullah}
\author[address3]{Guy Houser}
\author[address]{Marco Salviato\corref{cor1}}

\address[address]{William E. Boeing Department of Aeronautics and Astronautics, University of Washington, Guggenheim Hall, Seattle, Washington 98195-2400, USA}
\address[address2]{Department of Materials Science and Engineering, University of Washington, Roberts Hall, Seattle, Washington 98195-2400, USA}
\address[address3]{Composite Recycling Technology Center, 2220 W 18th St, Port Angeles, Washington 98363-1521, USA}

\begin{abstract}
\linespread{1}\selectfont

In this study, the effects of prepreg out time on the mechanical and fracture properties of Discontinuous Fiber Composites (DFCs) are investigated. Carbon fiber prepregs are aged at $0\times$, $1\times$, $2\times$, and $3\times$ the out life in an environmental chamber at constant temperature and humidity. Degree of cure is measured via Differential Scanning Calorimetry (DSC) while tension, compression, and shear tests are performed to investigate the effects that aging has on these mechanical properties. For the first time, Mode I intra-laminar fracture and its size effect are also investigated by means of fracture tests on geometrically-scaled Single Edge Notch Tension (SENT) specimens.

From the tension, compression, and shear experiments it is seen that the out time has no effect on the elastic moduli. However, the strength increases with increasing age of the specimens for all the loading conditions. The percent increase compared to the non-aged material ranges from $15\%$ to $33\%$. This is likely caused by plasticization of the matrix with age, allowing for higher energy absorption.

More complex trends are reported for the SENT specimens for all the sizes. It is found that the fracture energy and characteristic length initially decrease with age, and then finally increase for the longest out time. This trend is owed to two factors with countering effects on the fracture behavior: $1)$ the increase of the average number of platelets with increasing aging due to increase in resin viscosity, and $2)$ the plasticization of the matrix with aging.

The results from this study suggest that Discontinuous Fiber Composites (DFCs) made from reused materials can have equal, if not better, performance than non-aged DFCs. The experimental data presented in this work can be used as a baseline to design DFC composite components made from repurposed prepreg scrap and waste.

\end{abstract}


\end{frontmatter}

\section{Introduction}
The use of polymer composites in the aerospace, automotive, and energy sectors has seen a tremendous increase in the last few decades \cite{das2019preparation, harris2002design, othman2018application}. To give an idea, the U.S. composite end products market was valued at \$26.7 billion in 2019 and is forecast to grow to \$33.4 billion by 2025 \cite{Mazumdar2020,ACMAsite}. Similar trends are reported worldwide. 

While these numbers give an idea of the key role that composites are playing in the global economic growth, they also show that sustainability is becoming an increasingly important issue considering that hard-to-recycle thermosetting composites represented 82\% of the U.S. composite market in 2019 \cite{Mazumdar2020,ACMAsite}. Fostering sustainable practices and finding ways to reuse and recycle thermoset composite prepreg waste is imperative to mitigate the environmental impact of this industry and enable the expected growth of the market across the wind, aerospace, automotive, and construction industries.

Following \cite{NuttAging}, thermoset composite prepreg waste can be broadly classified into uncrosslinked waste produced during manufacturing and crosslinked waste from components reaching end of life. Efforts to handle the former material stream can be defined as ``reuse" while efforts to handle the latter are commonly defined as ``recycle". The overarching goal of recycling is to reclaim fibers from cured waste by getting rid of the matrix. Through the years, several interesting techniques for recycling have been developed including pyrolisis (see e.g. \cite{naqvi2018critical}), matrix digestion (see e.g. \cite{ma2018chemical}), and depolymerization (see e.g. \cite{kuang2018recycling}). Typically, depending on the technique, reclaimed fibers can have comparable mechanical properties to virgin fibers. However, this is often achieved at the expenses of fiber continuity. In fact, reclaimed fibers often take the form of tangled, discontinuous arrays which undermines the manufacturing of high-quality composite laminates requiring long, continuous fibers to achieve high mechanical performance \cite{pickering2006recycling, goodship2009management, witik2013carbon}. Different from recycling, reusing aims at repurposing both the fibers and the uncured matrix in their original configuration to manufacture new composite parts \cite{NuttAging}. In this case, the material sources include e.g. ply cutter waste, end-of-roll material, and out-of-spec material beyond out-life or storage life. For instance, the aerospace industry follows very strict requirements on material certification and the costs of re-certifying an out-of-spec material would be prohibitive, leading to significant amounts of prepreg waste. Except for few cases in which out-of-spec material is used internally for R\&D or donated to research institutions, most prepreg scrap and waste ends in landfills, posing an environmental challenge. It is clear that this situation will not be sustainable with the future growth of the composite market unless prepreg waste reuse becomes common practice.

\begin{figure}[htb!]
    \centering
    \includegraphics[width=1\textwidth]{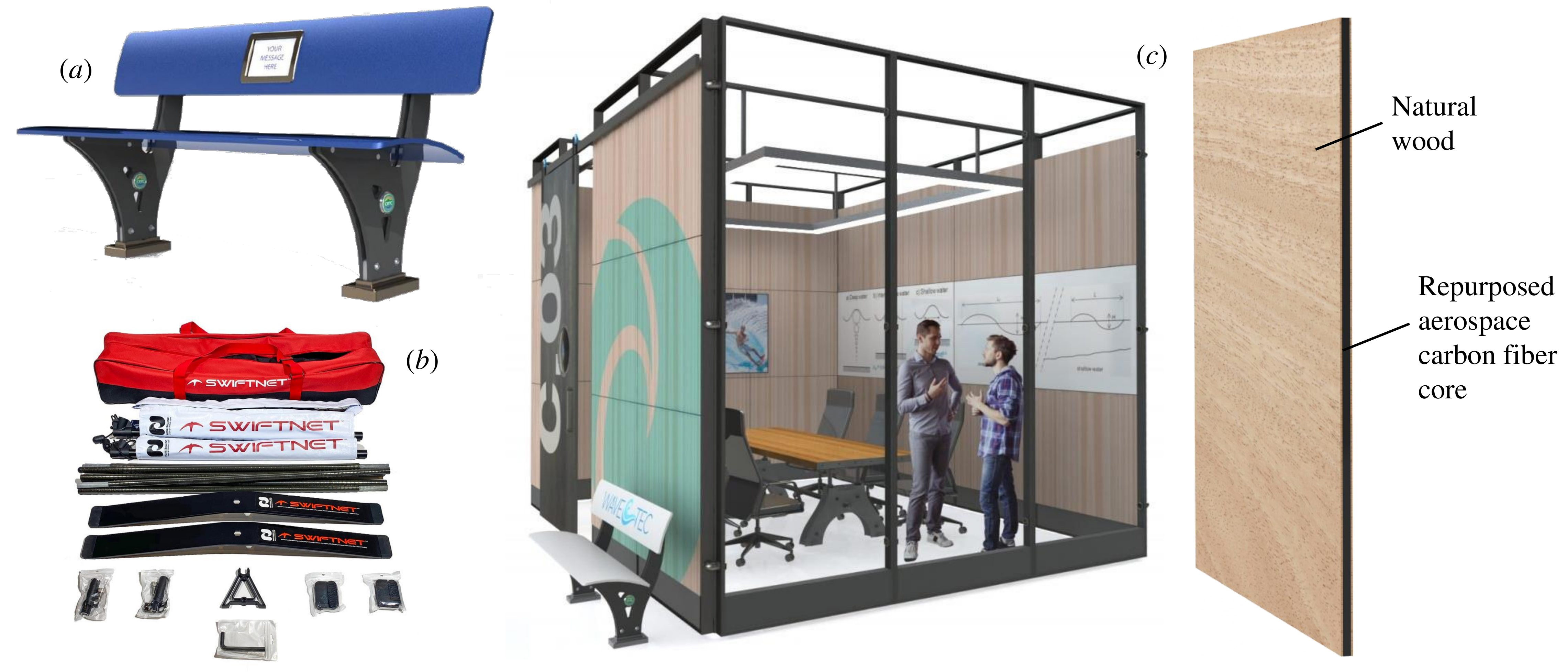}
    \caption{Typical products made by CRTC using repurposed aerospace carbon fiber composites: (a) carbon fiber bench, (b) portable pickle-ball net, (c) advanced laminated panels featuring natural wood and $85\%$ in weight of recycled carbon fibers \cite{CRTC}.}
    \label{fig:CRTC Examples}
\end{figure}

Developing technologies for the reuse of prepreg scrap and waste is one of the core missions of the Composite Recycling Technology Center (CRTC) \cite{CRTC} located in Port Angeles, WA, USA. Since its establishment the CRTC has developed unique products made from repurposed aerospace carbon fiber prepreg with applications ranging from the sport goods industry to construction and even defense. Figure \ref{fig:CRTC Examples}a, for instance, shows a corrosion-resistant bench made using a combination of reused scrap roll form and Discontinuous Fiber Composites (DFCs). Figure \ref{fig:CRTC Examples}b shows Swiftnet\textsuperscript{TM} a lightweight, portable pickle-ball net while Figure \ref{fig:CRTC Examples}c shows a novel Advanced Cross Laminated Timber Panel (ACLT) featuring natural wood and $85\%$ in weight of repurposed aerospace carbon fibers.

\begin{figure}[htb!]
    \centering
    \includegraphics[width=1\textwidth]{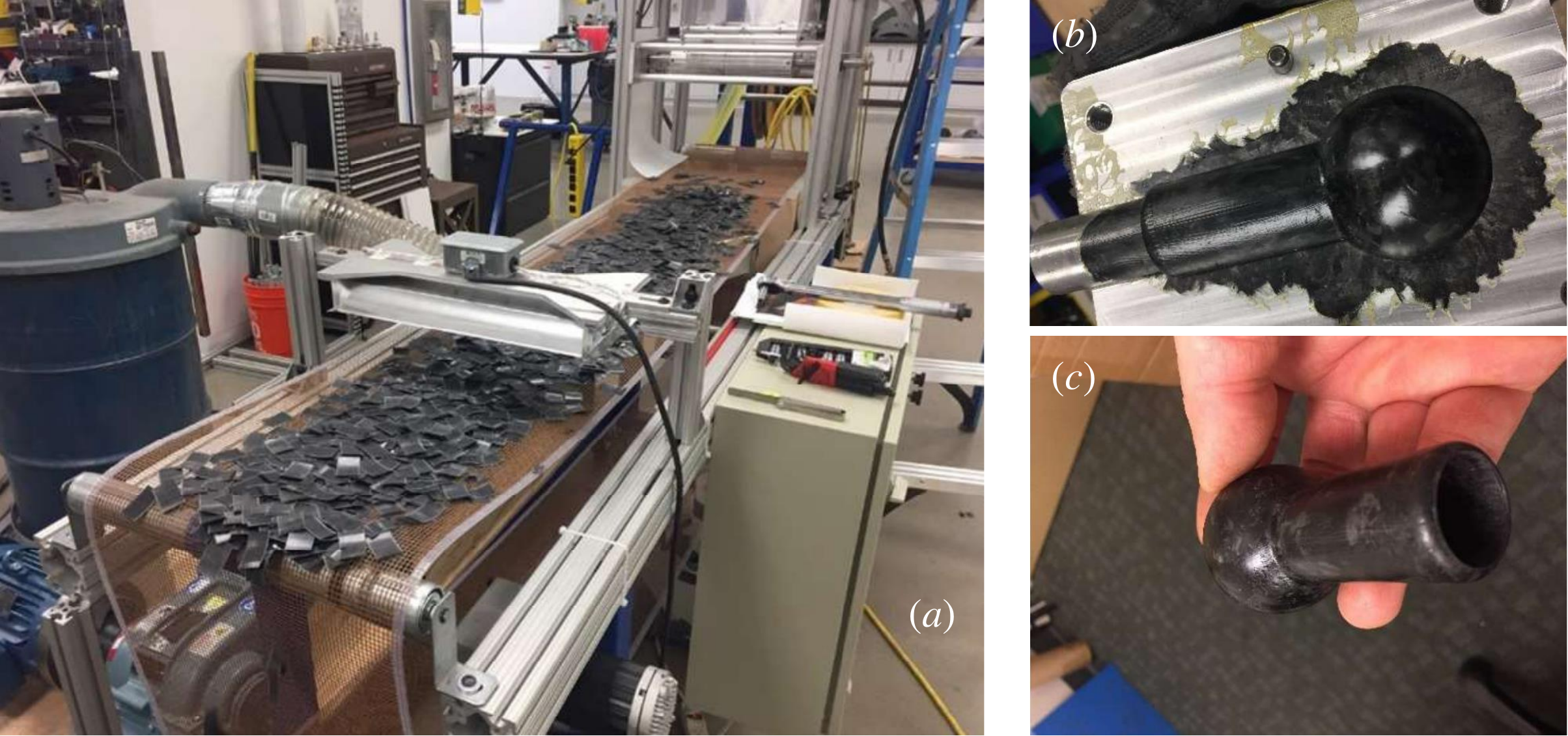}
    \caption{Discontinuous Fiber Composites made from repurposed aerospace carbon fiber prepregs: (a) platelets are cut from scrap prepregs to the desired shape and size, (b) compression molding is used to manufacture a ball joint, (c) the net-shape product is extracted. Thanks to this approach very complex shaped parts can be manufacturing without machining.}
    \label{fig:DFC}
\end{figure}

Among the several technologies utilized by the CRTC, using prepreg scrap and waste to manufacture discontinuous fiber composites (also known as chopped fiber composites) is one that has shown great promise. As Figures \ref{fig:DFC}a-c show, Discontinuous Fiber Composites (DFCs) are made by cutting rectangular platelets from scrap prepregs. The size and aspect ratio of the platelets depend on the particular part to manufacture but platelet widths range from hundreds to few millimetre. The platelets can then be used to fill a mold and can be compression molded to make very complex shape parts. Figures \ref{fig:DFC}b,c show an example of a ball joint made by CRTC using repurposed prepreg. It is worth noting that, thanks to the unique characteristics of DFCs, even a complex part such as the one shown in the figure can be manufactured without additional machining. This characteristic alone might enable DFCs made from reused prepregs to reach markets that are not easily accessible by advanced unidirectional composites due to the manufacturing costs.

DFC parts made from virgin prepregs have already been shown to have outstanding mechanical properties, comparable to aluminum and quasi-isotropic carbon fiber laminates \cite{SeungPlateletSE, SeungThicknessSE}. However, it is possible that some of the properties of DFCs made from reused materials might be degraded by the aging of the prepregs. Hence, considering how difficult it would be to control the ageing of this material stream, the question to answer is: how much does ageing affect the mechanical performance of DFCs made from repurposed prepregs? Answering this question is the goal of the present study which presents one of the most comprehensive investigations into the effects of aging on the tensile, compression, and shear strength, and global mode I fracture energy of DFCs made from reused material. Understanding the link between aging and mechanical performance is quintessential to formulate design guidelines which can pave the way for the use of DFCs made from repurposed prepregs also in secondary structural applications. 

\section{Material preparation and test description}
\subsection{DFC manufacturing procedure}
In this study, Toray T$700$G-$12$k/$2510$ prepreg was aged and used 
to manufacture DFC laminates. To follow the standard conditioning atmosphere stated in ASTM D$4332$ \cite{ASTMD4332}, an in-house environmental chamber was made to keep the temperature and humidity at $23\pm2^{\circ}$C and $50\pm5\%$RH respectively. An indoor grow tent was used to isolate the prepreg, and an off-the-shelf temperature and humidity controller was used to control a system composed of an Air Conditioner (AC), heater, humidifier, and dehumidifier. The prepreg was aged to $0\times$, $1\times$, $2\times$, and $3\times$ the out life of the prepreg ($0$, $28$, $56$, and $84$ days). 
The manufacturing method detailed by Ko et al. \cite{SeungSAMPE, SeungPlateletSE} was used to make the DFC laminates. Platelets were cut using a fabric cutter to the desired size of $50 \times 8$ mm. The weight was controlled to attain the desired thickness of $3.3$ mm. In addition to the DFC specimens, a quasi-isotropic laminate using $[45/90/-45/0]_{3s}$ was made using non-aged material to serve as an additional benchmark to evaluate the performance of the aged DFCs.

\subsection{Analysis of crosslinking degree}
To quantify the progression of cure while the prepreg ages, Differential Scanning Calorimetry (DSC) was used to obtain the enthalpy of the prepreg. These tests were done following ASTM D$3418$ \cite{ASTMD3418} using a Mettler Toledo DSC $3$+. For the first week of aging DSC tests were performed daily. After the first week, DSC was performed once a week for $28$ days (which corresponds to one out life). Then, DSC was performed every other week until $3\times$ the out life was reached after which DSC testing was not performed until $7\times$ out life. After this, DSC was performed at $9\times$ the out life and at every $n\times$ out-lives until $16\times$ out-lives ($448$ days) was reached. This time is longer than any previous study found by the authors in the open literature. 

For each DSC analysis, four samples of $10$ mg were tested in aluminum pans. A ramp rate of $10^{\circ}$C/min was used to sweep from $23$ to $280^{\circ}$C. Nitrogen was used as the purge gas and had a flow rate of $10$ mL/min. To calculate the enthalpy, a MATLAB code was developed to calculate the area under the specific heat flow vs time curve. During the later parts of the tests, glass transition temperature (T$^{\circ}_g$) could also be calculated from these scans. To calculate the T$^{\circ}_g$ Mettler Toledo's STARe software was used which follows ASTM E$1356$ \cite{ASTME1356}.

\subsection{Mesostructure}
An investigation into the mesostructure of aged DFCs was also performed. To this end, small sections of $25$ mm were cut from the laminates. For out times $0\times$ and $2\times$, $6$ specimens were investigated, for out time $1\times$, $5$ specimens were investigated, and for out time $3\times$, $4$ specimens were investigated. The samples were encased into epoxy pucks made using Allied High Tech's QuickCure Acrylic to be used in a Struers RotoPol 21 and a RotoFoce 3 auto-polisher. Once polished, an Olympus BX$50$ microscope was used to inspect the mesostructure. A magnification of $5\times$ was used to find the constituent content, thickness of platelets, and the number of platelets through the thickness. Stitching of the images was done using ImageJ which was also used to find the constituent content. To find the number of platelets though the thickness and the thickness of the platelets, 7 evenly spaced sections within the $25$ mm length were investigated. The first and last sections were $1$ mm away from the ends of the specimen. A MATLAB code was developed to find the thickness of platelets, and the number of platelets through the thickness \cite{matlab}.

\subsection{Mechanical and fracture tests}
\subsubsection{Data acquisition}
Tests were performed using an Instron $5585$H $250$ kN electro-mechanical load frame with an Interface $1210$ACK-$50$kN-B load cell, and with the $250$kN Instron load cell for the largest size tested for the size effect testing. 
The load cells recorded with a sampling frequency of $10$ Hz while Digital Image Correlation (DIC) was used to obtain the strain values needed. Images were captured using a Nikon D$5600$ DSLR camera with Nikon AF micro $200$ mm and Sigma $135$ mm DG HSM lenses with a sampling rate of $1$ Hz. GOM Correlate \cite{gom} was used to process the photos and obtain the strain values. The number of specimens tested for each loading case can be found in Tables \ref{tab:Ten}-\ref{tab:SizeEffect}.

\subsubsection{Tension}
Tensile data for the quasi-isotropic and non-aged DFCs made from the same material system tested in this work was taken from tests performed by Ko et al. \cite{SeungSAMPE}. For the aged samples, tensile tests were performed based on ASTM D$3039$ \cite{ASTMD3039}. Specimens were made to have a width of $25.5$ mm, which corresponds to about three times the width of a platelet, and a gauge length of $140$ mm. The tabs were made with $1/8$ in. thick garolite and JB-Weld Cold Weld Steel Reinforced Epoxy and had a length of $55$ mm with a bevel of $8^{\circ}$. A displacement rate of $2$ mm/min was maintained for all the tests. 

\subsubsection{Compression}
Compression tests were performed following ASTM D$3410$ \cite{ASTMD3410}. The samples were made to have a width of $12.5$ mm, a gauge length of $25$ mm and $50$ mm tabs. All compression tests were performed using the Instron load frame and a Wyoming Modified IITRI Compression test fixture. A displacement rate of $1.5$ mm/min was used.

\subsubsection{Shear}
Shear tests were performed according to ASTM D$5379$ \cite{ASTMD5379}. An Omax $2652$ water jet cutter was used to cut the specimens to the required geometry. Specimens were tabbed using $1/8$ in thick garolite and Solvay FM $94$k adhesive film. During the tests, the specimens were clamped in the tab area to prevent crushing and twisting and tested using the Instron load frame and a Wyoming Iosipescu test fixture. A displacement rate of $1$ mm/min was used. A study was performed on the quasi-isotropic shear specimen to find the optimum width to average the shear strain obtained from DIC. This width was found to be $1$ mm.

\subsubsection{Size effect tests}
For the aged coupons, the experimental procedures used in this study followed \cite{SeungPlateletSE}. This source also provided size effect test data for quasi-isotropic and non-aged DFCs. The smallest size effect specimen had a base dimension of $20\times44.5$ mm ($D\times L_g$) and a target thickness of $3.3$ mm. The gauge areas of the specimens were geometrically scaled following a $2$D scaling of $1:2:4$. Tabs were made with $1/8$ in. thick garolite and JB-Weld Cold Weld Steel Reinforced Epoxy with a length of $38$ mm for all specimens. To create the notch, a diamond-coated razor blade saw was used. The blade thickness was $0.2$ mm. The initial notch length, $a_{\circ}$, was kept at a constant ratio of $D/5$. 
A constant strain rate of $0.2\%$/min was used for all specimens.

\section{Experimental results}
\subsection{Crosslinking evolution with aging}
\begin{figure}[htb!]
    \centering
    \includegraphics[width=0.9\textwidth]{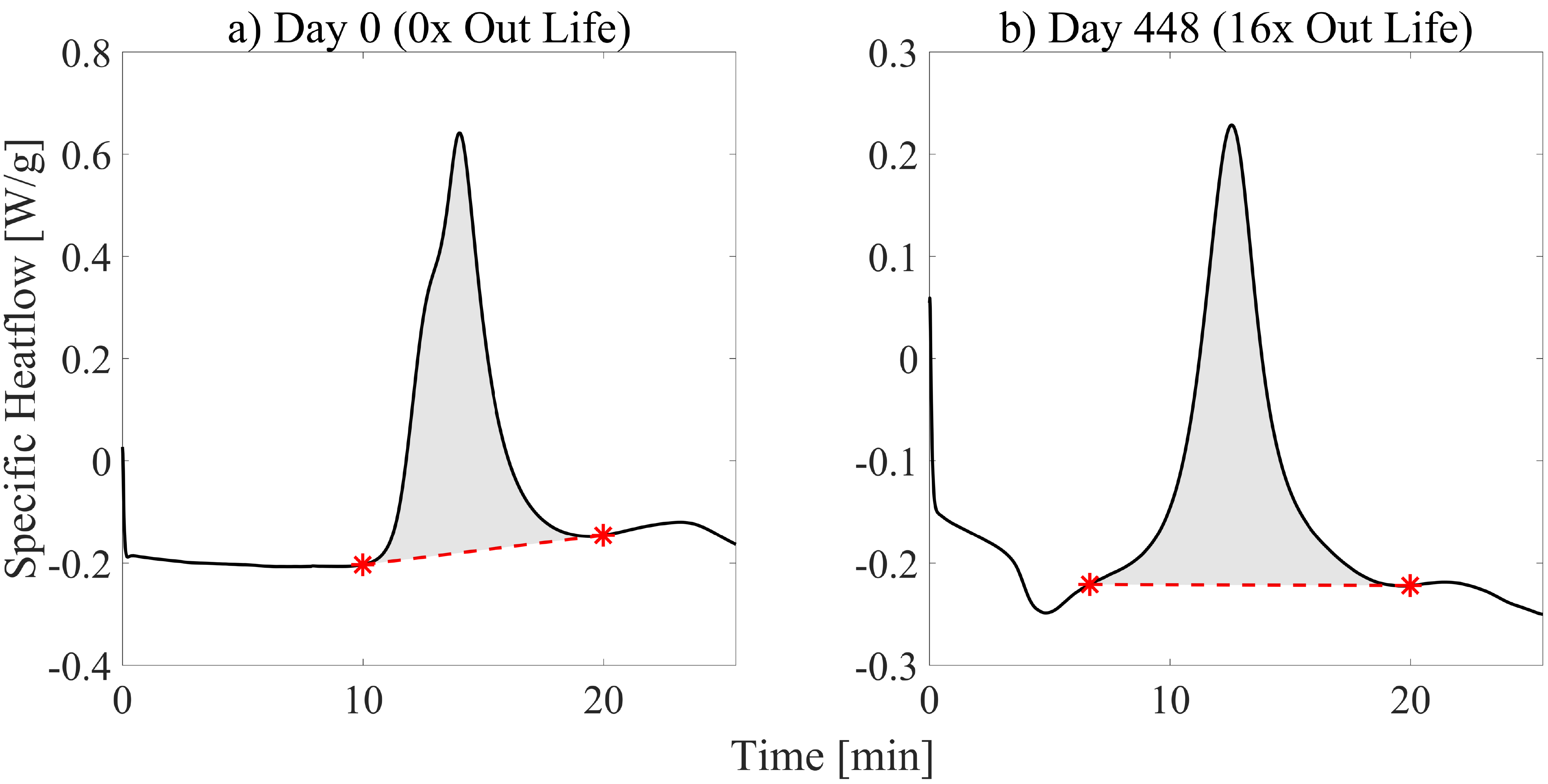}
    \caption{Representative heat flow plots from DSC tests for out times of a) 0x (0 days) and b) 16x (448 days)}
    \label{fig:HeatFlow}
\end{figure}

In fig. \ref{fig:HeatFlow}, two representative heat flow curves from DSC testing are shown. The red stars on the plot indicate the bounds used to calculate the enthalpy of the prepreg. Here, the bounds were selected manually by finding the transition points of the exothermic cure peak. It can be seen in fig. \ref{fig:HeatFlow}b that the first bound can be hard to determine due to the bump in the heat flow caused by the glass transition temperature. A good way to determine the location of the first bounds is by looking at previous out times where the glass transition temperature does not appear on the curve, as well as interpolating from the straight parts of the curve at the beginning and end of the curve.

\begin{figure}[htb!]
    \centering
    \includegraphics[width=0.9\textwidth]{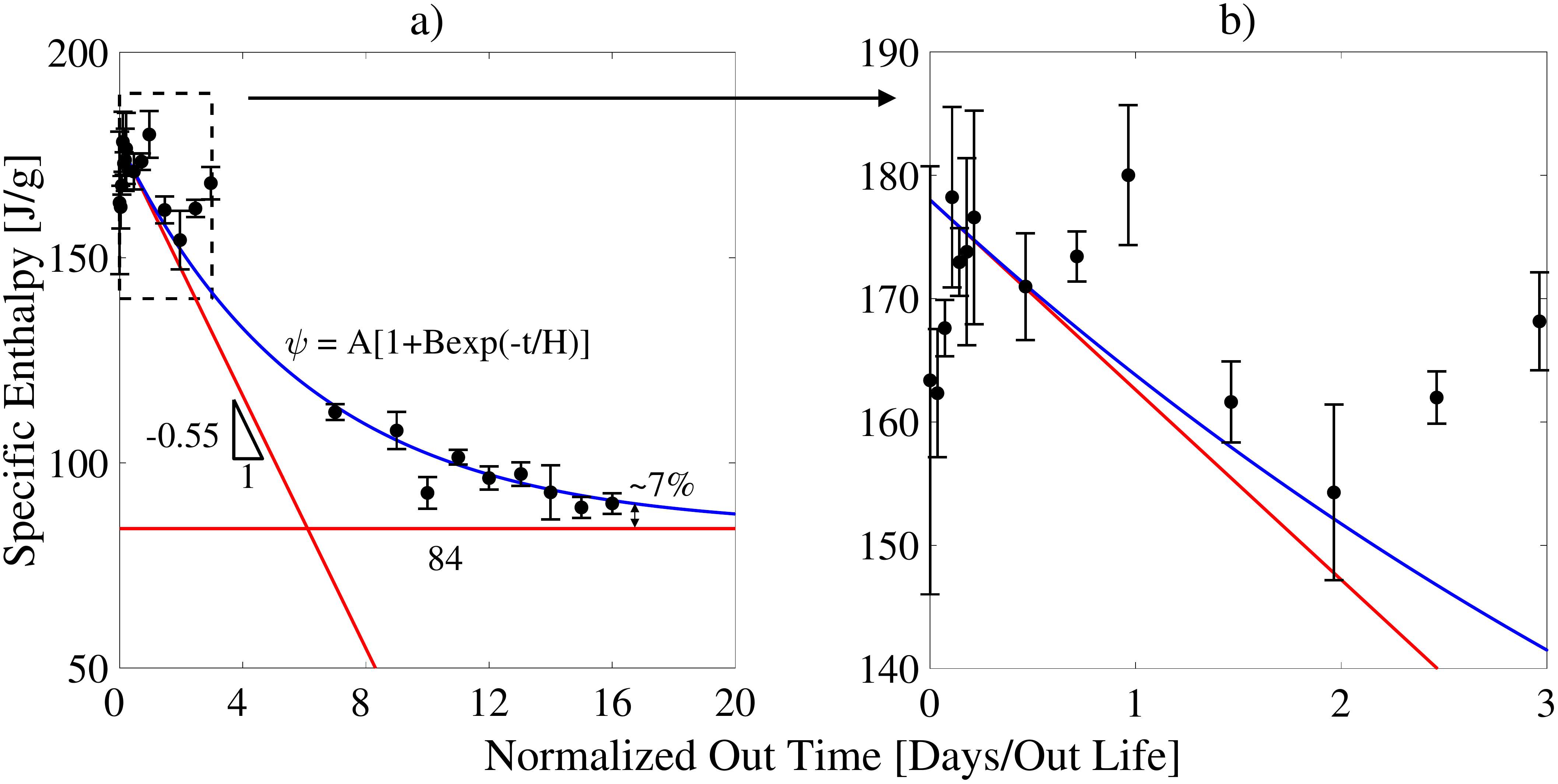}
    \caption{Results from DSC analysis. a) Enthalpy vs the normalized out time, and b) a zoomed in view of the enthalpy vs normalized out time plot showing the out time window the mechanical testing was performed.}
    \label{fig:DSCResults}
\end{figure}

In fig. \ref{fig:DSCResults}a and b, the results from the enthalpy calculations are shown from which it can be seen that the variation in enthalpy found was very small. On average, the variation in enthalpy at one out time was $3.38\%$. There were two outliers where the variation were large. These were day $0$ ($0\times$ out life) and day $392$ ($14\times$ out life) with a variation of $10.62\%$ and $7.07\%$ respectively. If the outliers are taken out, the average variation drops to $2.86\%$. This low variability validates the low sample count used in this study. 

\begin{figure}[htb!]
    \centering
    \includegraphics[width=0.5\textwidth]{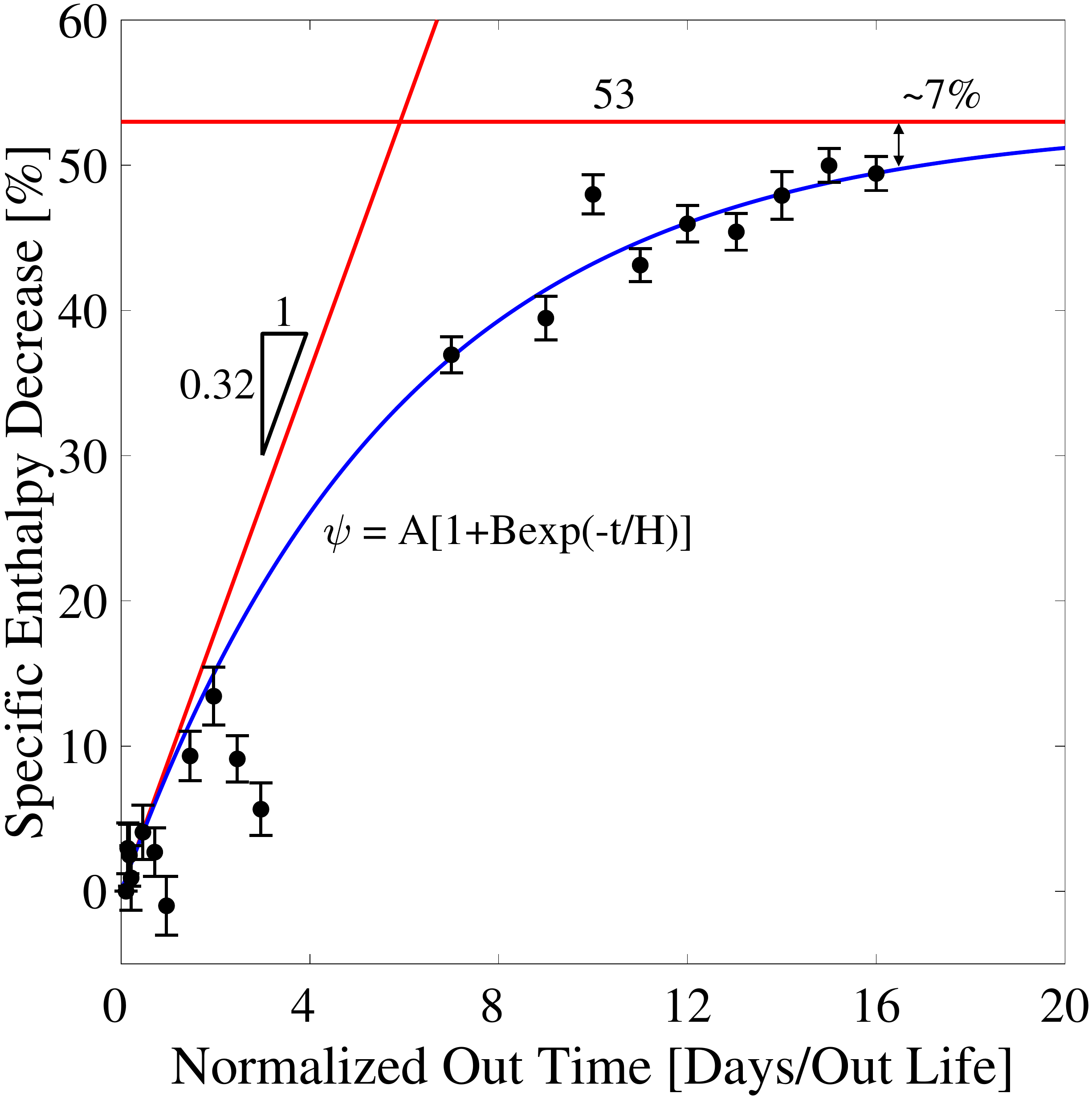}
    \caption{Percent decrease of enthalpy vs the normalized out time, detailing the progression of curing compared to the ``as purchased'' enthalpy.}
    \label{fig:DoC}
\end{figure}

It can be seen in fig. \ref{fig:DSCResults}b, that at the beginning there is an increase in enthalpy from day $0$ to day $4$. This is because the initial samples were taken from the edge of the prepreg roll and each day was a little closer to the center of the roll. Starting from day $4$ the samples for DSC analysis were taken from the center of the roll. For this reason, when calculating the percent decrease of enthalpy in fig. \ref{fig:DoC} and when performing the linear fit, the initial enthalpy is taken to be the one calculated on day $4$.

\begin{figure}[htb!]
    \centering
    \includegraphics[width=0.5\textwidth]{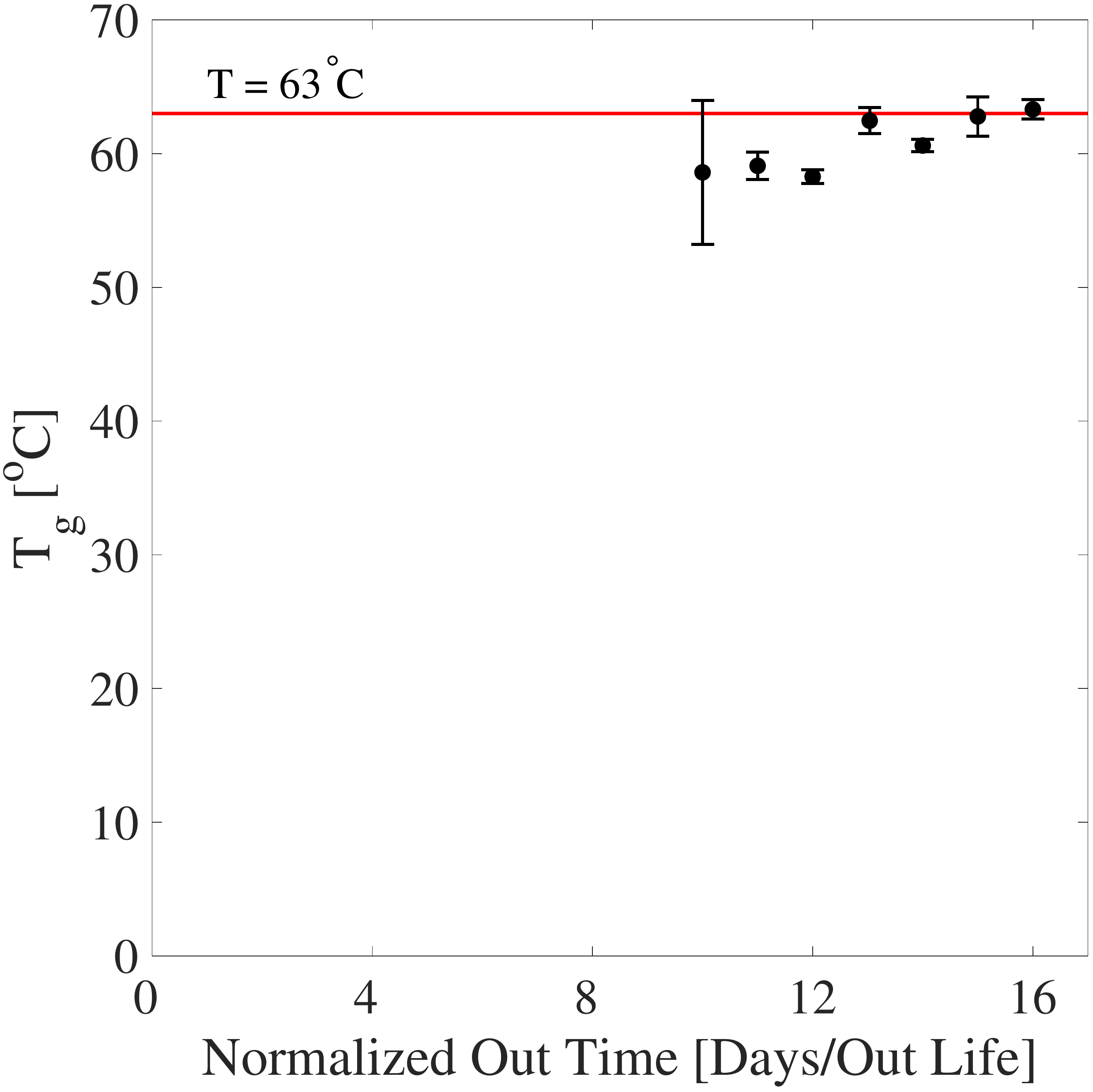}
    \caption{Glass transition temperature (T$_g$) for out times 10-16x (280-448 days).}
    \label{fig:Tg}
\end{figure}

The data in fig. \ref{fig:DSCResults} and \ref{fig:DoC} were fit using the following equation:
\begin{equation}
    \Psi=A\left[1+B\exp\left(-\frac{t}{H}\right)\right]
\end{equation}
where $A$ is the horizontal asymptote, the initial slope is $-AB/H$, and the y-intercept is $A(1+B)$. At the early stages of aging the decrease in enthalpy is approximately linear. Then the enthalpy starts to asymptotically approach a value of $84$ J/g, indicating a saturation of crosslinking. It can be seen in the figures, that more crosslinking can occur, but the time required to complete the curing would be extremely large. The final enthalpy measured has about a $7\%$ difference from the asymptote, and to get to a difference of $1\%$, an additional $358$ days of aging is needed. This would correspond to the percent enthalpy decrease to go from $49.4\%$ to $52.6\%$. At this point the evolution of crosslinking is marginal. A similar saturation can also be seen in the glass transition temperature (T$_g$) in fig. \ref{fig:Tg}. From $10\times$ the out life ($280$ days) the T$_g$ can be seen in the heat flow curves. The T$_g$ appears to be saturating at about $63^{\circ}$C. This shows that if the prepreg is left out past this time there should be no further progression of ambient curing. It should be noted that the exact degree of cure cannot be obtained because of the unknown initial enthalpy or degree of cure of the epoxy. 
Thus the percent decrease stated here is related to a progression of cure since the ``as purchased'' state rather than the degree of cure of the prepreg.


\subsection{Mesostructure}
\begin{figure}[htb!]
    \centering
    \includegraphics[width=0.5\textwidth]{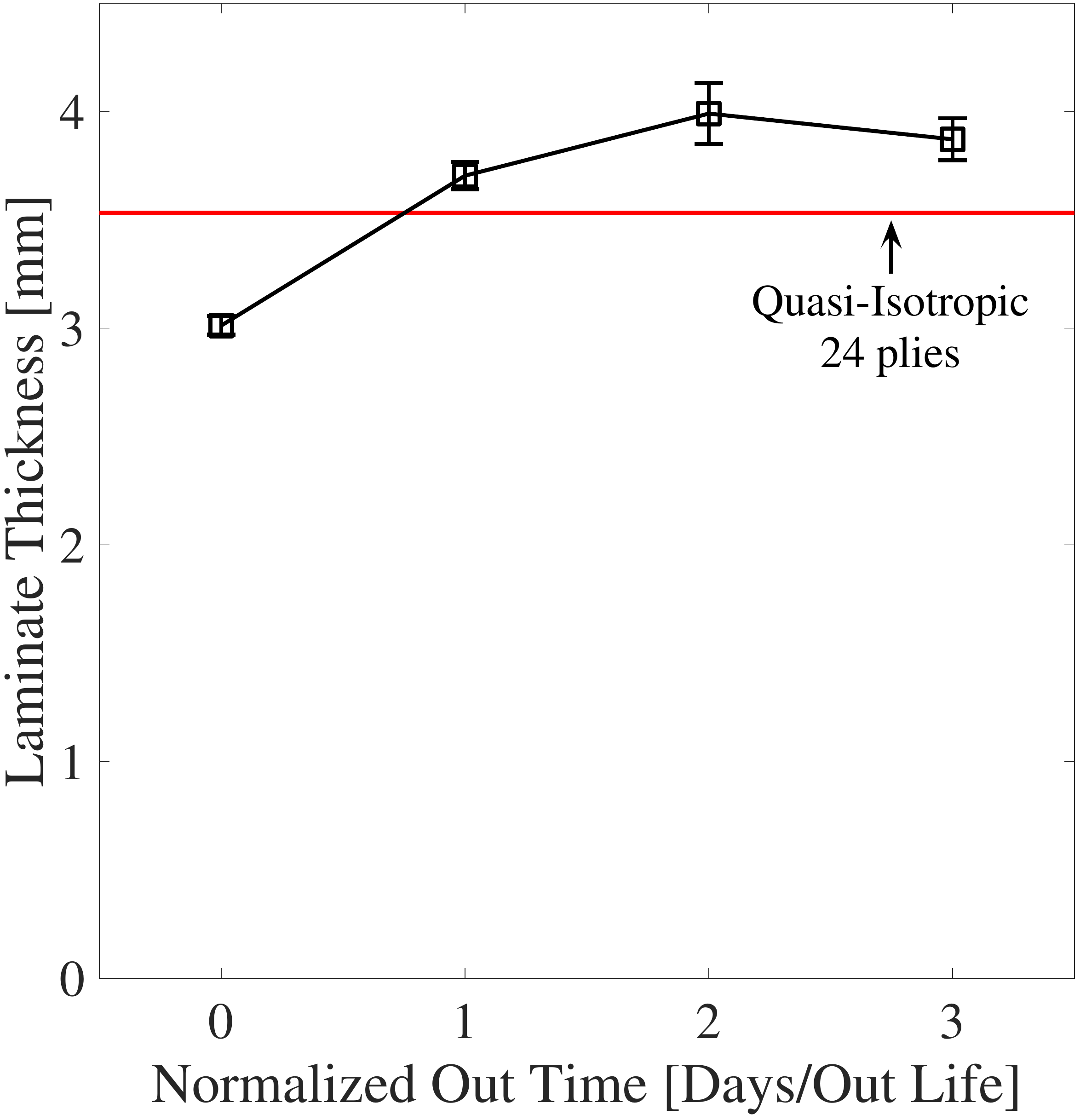}
    \caption{Thickness of non-aged quasi-isotropic and DFC laminates at various out times.}
    \label{fig:LamThick}
\end{figure}

Figure \ref{fig:LamThick} shows the laminate thickness measured from all the tested specimens. It can be seen that there was an increase in the specimen thickness with out time. The increase from the non-aged to the aged specimen was between $23\%$ and $33\%$ while there was much less change in thickness between the aged specimens. 

\begin{figure}[htb!]
    \centering
    \includegraphics[width=0.5\textwidth]{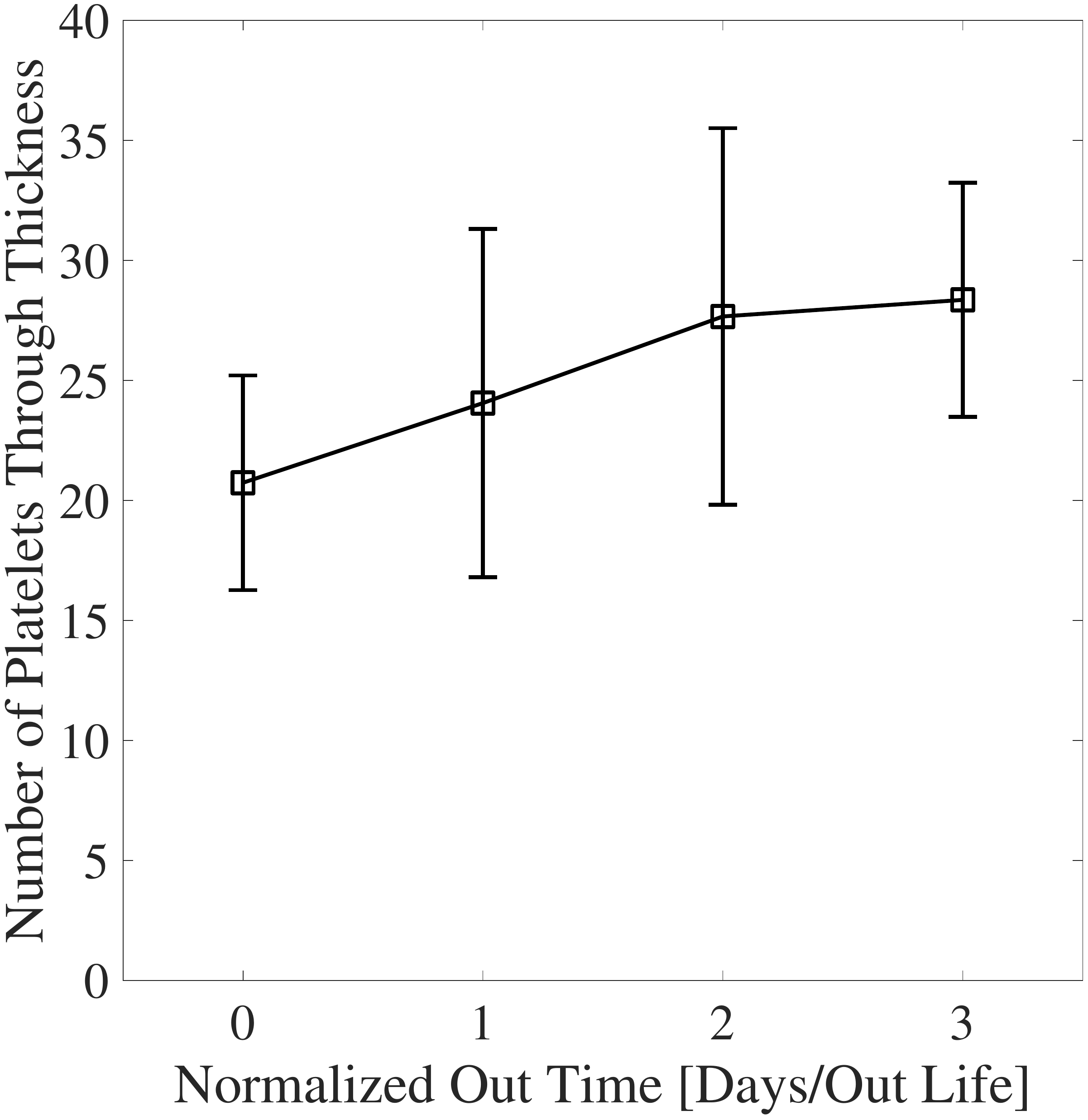}
    \caption{Number of platelets through the thickness of DFC laminates at various out times.}
    \label{fig:NumPlatelet}
\end{figure}

\begin{figure}[htb!]
    \centering
    \includegraphics[width=0.5\textwidth]{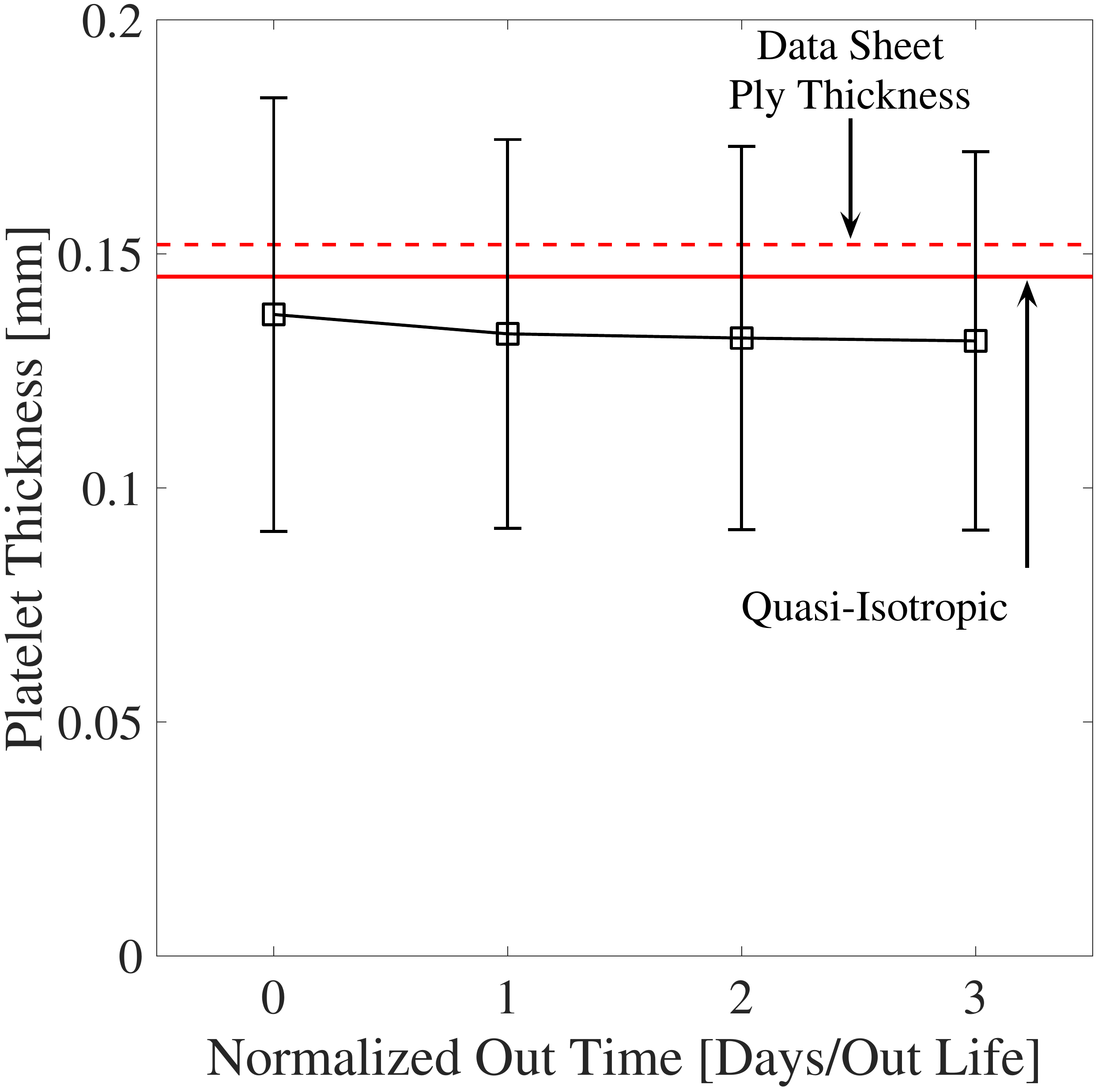}
    \caption{Platelet thickness at various out times. The dashed line is the cure ply thickness stated in the T$700/2510$ material data sheet, and the solid line was found by analysing one quasi-isotropic specimen.}
    \label{fig:Platelet-Thick}
\end{figure}

\begin{figure}[htb!]
    \centering
    \includegraphics[width=0.5\textwidth]{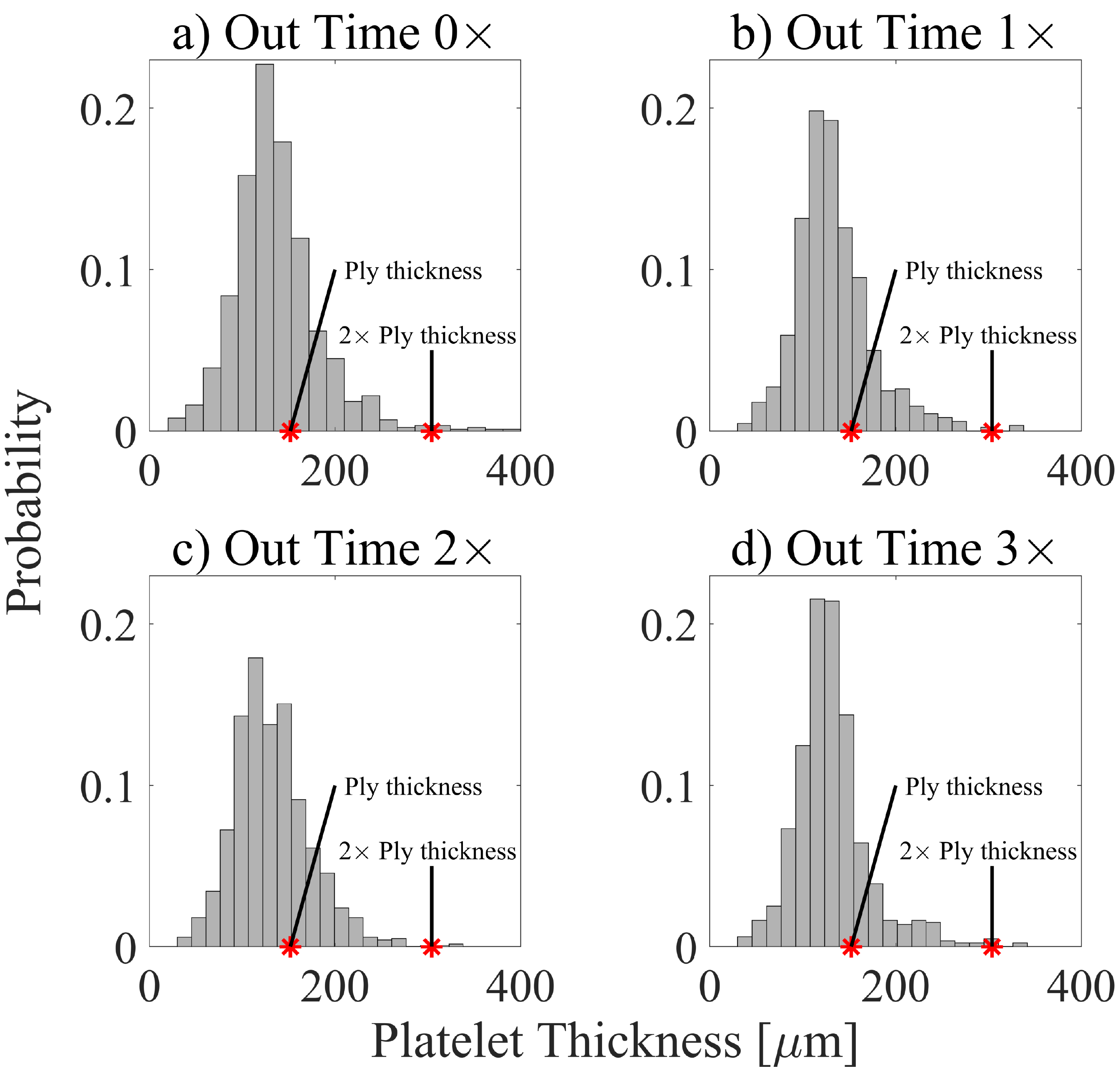}
    \caption{Histogram of the platelet thickness for all out times. The red stars in the histogram represents multiples of the cure ply thickness stated in the T$700/2510$ material data sheet.}
    \label{fig:Platelet-Thick-Hist}
\end{figure}

To understand the thickness increase, microscope images were taken and analysed to observe any changes in the mesostructure. With reference to Figure \ref{fig:NumPlatelet} and table \ref{tab:Meso}, it can be noted that the number of platelets through the thickness increased with the out time. This can be explained by looking at the thickness of the platelets reported in Figure \ref{fig:Platelet-Thick} and table \ref{tab:Meso}. In fact, it can be noted that the platelets in the DFC specimens are slightly thinner than those of the plies in the quasi-isotropic laminate. However, the platelet thickness does not change with the increase in out time. Figure \ref{fig:Platelet-Thick-Hist} shows the probability density of the platelet thicknesses constructed leveraging $800-1160$ platelet observations for each out time. As can be noted, there are only slight changes in the peaks of the probability distributions with a change in out time, confirming that the mesostructure does not change significantly with increasing out time. 

\begin{figure}
    \centering
    \begin{subfigure}[b]{\textwidth}
        \centering
        a) Area with void \\
        \includegraphics[width=1\textwidth]{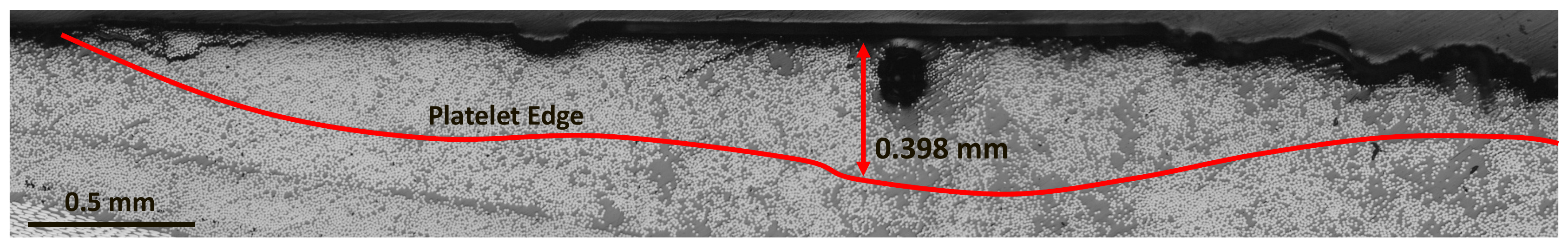}
    \end{subfigure}
    \hfill \\
    \begin{subfigure}[b]{0.5\textwidth}
        \centering
        b) Resin rich area
        \includegraphics[width=1\textwidth]{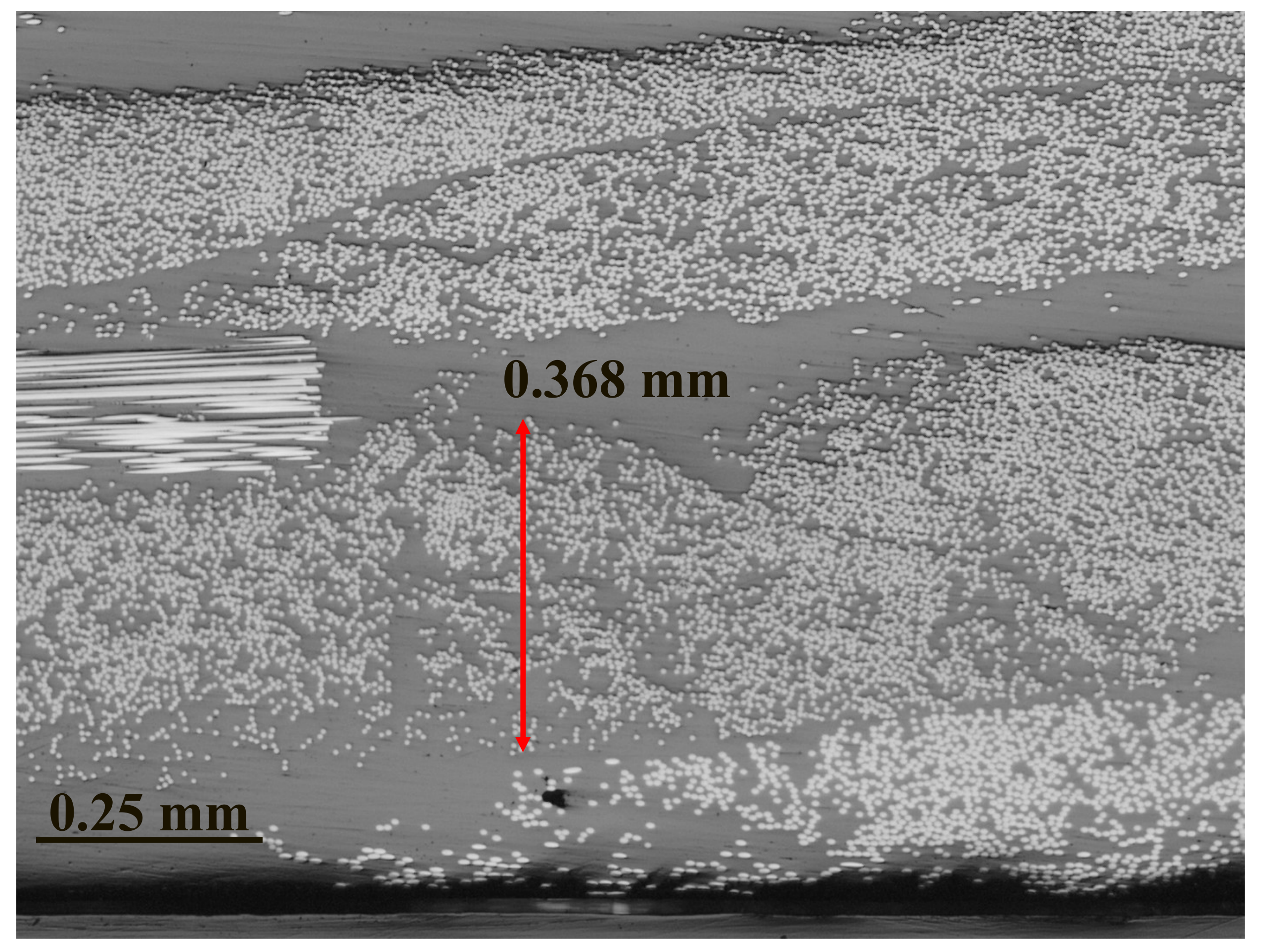}
     \end{subfigure}
    \caption{Microscope images of two DFC specimens. a) Shows the increase in platelet thickness caused by a void. b) Shows the increase in platelet thickness caused by lack of platelets when compared to the average number of platelets through the thickness.}
    \label{fig:Microscopy}
\end{figure}

Figure \ref{fig:Microscopy} shows areas where the platelet thickness is significantly different from the average value due to defects. Fig. \ref{fig:Microscopy}a, for instance, shows fibers rearranging around a void causing an increase in thickness of the platelet. On the other hand, fig. \ref{fig:Microscopy}b shows an area where there are fewer platelets compared to the average number of platelets and the remaining volume must be filled with resin or voids. 
Due to the resin filling the volume where there are no platelets, the thickness of the platelet that is in the volume increases.

\begin{figure}[htb!]
    \centering
    \includegraphics[width=0.9\textwidth]{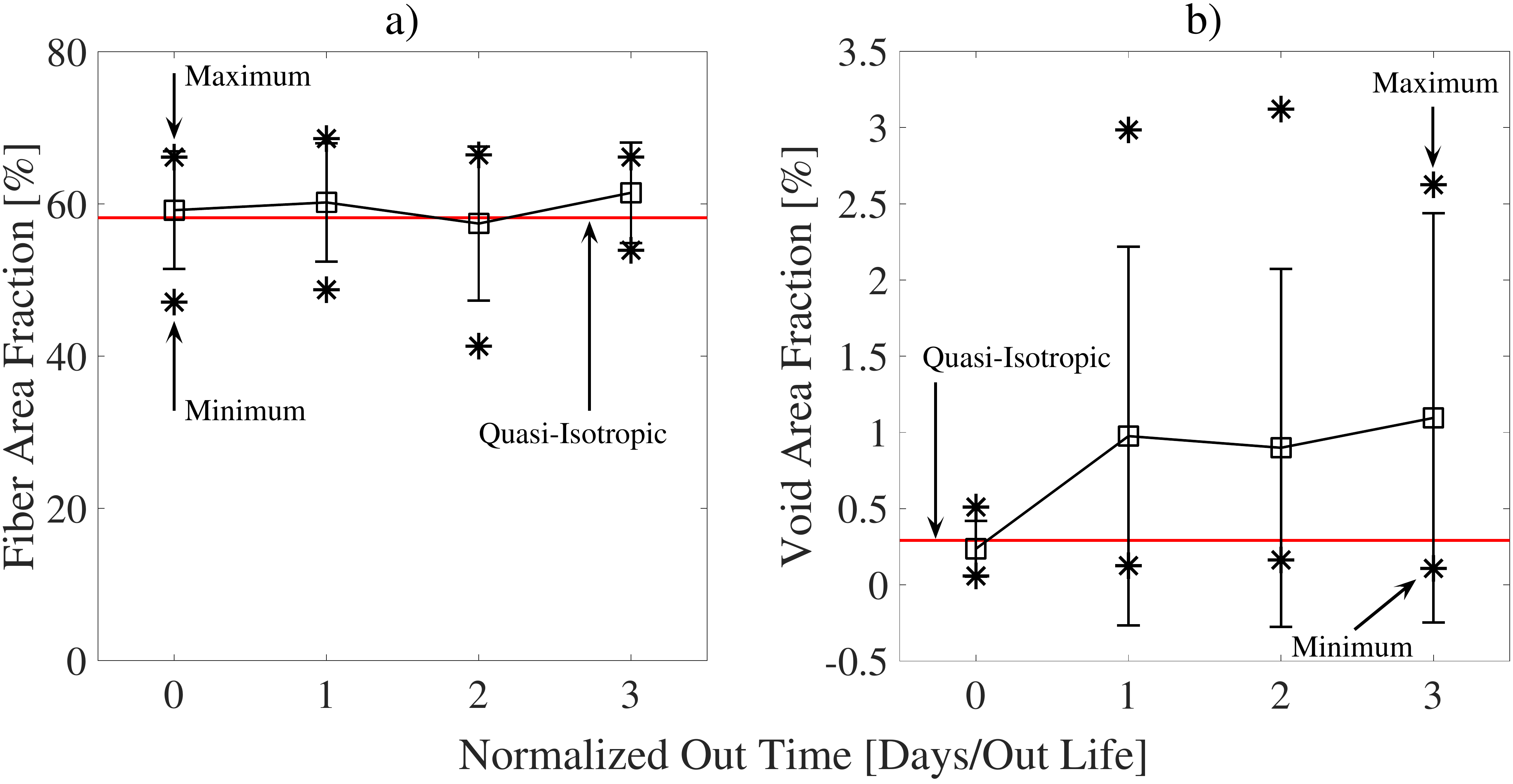}
    \caption{a) Fiber and b) void area fractions for a non-aged quasi-isotropic laminate, and DFCs at various out times. The stars represent the maximum and minimum area fraction of the DFC laminate at various out times.}
    \label{fig:fiber-void-area-frac}
\end{figure}

Figure \ref{fig:fiber-void-area-frac} and table \ref{tab:Meso} provide information on the fiber and void area fractions. It can be seen from fig. \ref{fig:fiber-void-area-frac}a, that the fiber area fraction does not change with age, and that the fiber area fractions of the DFC specimens are equal to those of the quasi-isotropic laminate. Moreover, fig. \ref{fig:fiber-void-area-frac}b, shows that the non-aged DFCs have a void area fraction similar to that of the quasi-isotropic layup, and that the void area fraction of DFCs increases with age. In fact, the quasi-isotropic and non-aged DFCs have a void volume fraction of $0.29\%$ and $0.24\%$ respectively while all the aged DFCs have a void area fraction of about $1\%$. 

The void volume fraction does not change between out times, and the maximum and minimum void content also do not change between the aged DFCs. The minimum void content observed was $0.13\%, 0.16\%,$ and $0.11\%$ for out times $1\times, 2\times$, and $3\times$ respectively while the maximum void content observed was $2.98\%, 3.12\%,$ and $2.63\%$. The variation of the aged specimens is much higher than the quasi-isotropic and non-aged specimens.

\subsection{Tension}\label{Tension}
\begin{figure}[htb!]
    \centering
    \includegraphics[width=1\textwidth]{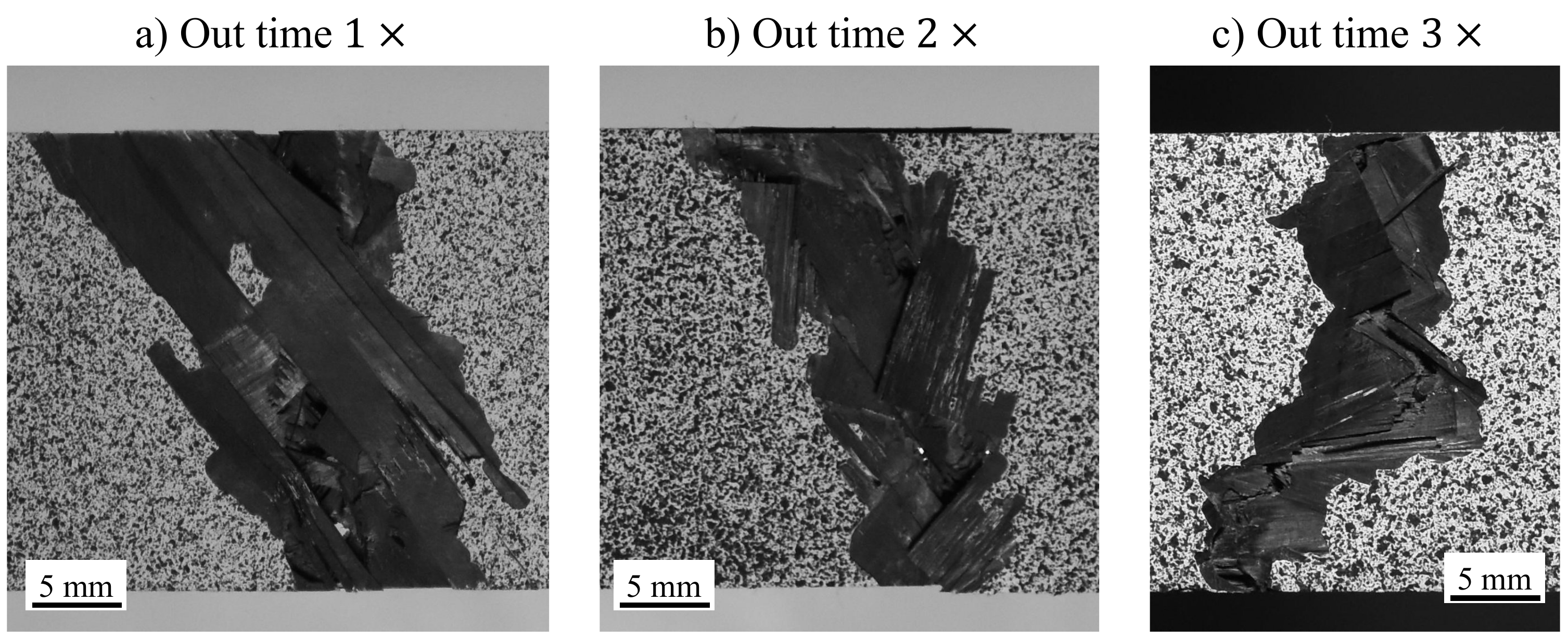}
    \caption{Fracture surfaces of out times $1\times$, $2\times$, and $3\times$ tension specimen, which mainly exhibit matrix damage and debonding, with some fiber damage.}
    \label{fig:TenFrac}
\end{figure}

Figure \ref{fig:TenFrac} shows examples of fracture surfaces of tensile coupons of out times $1\times$, $2\times$, and $3\times$. Similar to non-aged specimens seen in previous papers \cite{SeleznevaTenCompShear, Feraboli1, Feraboli2, Takahashi1, Takahashi2}, the failure path can be seen to go around platelets or within a platelet. Fiber breakage still occurs, however its frequency is much lower in comparison to delamination and matrix damage since the failure attempts to take the path of least resistance thus avoiding fiber breakage. There was no perceivable difference in failure mechanisms between out times.

\begin{figure}[htb!]
    \centering
    \includegraphics[width=1\textwidth]{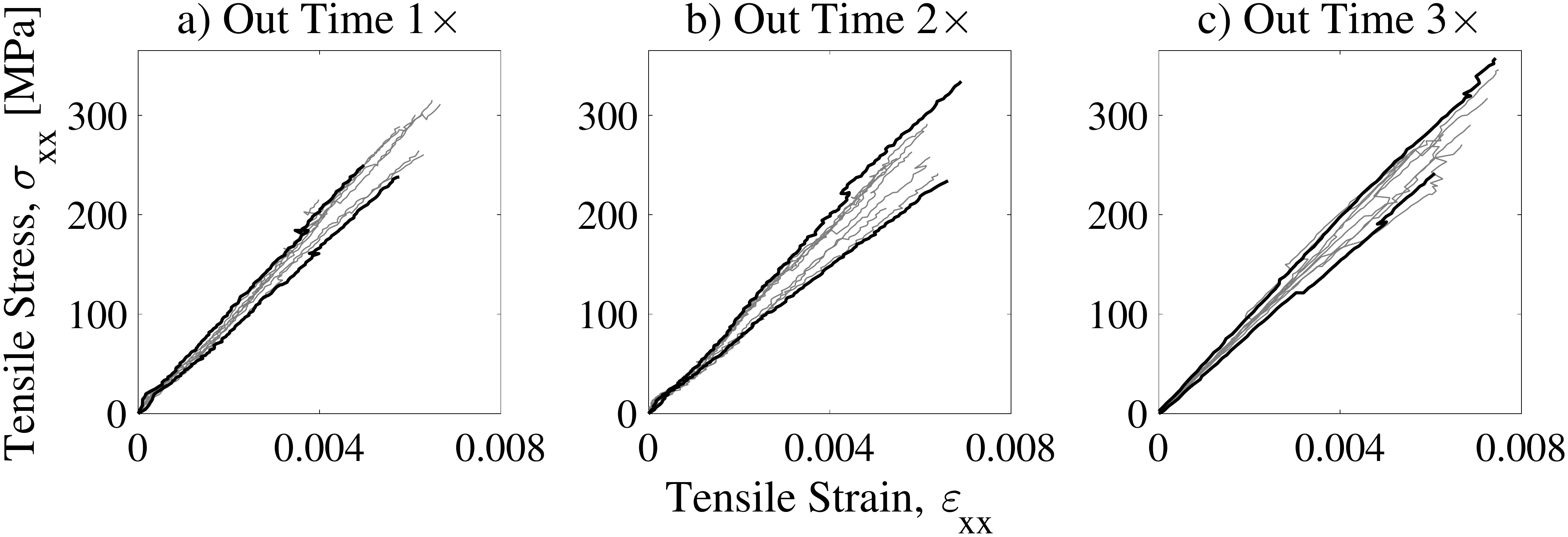}
    \caption{Stress vs strain curves for out times $1\times$, $2\times$, and $3\times$ tension specimen.}
    \label{fig:Ten-stress-train}
\end{figure}

Figure \ref{fig:Ten-stress-train} shows the stress vs strain curves for out time $1\times$, $2\times$, and $3\times$ DFC specimens. It can be seen that the behavior was linear up to the peak and that there was progressive damage for all the specimens. For the lowest modulus specimens shown in fig. \ref{fig:Ten-stress-train}c, it can be seen that at about $50\%$ the ultimate strength there was damage, however the specimen was still able to carry load. This is a property of DFCs: where there is an area with less favorable fiber orientation there will be damage, however the areas with favorable orientation can carry more load.

\begin{figure}[htb!]
    \centering
    \includegraphics[width=0.9\textwidth]{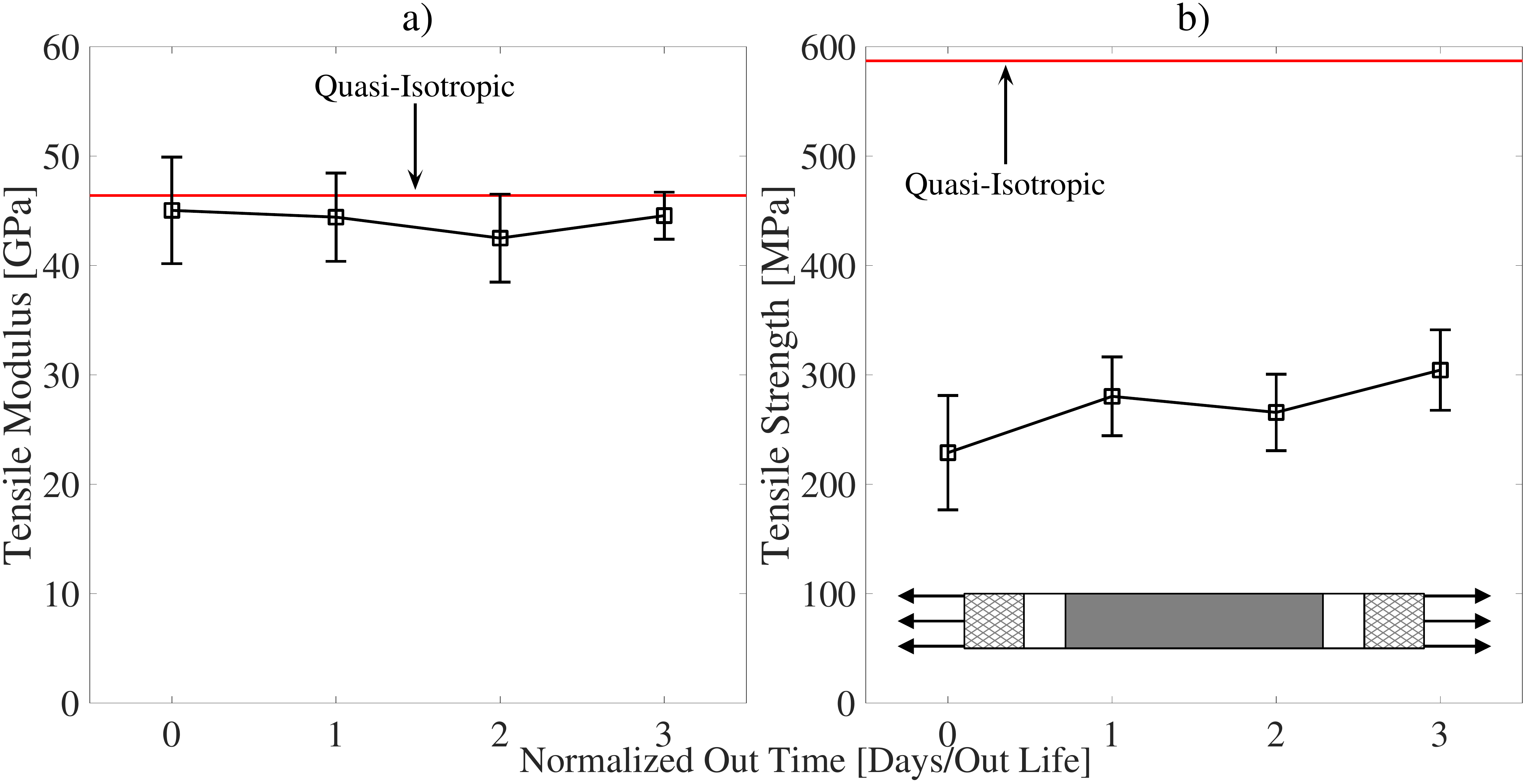}
    \caption{Effects of out time on the tensile a) modulus and b) strength. Quasi-isotropic and non-aged DFC results are taken from Ko et al. \cite{SeungSAMPE}.}
    \label{fig:Ten}
\end{figure}

Figure \ref{fig:Ten} and table \ref{tab:Ten} provide information on the tensile modulus and strength for the quasi-isotropic and DFC specimens. It can be seen in fig. \ref{fig:Ten}a that the modulus does not change significantly with age and there is little difference when compared to the quasi-isotropic value. The modulus for the DFC specimens remains at $92-97\%$ that of the quasi-isotropic modulus. The largest difference in modulus occurs at out time $2\times$, which is $91.6\%$ that of the quasi-isotropic modulus. However, this modulus still falls within one standard deviation of the non-aged DFC specimens. 

In fig. \ref{fig:Ten}b, it can be seen that the strength of the DFC specimens can be from $39\%$ to $52\%$ that of the quasi-isotropic laminate. It is interesting to note that there is an increase in tensile strength of the DFC specimens when the prepreg is aged. After aging for one out time, the aged DFCs exhibit a strength $28\%$ higher than the non-aged sample. After the initial increase, little change in strength can be noted past for longer aging times.

\subsection{Compression}
\begin{figure}[htb!]
    \centering
    \includegraphics[width=0.5\textwidth]{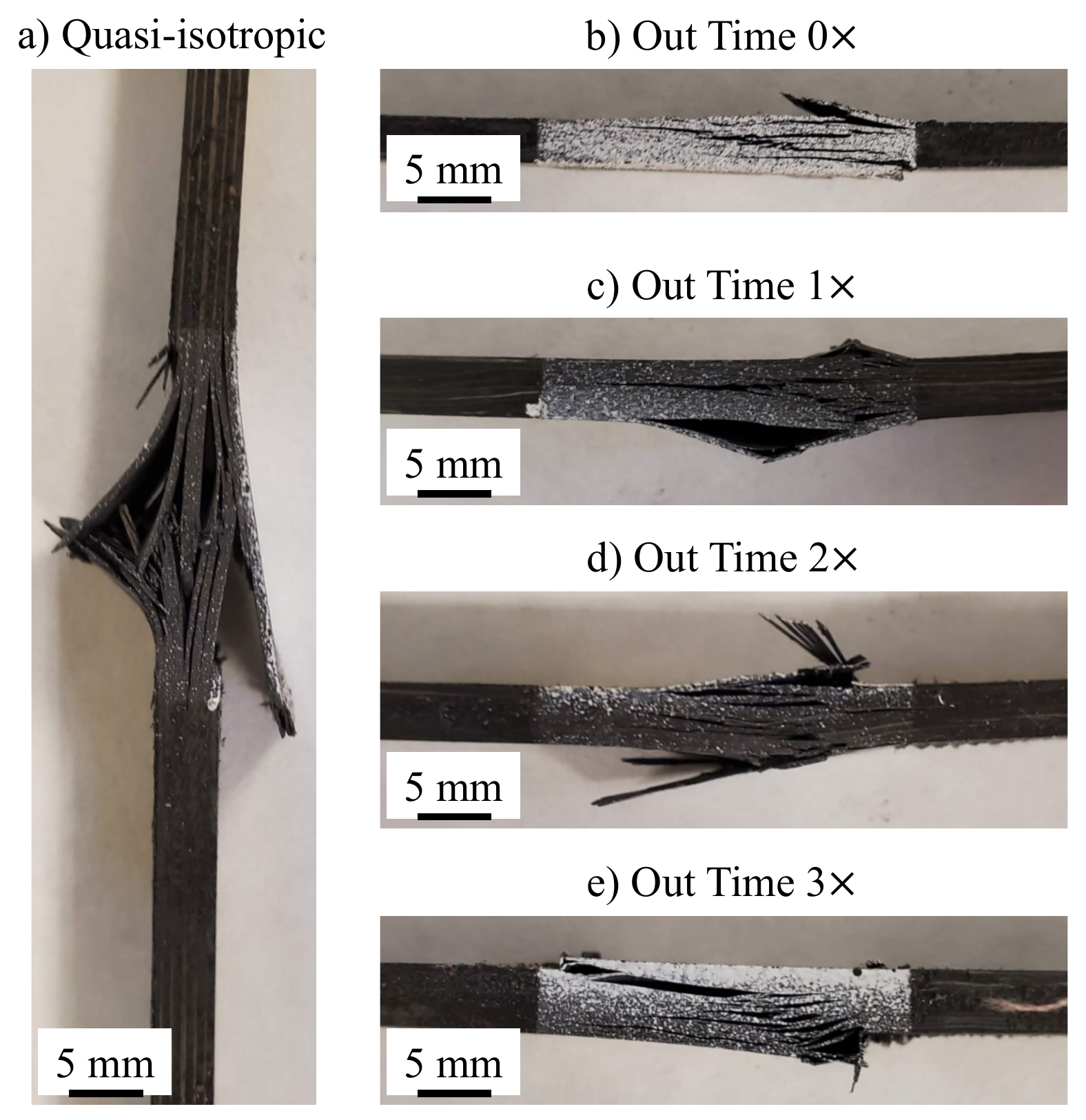}
    \caption{Fracture surfaces of quasi-isotropic, and DFC compression specimen out times $0\times$, $1\times$, $2\times$, and $3\times$. DFC specimens mainly exhibit matrix damage and debonding, with some fiber damage.}
    \label{fig:CompFrac}
\end{figure}

Figure \ref{fig:CompFrac} shows examples of fracture surfaces of quasi-isotropic, and DFC compression specimens for out times $0\times$, $1\times$, $2\times$, and $3\times$. It can be seen that the failure of DFC specimens is much less brittle when compared to the quasi-isotropic specimens. There is still fiber failure at the outer surfaces of the DFC specimen at various out times, however most of the damage is through the thickness. In fact, similar to the tension specimens, the majority of the damage is from the matrix and platelet debonding. There was no perceivable difference in failure mechanisms between out times and it is worth mentioning that similar failure for non-aged compression specimens were seen in previous papers \cite{SeleznevaTenCompShear, Feraboli1, Takahashi1}.

\begin{figure}[htb!]
    \centering
    \includegraphics[width=0.8\textwidth]{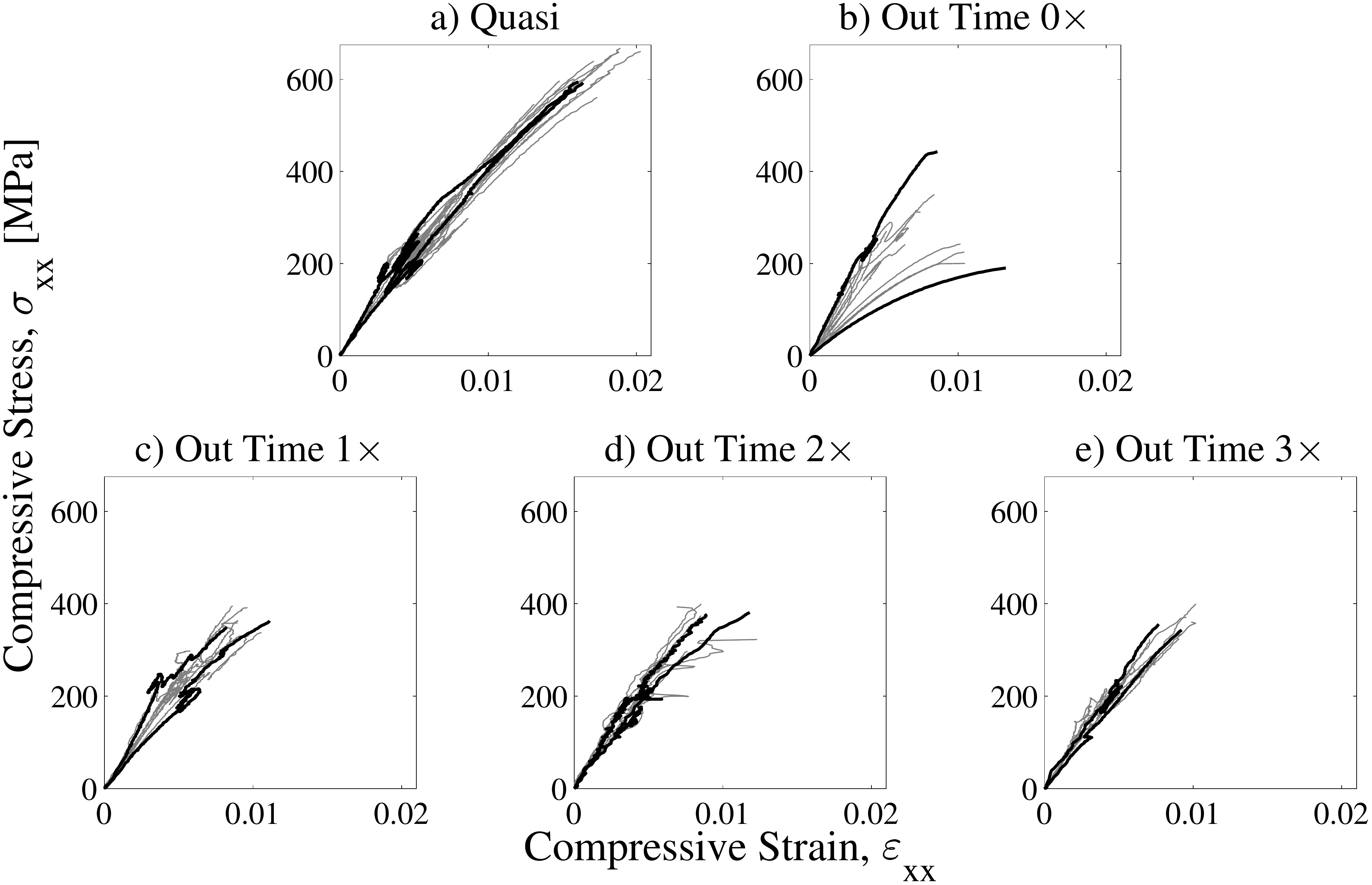}
    \caption{Stress vs strain curves for quasi-isotropic and out times $0\times$,$1\times$, $2\times$, and $3\times$ compression specimens.}
    \label{fig:Comp-stress-train}
\end{figure}

Figure \ref{fig:Comp-stress-train} shows the stress vs strain curves for quasi-isotropic, and out time $0\times$, $1\times$, $2\times$, and $3\times$ DFC specimens. It can be seen that when the nominal stress was at about $200$ MPa, that there could be unloading, this is caused by the fact that the compression fixture grips are wedges \cite{ASTMD3410} and must readjust at this point. It can be seen from the curves that the compression specimens exhibit varying degrees of non-linearity. The out time $0\times$ specimen exhibited the most non-linear behavior and as the out time increased the non-linear behavior decreased.

\begin{figure}[htb!]
    \centering
    \includegraphics[width=0.9\textwidth]{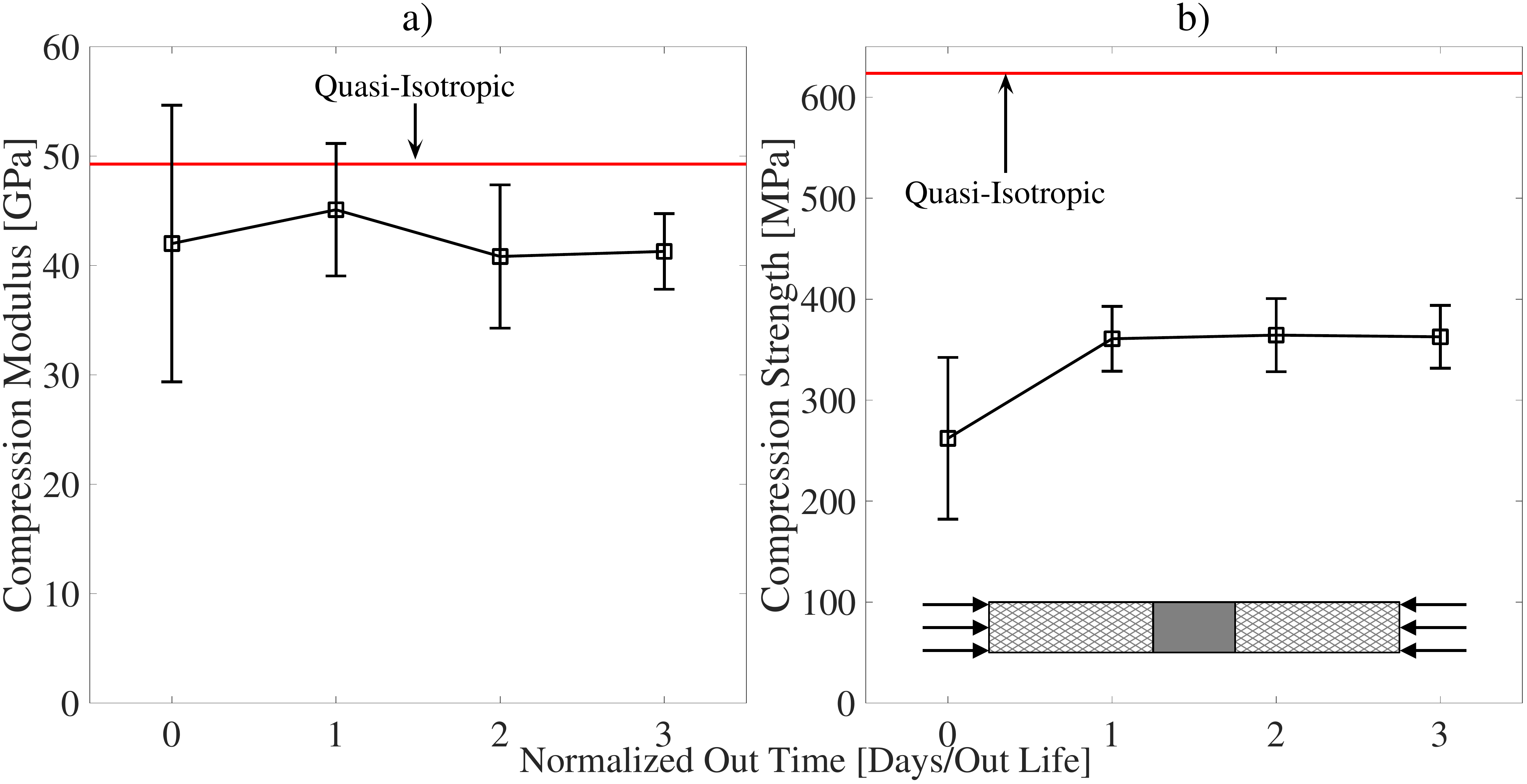}
    \caption{Effect of out time on the compression a) modulus and b) strength.}
    \label{fig:Comp}
\end{figure}

Figure \ref{fig:Comp} and table \ref{tab:Comp} provide the compression modulus and strength for the quasi-isotropic and DFC specimens. It can be seen in fig. \ref{fig:Comp}a that the modulus does not change significantly with age. There is also little difference in the modulus when compared to the quasi-isotropic value, but there is more of a difference between the DFC and quasi-isotropic modulus than what was seen with the tensile modulus. The modulus for the DFC specimens remain at $83-92\%$ that of the quasi-isotropic modulus with the largest difference in modulus occurring at out time $2\times$, which is $82.9\%$ that of the quasi-isotropic modulus. In any case, this modulus still falls within one standard deviation of the non-aged DFC specimens confirming a minor effect of aging on the elastic behavior of DFCs.

In fig. \ref{fig:Comp}b, it can be seen that the strength of the DFC specimens is from $42\%$ to $58\%$ that of the quasi-isotropic laminate. It can also be seen that, there is an increase in compressive strength of the DFC specimens when the prepreg is aged with a difference as high as a $33\%$ between aged and non-aged DFCs. This is similar to what is seen for the tension specimen.

\subsection{Shear}
\begin{figure}[htb!]
    \centering
    \includegraphics[width=1\textwidth]{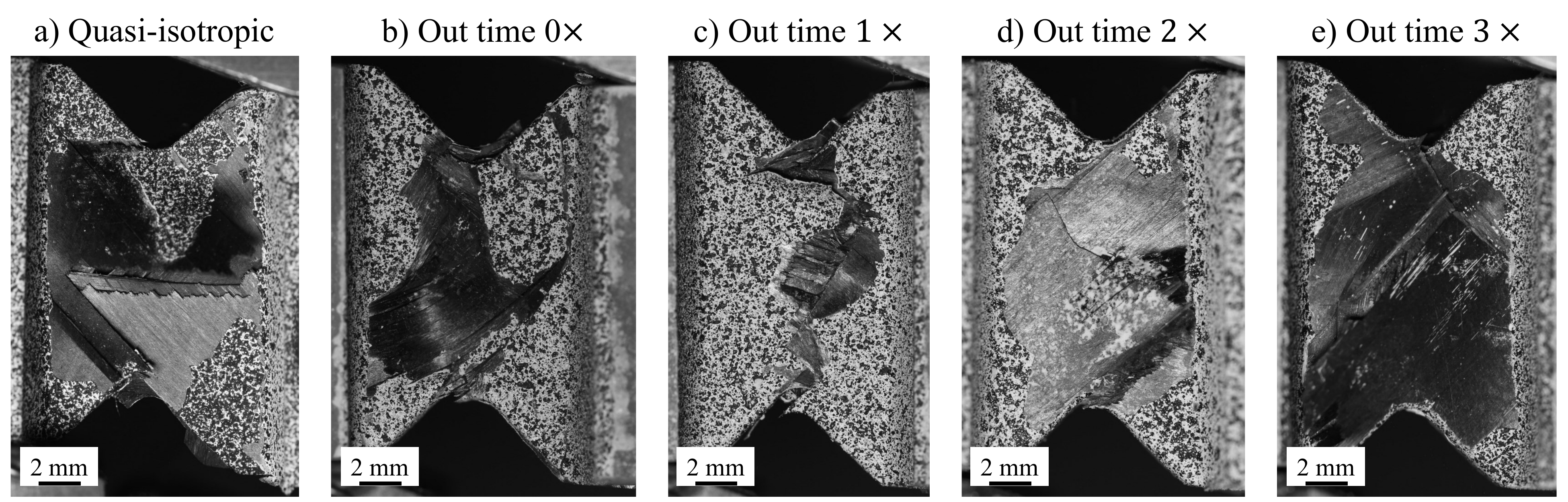}
    \caption{Fracture surfaces of quasi-isotropic, and DFC shear specimens out times $0\times$, $1\times$, $2\times$, and $3\times$. DFC specimens mainly exhibit matrix damage and debonding, with some fiber damage.}
    \label{fig:ShearFrac}
\end{figure}

Figure \ref{fig:ShearFrac} shows examples of fracture surfaces of quasi-isotropic, and DFC shear specimens for out times $0\times$, $1\times$, $2\times$, and $3\times$. It can be noted that the quasi-isotropic samples had fiber breakage at the surface $45^\circ$ ply. However, initiation of failure occurred at the notch and propagated along the fibers. For DFC specimens, the majority of the failure occurred at the platelet boundaries, similar to the tension and compression specimens. There was no perceivable difference in failure mechanisms between out times and it is worth mentioning that similar failures of non-aged shear specimen were reported previously by Selezneva et al. \cite{SeleznevaTenCompShear}.

\begin{figure}[htb!]
    \centering
    \includegraphics[width=0.9\textwidth]{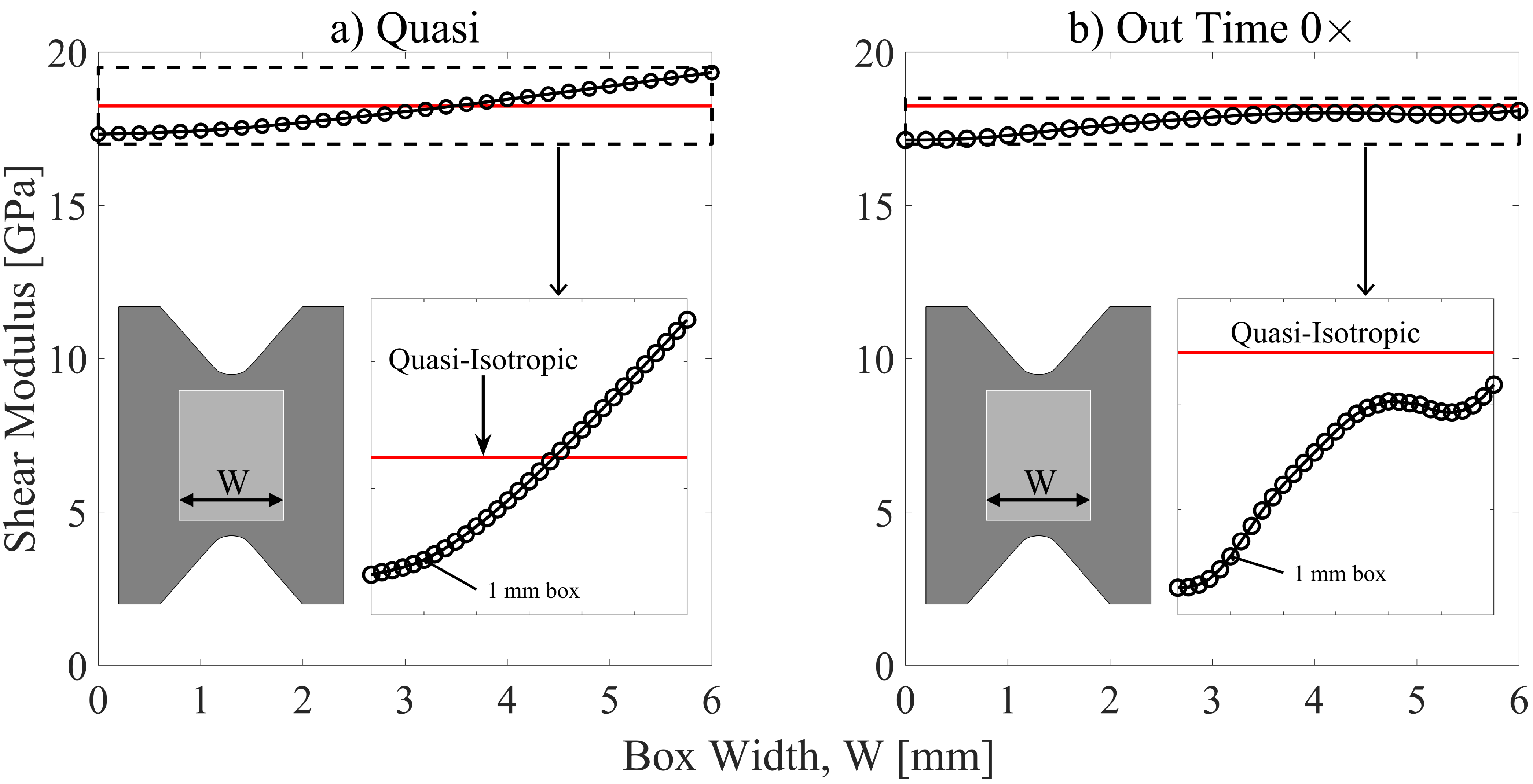}
    \caption{Change of calculated shear modulus when changing the the width of the box where the shear strain from DIC analysis is averaged. The box has a height such that it is $1$ mm from the notch. The red line is the average shear modulus of the quasi-isotropic specimens.}
    \label{fig:ShearBox}
\end{figure}

For the calculation of the shear strains during the tests, a thorough investigation was performed to identify the area of analysis providing the most consistent and accurate results. Figures \ref{fig:ShearBox}a,b show the effect of the size of the area used for the calculation of the shear modulus. It can be seen in fig. \ref{fig:ShearBox}a that the shear modulus does not change much when the box width is less than $1$ mm. But, after this point the modulus starts to increase rapidly. At $1$ mm, the percent increase of the shear modulus calculated is only $0.68\%$. The increase after $1$ mm is from the fact that only the center of the specimen is in pure shear. When looking at a DFC specimen, fig. \ref{fig:ShearBox}b, a similar trend can be seen. At $1$ mm the percent increase in modulus is only $0.87\%$. Even though the area of pure shear will be the same for the DFC and quasi-isotropic specimen, the modulus change within that width will look different for DFC specimens, since DFCs' mesostructure is inhomogeneous. This is confirmed by the fact that the shear modulus increase within the $1$ mm box is different when compared to the quasi-isotropic layup. Another observation showing the inhomogeneous mesostructure of DFCs' is that, when the box width is between about $4-5$ mm, the modulus decreases which does not happen for the quasi-isotropic specimen. From the quasi-isotropic layup the shear strain does not change within the $1$ mm area at the center, thus it can be assumed that this area is in pure shear and is a good area to average shear strain over for all the specimens. This size was used for the analysis of all the shear specimens investigated in this work.

\begin{figure}[htb!]
    \centering
    \includegraphics[width=1\textwidth]{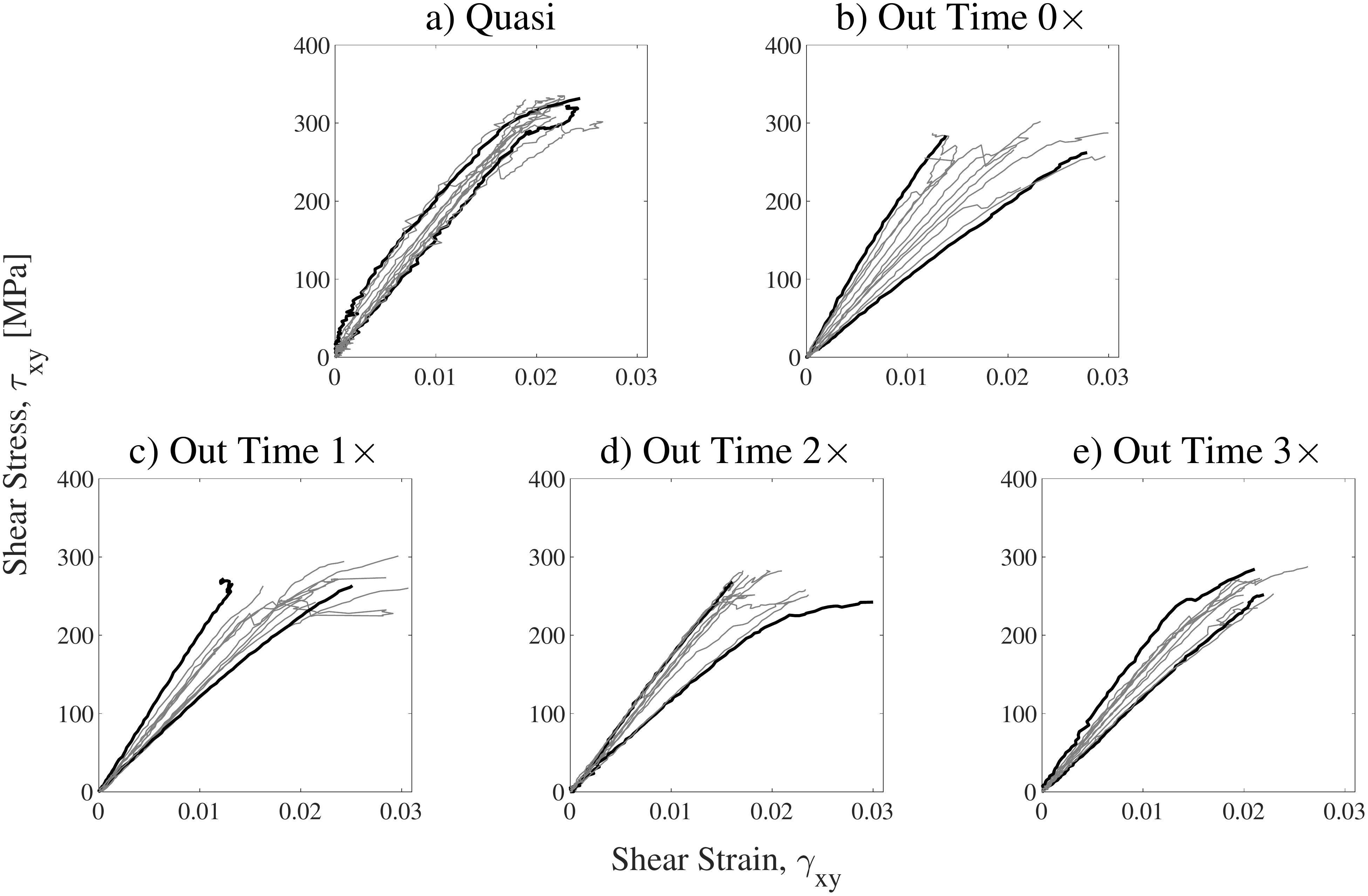}
    \caption{Stress vs strain curves for quasi-isotropic and out times $0\times$, $1\times$, $2\times$, and $3\times$ shear specimens.}
    \label{fig:Shear-stress-strain}
\end{figure}

Figure \ref{fig:Shear-stress-strain} shows the stress vs strain curves for quasi-isotropic, and DFC samples made from aged prepregs with out time $0\times$, $1\times$, $2\times$, and $3\times$. The post peak behavior is not captured here because after the peak is reached, the DIC becomes too distorted to give accurate results. However it should be noted that most of the specimens exhibited progressive damage and did not fail when the ultimate strength was reached. This is because when one ply or platelet fails there are still others that are able to carry load. There were many specimens that were linear up to the peak, however for the quasi-isotropic specimen and all the DFC specimens, there were $3$ out of $12$ that exhibited non-linearity prior to the peak. The nonlinear behavior can be ascribed to sub-critical damage dissipating energy before reaching the ultimate load.

\begin{figure}[htb!]
    \centering
    \includegraphics[width=0.9\textwidth]{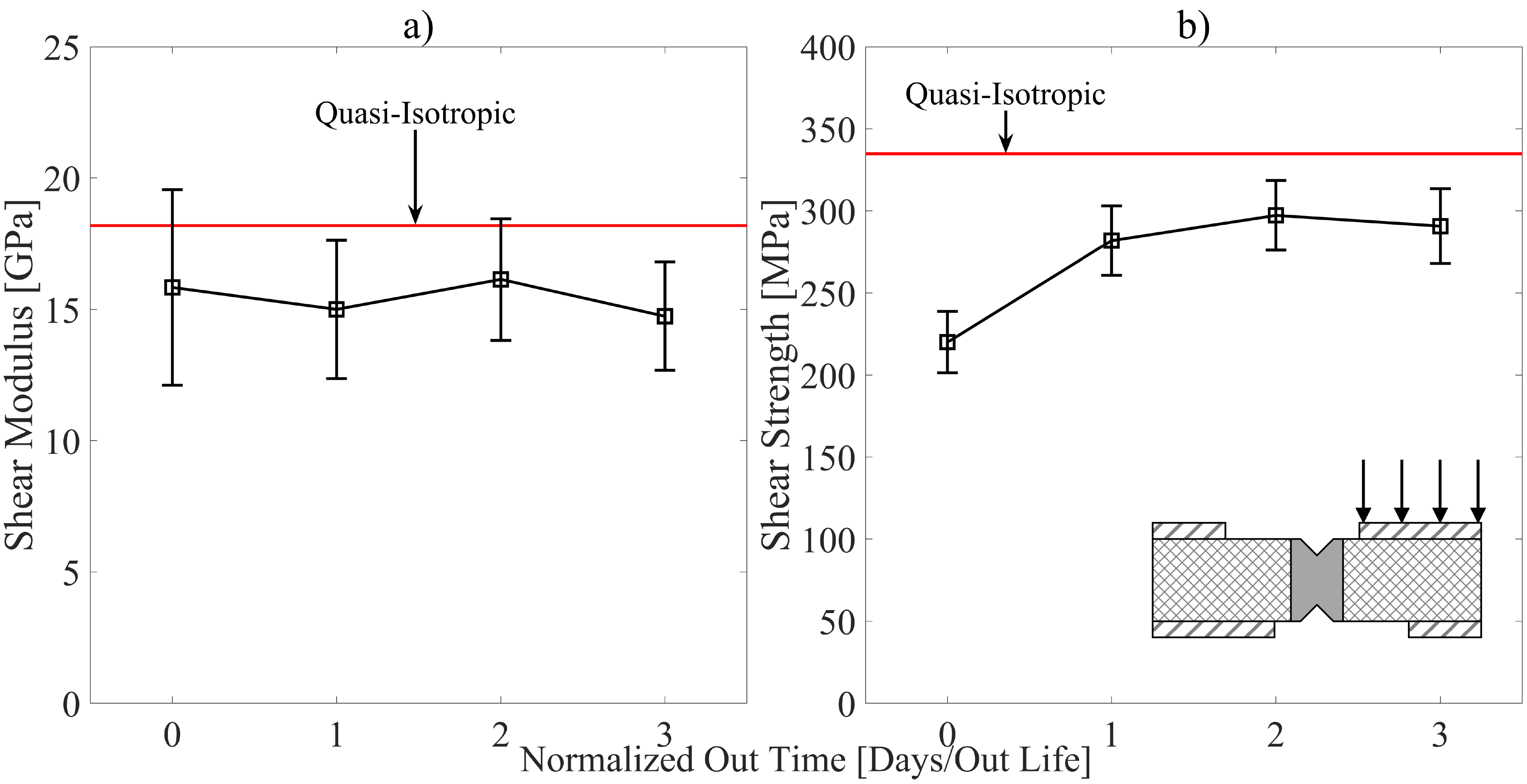}
    \caption{Effect of out time on the shear a) modulus and b) strength.}
    \label{fig:Shear}
\end{figure}

Figure \ref{fig:Shear} and table \ref{tab:Shear} show the shear modulus and strength for the quasi-isotropic and aged DFC specimens. As for the case of tension and compression, fig. \ref{fig:Shear}a shows that also the shear modulus does not change significantly with age. There is also little difference in the modulus when compared to the quasi-isotropic value, but the difference between the DFC and quasi-isotropic modulus is larger than what was seen with the tensile modulus. The modulus for the DFC specimens remains at $81-89\%$ that of the quasi-isotropic modulus. The largest difference in modulus occurs at out time $3\times$, which is $81.06\%$ that of the quasi-isotropic modulus although this modulus still falls within one standard deviation of the non-aged DFC specimens.

In fig. \ref{fig:Shear}b, it can be seen that the strength of the DFC specimens is $66-89\%$ that of the quasi-isotropic laminate. This is higher than that of the tension and compressive strengths. In fact, for tension and compression the strength on non-aged DFC was about $40\%$ that of the quasi-isotropic specimen, but it is $66\%$ for shear. However the increase in strength of the aged specimens compared to the non-aged ones is similar to the one reported for the tension and compression specimens. There was as high as a $30\%$ difference for the non-aged to aged DFC specimen. Similar to the tension and compression tests there was little change in strength past out time $1\times$.

\subsection{Size effect tests}
\begin{figure}[htb!]
    \centering
    \includegraphics[width=1\textwidth]{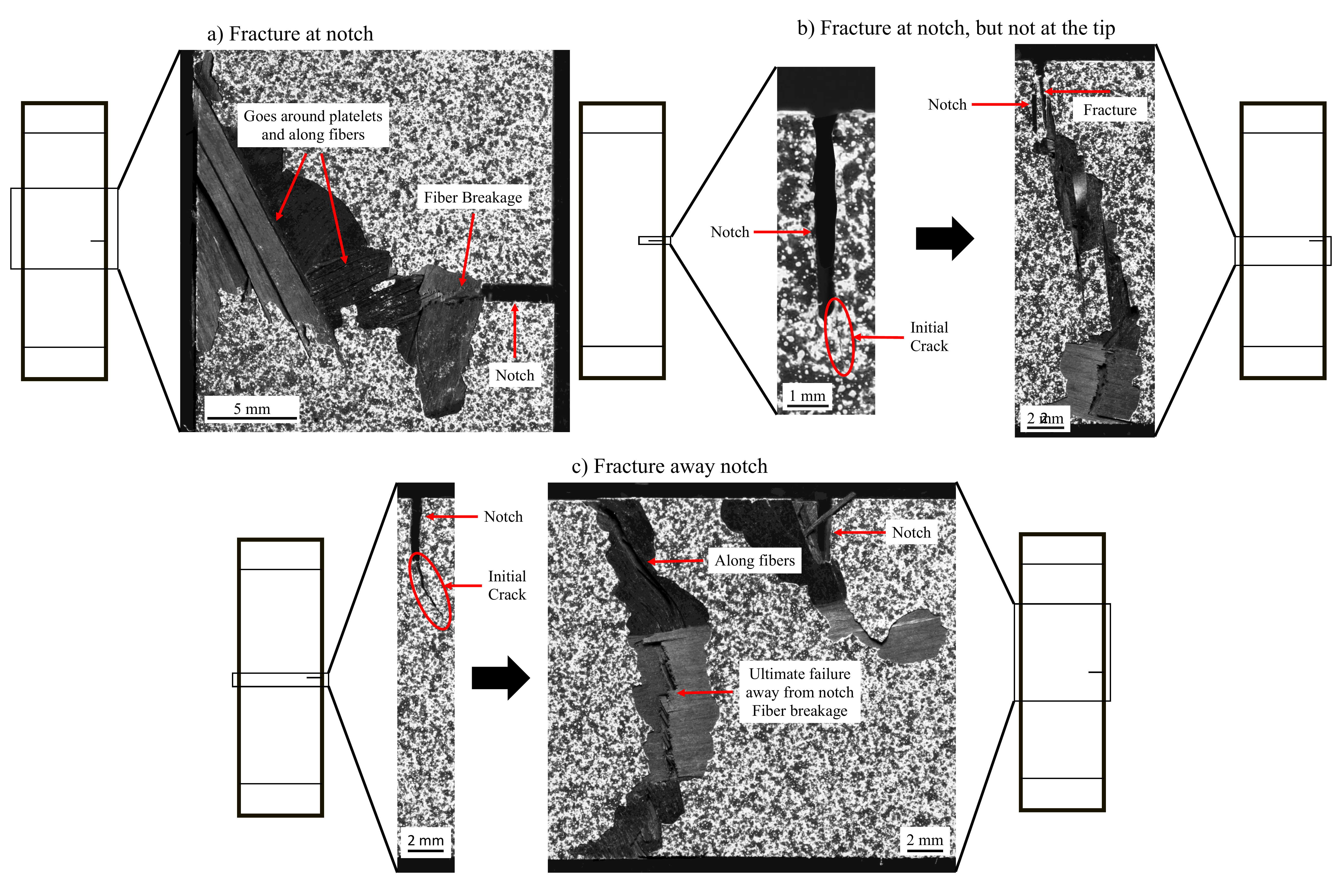}
    \caption{Three different types of fracture surfaces of SENT specimens, which mainly exhibit matrix damage and debonding, with some fiber damage.}
    \label{fig:SEFrac}
\end{figure}

Figures \ref{fig:SEFrac}a-c show the three types of fracture surfaces of SENT Discontinuous Fiber Composite specimens. Figure \ref{fig:SEFrac}a shows a fracture that occurred at the notch. Similar to three mechanical tests, matrix damage and debonding were much more common than fiber breakage.
Figure \ref{fig:SEFrac}b shows a fracture that happened at the notch, but not at the tip. This type of failure occurred in two specimens, one in out time $1\times$, size $3$, and one in out time $2\times$, size $2$. There was initial damage at the notch tip, however due to the size of the fracture process zone (FPZ), damage happened to platelets not directly touching the notch tip. Thus when the platelet failed, the fracture path was not connected to the notch tip. 

Figure \ref{fig:SEFrac}c shows fracture that occurred away from the notch. It can be seen that there was damage at the notch before ultimate failure, however the ultimate failure occurred away from the notch. It was seen for out time $1\times$, $1$ out of $8$ specimen failed away from the notch, and for out time $2\times$ and $3\times$, $2$ out of $8$ specimens failed away from the notch. For the two larger sizes all the specimen failed at the notch. The specimens that failed away from the notch or at the notch but not at the tip, showed no difference in nominal strength when compared to specimen that failed at the notch. Previous studies \cite{SeungSAMPE, SeungPlateletSE, SeungThicknessSE, FeraboliNotch, QianNotch} have seen similar fracture behaviors presented here in non-aged DFCs.

\begin{figure}[htb!]
    \centering
    \includegraphics[width=1\textwidth]{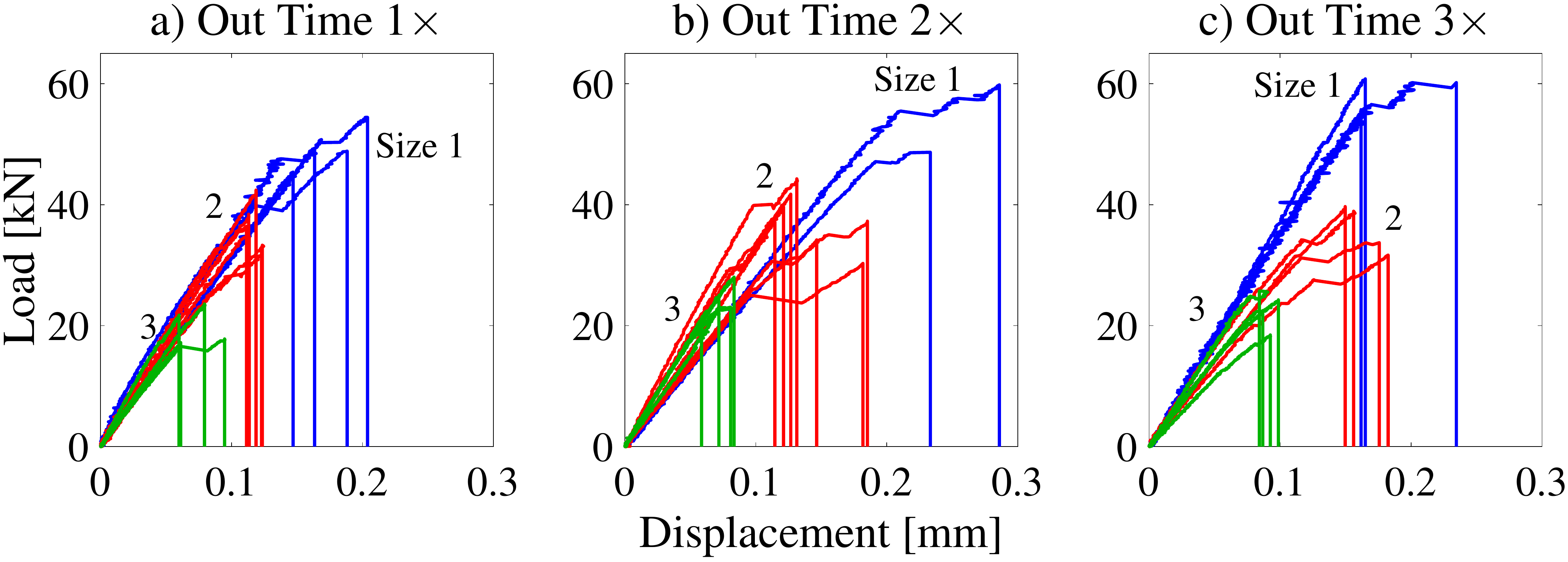}
    \caption{Load vs displacement curves for out times $1\times$, $2\times$, and $3\times$ SENT specimens.}
    \label{fig:SE-load-disp}
\end{figure}

Figure \ref{fig:SE-load-disp} shows the load vs displacement curves for out time $1\times$, $2\times$, and $3\times$ DFC specimens. The displacements used for the load vs displacement curves were found using DIC. The nominal displacement was calculated by averaging the relative displacement between two horizontal lines spanning the width of the specimen. The distance between the lines and the notch was taken to be $1.2D$. This was done to remove the effects of the compliance of the machine. The stiffness of each specimen was seen to be relatively the same. Most of the specimens had a linear behavior up to the peak, however, some specimens in out time $3\times$ size $3$ exhibited minor non-linearity. This nonlinear behavior comes from damage in the FPZ.

\begin{figure}[htb!]
    \centering
    \includegraphics[width=1\textwidth]{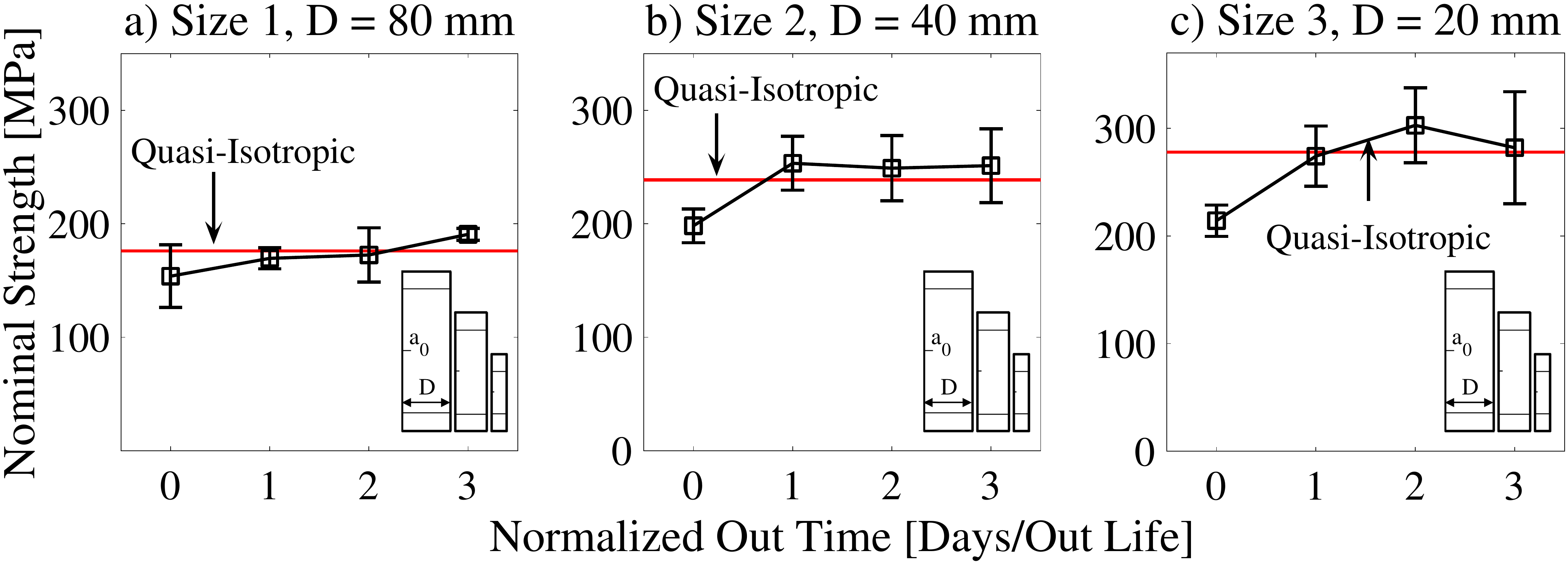}
    \caption{Nominal strength ($\sigma_N = P_c/D/t$) of SENT specimen vs normalized out times for a) Size $1$, b) Size $2$, and c) Size $3$ specimen. Quasi-isotropic and non-aged DFC results are taken from Ko et al. \cite{SeungPlateletSE}.}
    \label{fig:SE-Strength}
\end{figure}

Following e.g. \cite{SalviatoTextile, SeungPlateletSE, SeungThicknessSE}, the nominal strengths of the specimens were defined as $\sigma_{Nc}=P_c/tD$, where $P_c$ is the peak load found during tests, $t$ is the specimen thickness, and $D$ is the specimen width. Figure \ref{fig:SE-Strength} and table \ref{tab:SizeEffect} report the nominal strength for the DFC and quasi-isotropic specimens. It can be seen that there was an increase in strength with aging, similar to the mechanical tests. In this case, the aged DFC specimens have a nominal strength similar to that of the quasi-isotropic specimens. At the same time a size effect can be seen where the strength of the SENT specimen increases with a decrease in specimen width.

\section{Size Effect Analysis}
In this study, Type $2$ size effect of aged DFCs is investigated. This is quintessential to obtain an accurate estimation of the fracture energy of the material, which is an important measure of the capability of the material of resisting crack propagation by dissipating energy. 

Type $2$ size effect deals with structures featuring a stress-free crack or notch \cite{Salbook,Baz04}. Due to the complex mesostructure of DFCs, significant stress redistribution occurs during the damage process \cite{SeungPlateletSE, SeungThicknessSE}. This redistribution is characterized by a nonlinear Fracture Process Zone (FPZ) whose size is proportional to the size of the largest inhomogeneity of the material. For increasing structure sizes, the percentage of the structure subject to the stress redistribution occurring in the FPZ gets smaller and smaller, leading to significant size effects. Typically, large quasibrittle structures experience limited stress redistribution prior to failure and they behave in a rather brittle manner. In contrast, small structures compared to the characteristic size of the FPZ incur 
significant nonlinear stress redistribution which reduce the severity of the notch and generally lead to a more quasiductile behavior. These size effects have been confirmed for a number of quasibrittle materials including e.g. fiber and 2D/3D textile composites \cite{SalviatoTextile, LiQia21, SalKir19, SalEsn16}, polymers \cite{QiaSal19}, nanocomposites \cite{MefQia17, QiaSal19b}, concrete \cite{BazPfe87, CarCus19}, metals \cite{NguDon21}, and many other materials \cite{Salbook}.

Equivalent fracture mechanics, which was pioneered by Irwin \cite{Irwin} and extended to quasibrittle materials \cite{Bazant}, can be used to account for these effects provided that the FPZ is still not too large compared to the structure size. Towards this goal, an additional effective FPZ length, $c_f$, is added to the original crack length, $a_0$. The additional length is added such that the resultant stress of the effective crack equals the ones related to the cohesive stresses in the FPZ.

The effective FPZ size, $c_f$, depends on how the elastic energy in the FPZ is being dissipated. In DFCs, some mechanisms for energy dissipation are fiber fracture, platelet pullout, platelet delamination, and matrix microcracking \cite{SeungPlateletSE, SeungThicknessSE}. These mechanisms are all influenced by the platelet geometry, orientation, and the number of platelets through the thickness. Thus, the mesostructure must be properly accounted for to capture the fracture behavior and the effects of the nonlinear FPZ.

If the mesostructure and quasibrittle softening laws are calibrated properly, the progressive damage in a structure can be modeled explicitly and the $c_f$ can be predicted. However, with equivalent fracture mechanics, $c_f$ can be found experimentally through size effect analysis. Doing so means the progressive damage does not need to be modeled explicitly, but the effects of the mesostructure still needs to be captured. Through finite element modeling, the effects of the mesostructure on the elastic strain energy can be captured. By combining the size effect experiments and finite element modeling the fracture behavior of DFCs can be characterized. The following sections provide an explanation of the analytical and computational framework used.

\subsection{Size Effect Law (SEL)}
As stated previously, to account for the nonlinear Fracture Process Zone (FPZ), an equivalent crack length is used: 
\begin{equation} \label{eq:crackL}
    a=a_0+c_f
\end{equation}
where $a_0$ is the original crack length, and $c_f$ is the effective FPZ length, which is treated as a material property.

From Linear Elastic Fracture Mechanics (LEFM), the energy release rate as a function of the dimensionless crack length is:
\begin{equation} \label{eq:G}
    G(\alpha) = \frac{\sigma_N^2D}{E^*}g(\alpha)
\end{equation}
with $\alpha=a/D$ being the dimensionless crack length, $\sigma_N=P/(tD)$ being the nominal stress, $E^*$ being the effective elastic tensile modulus, and $g(\alpha)$ being the dimensionless energy release rate.  The dimensionless energy release rate accounts for the effects of geometry on the energy release rate. For structures that are homogeneous, $g$ depends only on the geometry of the structure and is constant for geometrically-scaled specimens. However, DFC inhomogeneities, which depend on the platelet's geometry and are not geometrically scaled, can be comparable with the size of the structures. Thus different structure sizes may lead to significantly different energy release rates potentially making $g$ dependent on the structure size, $D$, and the thickness, $t$. We can write eq. \ref{eq:G} to account for the inhomogeneity of DFCs as follows:
\begin{equation} \label{Gd}
    G(\alpha,D) = \frac{\sigma_N^2D}{E^*}g(\alpha,D)
\end{equation}
where $g$ is now considered a function of both the dimensionless crack length and the characteristic length of the structure.

At the onset of fracture, the energy release rate, $G$, must be equal to the fracture energy, $G_f$, assumed to be a material property. By substituting eq. \ref{eq:crackL} into eq. \ref{eq:G}, $G_f$ can be expressed in terms of the effective crack length:
\begin{equation} \label{eq:Gf}
    G_f = G(\alpha_0+c_f/D,D) = \frac{\sigma_N^2D}{E^*}g(\alpha_0+c_f/D,D)
\end{equation}
Performing a Taylor expansion around $\alpha_0$ for a constant $D$ we get:
\begin{equation} \label{eq:Gftaylor}
    G_f = \frac{\sigma_N^2D}{E^*}\left[g(\alpha_0,D) + \frac{c_f}{D}\frac{\partial{g}}{\partial\alpha}(\alpha_0,D)\right]
\end{equation}
By rearranging eq. \ref{eq:Gftaylor}, Ba\v{z}ant's Size Effect Law (SEL) is obtained \cite{SEL}:
\begin{equation} \label{eq:SEL}
    \sigma_{Nc} = \sqrt{\frac{E^*G_f}{Dg(\alpha_0,D)+c_fg'_D(\alpha_0,D)}}
\end{equation}
where, $g'_D = [\partial{g}/\partial\alpha]_D$. Here, the subscript $D$ denotes that the partial differentiation is taken for a constant structure size. It should be noted that, in contrast to the traditional SEL \cite{SEL}, the dimensionless energy release rate in eq. \ref{eq:SEL} is a function of the structure size and can be calculated via stochastic finite element analysis. Unlike LEFM, eq. \ref{eq:SEL} depends both on the structure size and the material characteristic length. This is needed to capture the transition of the fracture behavior from quasi-ductile to brittle. It is worth noting that Eq. \ref{eq:SEL} can also be written as follows:
\begin{equation} \label{eq:SigFracConst}
    \sigma_{Nc} = \frac{\sigma_0}{\sqrt{1+D/D_0}}
\end{equation}
where, $\sigma_0 = \sqrt{E^*G_f/c_fg'_D(\alpha_0,D)}$ and $D_0 = c_fg'_D(\alpha_0,D)/g(\alpha_0,D)$ are the size effect constants depending on the structure geometry and FPZ size.

As discussed in following (Section \ref{sec: g and g'}), extensive computational studies were performed to investigate the dependence of the dimensionless energy release rate on the structure size $D$. It was found that, for the particular DFC systems investigated in this work the structure size, $D$, has no significant effect on the dimensionless energy release functions. This result agrees with previous work done by Ko et al. \cite{SeungSAMPE, SeungPlateletSE, SeungThicknessSE} which showed that, when the average number of platelets through the thickness of the structure is sufficiently large, the mesostructure becomes statistically homogeneous and the energy release rate has no dependence on structure size. Therefore, the dimensionless energy release rates were considered to be size independent in this work and the average $g$ and $g'_D$ values of all specimen sizes for a given out time were used in eq. \ref{eq:SEL}. However, it should be noted that for another platelet geometry or specimen thickness, $g$ and $g'_D$ may become dependent on the structure size and the dimensionless functions calculated for each size should be used instead of the averaged values.

\subsection{Fitting of the experimental data using SEL}
To obtain the size effect constants from experiments, linear regression analysis was performed as shown in fig. \ref{eq:linReg}. In fact, Eq. \ref{eq:SigFracConst} can be written in a linear form \cite{Bazant,Salbook}:
\begin{equation} \label{eq:linReg}
    Y = C+AX
\end{equation}
using the following transformation:
\begin{equation} \label{eq:linRegConst}
    X = D, \quad Y = \sigma_{Nc}^{-2}, \quad \sigma_0 = C^{-1/2}, \quad D_0 = \frac{C}{A}
\end{equation}

\begin{figure}[htb!]
    \centering
    \includegraphics[width=0.8\textwidth]{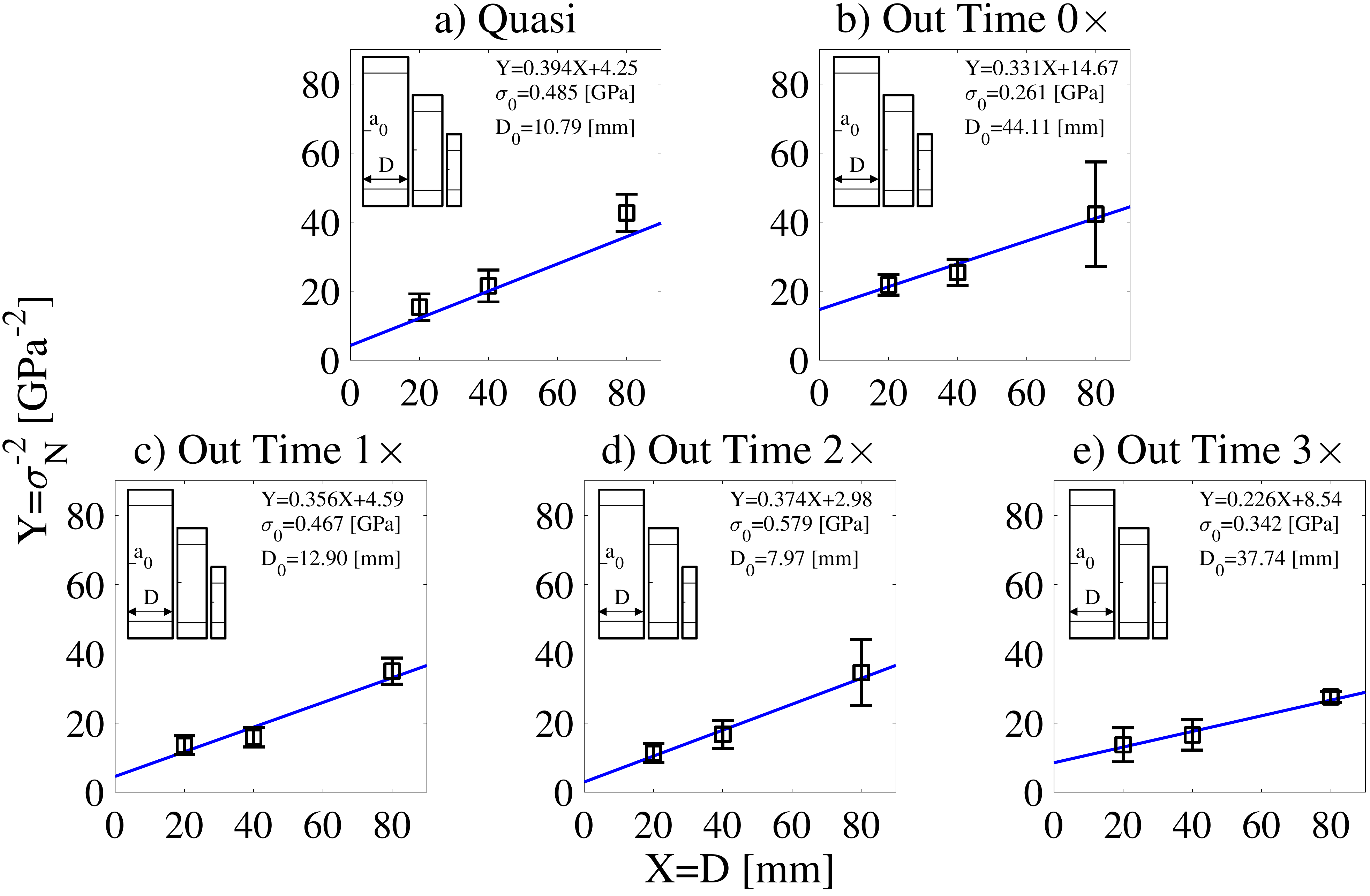}
    \caption{Linear regression analysis to find the size effect constants, $\sigma_0$ and $D_\circ$, for quasi-isotropic, non-aged DFC, and aged DFC specimen. Quasi-isotropic and non-aged DFC results are taken from Ko et al. \cite{SeungPlateletSE}}
    \label{fig:Lin-Reg}
\end{figure}

As can be noted from fig. \ref{fig:Lin-Reg} and eq. \ref{eq:linRegConst}, the size effect constants are obtained from the slope and the y-intercept providing the input for the construction of the size effect curves using eq. \ref{eq:SigFracConst}.

\begin{figure}[htb!]
    \centering
    \includegraphics[width=0.8\textwidth]{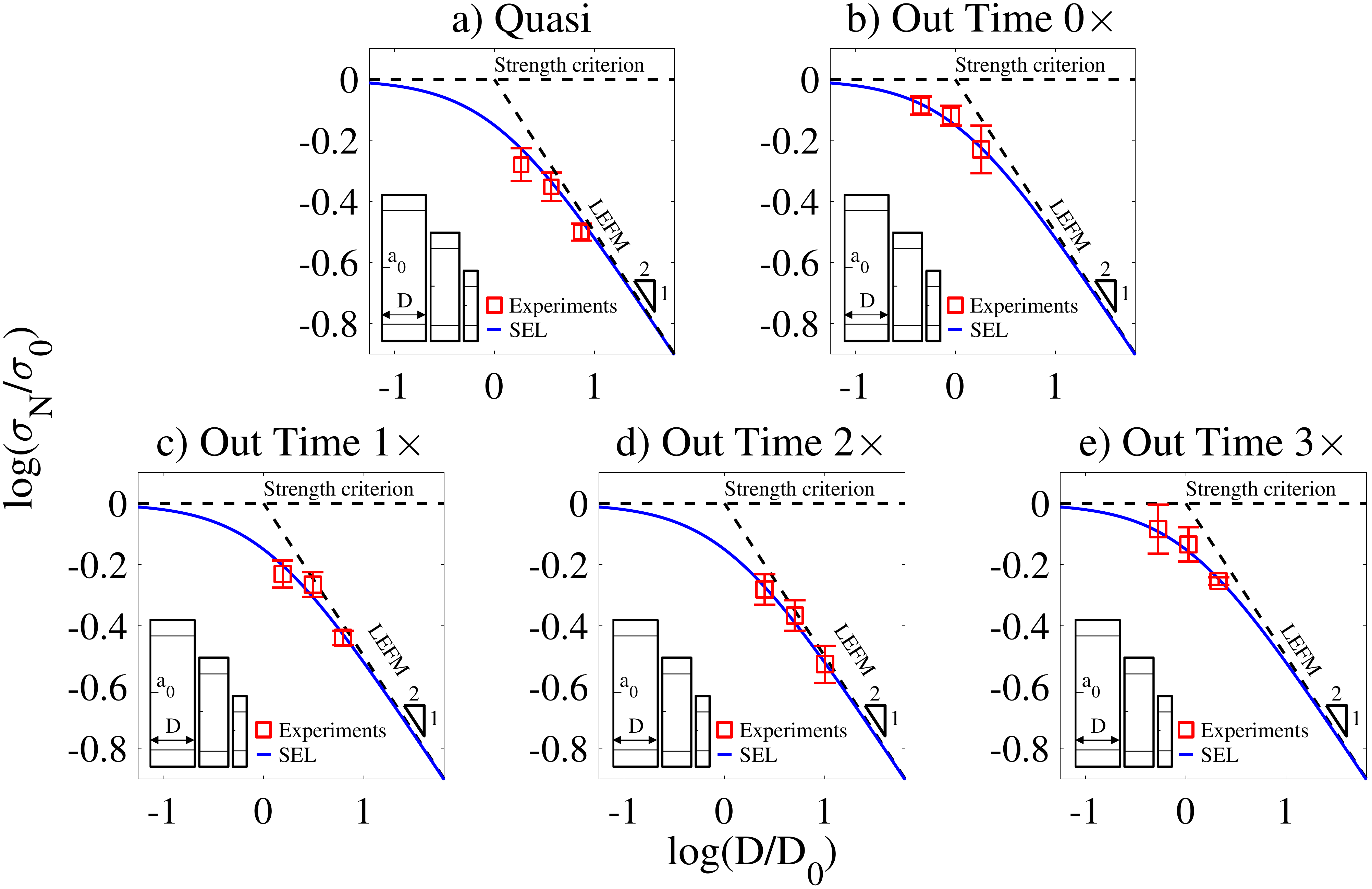}
    \caption{Normalized size effect curves for quasi-isotropic, non-aged DFC, and aged DFC specimens. Y-values are normalized to the size effect constant $\sigma_0$, and the x-values are normalized to the size effect constant $D_0$. Quasi-isotropic and non-aged DFC results are obtained from Ko et al. \cite{SeungPlateletSE}}
    \label{fig:SEL}
\end{figure}

Figure \ref{fig:SEL} shows the normalized size effect curves found using eq. \ref{eq:SigFracConst} and \ref{eq:linRegConst}. Here, the log of the nominal stress normalized to the size effect constant $\sigma_0$ is plotted against the log of the structural width $D$ normalized to the size effect constant $D_0$. As can be noted, all out times exhibit a deviation from LEFM. It can be seen that the specimens tested are in the transition between the horizontal asymptote, which dictates stress driven failure, and the asymptote with a slope of $-1/2$, where failure is fully energy driven and LEFM is valid. This can be attributed to the size of the Fracture Process Zone (FPZ) compared to the structure size. When the structure size is sufficiently small, the FPZ impacts the structural behavior and causes deviation from LEFM. As the structure size increases, the FPZ has less impact on the structural behavior and size effect can be captured using LEFM. It is interesting to note that out time $1\times$ and $2\times$ are closer to the LEFM region and are comparable to the quasi-isotropic specimen. Then once out time $3\times$ is reached, the specimen becomes comparable to out time $0\times$ and becomes more quasi-ductile.

\subsection{Brittleness number}
\begin{figure}[htb!]
    \centering
    \includegraphics[width=0.5\textwidth]{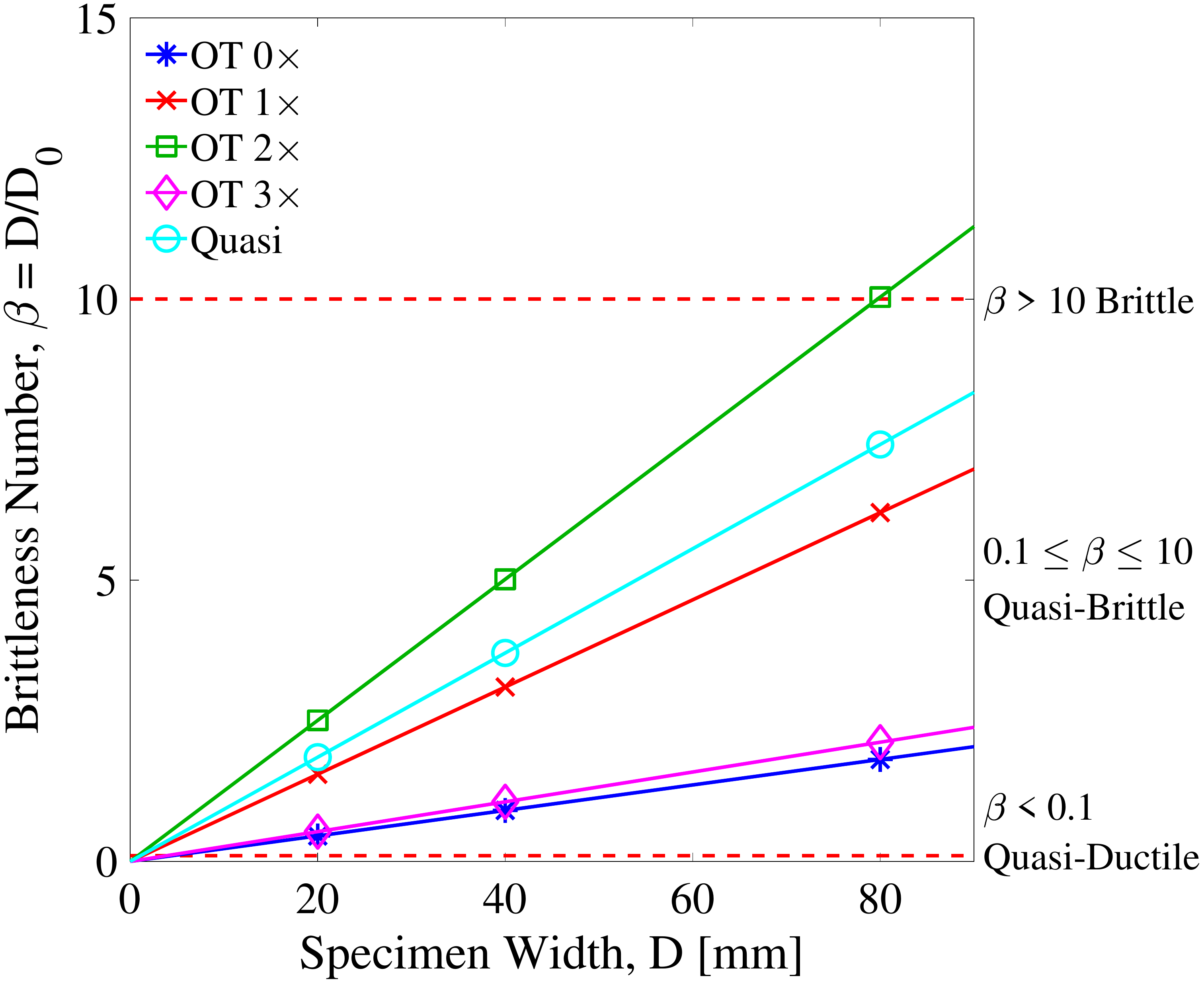}
    \caption{Brittleness number, $\beta$, vs the structure size for quasi-isotropic, non-aged DFC, and aged DFC specimens. Quasi-isotropic and non-aged DFC results are taken from Ko et al. \cite{SeungPlateletSE}}
    \label{fig:Brittleness}
\end{figure}

To compare the structural behavior of aged DFCs, non-aged DFCs, and traditional composites, a non-dimensional parameter called the brittleness number, $\beta$, can be used \cite{Bazant}. This number compares the brittleness of structures with similar geometry and size and is defined as the ratio between the characteristic size of the structure, $D$, and the size effect constant $D_0$. When $\beta$ is greater than $10$, then the behavior is brittle and LEFM is suitable to capture the fracture behavior. When $\beta$ is less than $0.1$, the behavior is quasi-ductile or perfectly plastic and a strength based failure criteria can predict the behavior. If $\beta$ is between these two points, the structure is quasi-brittle. Figure \ref{fig:Brittleness} shows $\beta$ for a quasi-isotropic laminate and DFC laminates at the various out times (the quasi-isotropic and non-aged DFC data is taken from Ko et al. \cite{SeungPlateletSE}). It can be seen that as aging increases the brittleness of DFCs increase up to a certain point. Once this point is met, the brittleness decreases and becomes comparable to the non-aged DFC. It can be seen that for out time $2\times$, $\beta$ is larger than the quasi-isotropic specimen, and for the largest size tested, the brittleness number leaves the quasi-brittle zone. This change in brittleness will change the damage tolerance of DFC material.

\subsection{Stochastic finite element model}
In order to obtain the fracture energy, $G_f$, and the FPZ length, $c_f$, the dimensionless energy release rate, $g(\alpha_0,D)$, and its derivative, $g'_D(\alpha_0,D)$, must be found. These values are highly dependent on the platelet constitutive properties and the random distribution of the platelets. To capture this, a stochastic finite element model is used.

\subsubsection{Mesostructure generation} \label{sec:MesoGen}
The mesostructure generation used in this work is an extension of the stochastic laminate analogy method proposed in \cite{MesoGen1, MesoGen2, MesoGen3}. A brief summary of the generation algorithm is provided here, and more details on the algorithm and its implementation can be found in \cite{SeungThicknessSE, SeungPlateletSE, KoSampe20, KoASC21}. In this study platelets are partitioned into grids of $1\times1$ mm.

The platelet generation algorithm can be divided into two parts. The first is a platelet distribution algorithm and the second is a thickness adjustment algorithm. In this study, the platelets are assumed to be perfectly randomly distributed and planar. This means that a uniform probability distribution is used for both the spacial component and orientation of the platelet, and the out-of-plane orientation is assumed to be zero. From the mesostructure investigation we have the average number of platelets through the thickness and the CoV of the laminates for the different out times which are used as inputs to the platelet generation algorithm. To achieve these parameters, saturation points and platelet limits are used to guide generation. Saturation points are set to be every three layers. The average number of platelets through the thickness must equal the current saturation point before moving on to the next saturation point. The platelet limits are used to prevent a higher concentration of platelets in certain areas. A higher concentration would lead to the average number of platelets to equal the saturation point, but there would be large areas of little to no platelets. These limits are taken to be the saturation points times the CoV.

Once the average number of platelets generated meets the one found from the mesostructure study, a thickness adjustment is performed. This is done to simulate the effects of resin flow, without modeling it explicitly. If the number of platelets through the thickness is greater than or equal to the average number of platelets found experimentally, then the thickness of the platelet will be evenly distributed to all the platelets. If the number of platelets though the thickness is less than the one found experimentally, then resin will fill the additional thickness needed. This is done to mimic resin flow. For more information about the platelet generation algorithm the reader is referred to previous articles published by the authors \cite{KoSampe20, KoASC21, SeungThicknessSE, SeungPlateletSE}.

\subsubsection{Computation of g($\alpha$) and g'($\alpha$)} \label{sec: g and g'}
The mesostructure that is generated from section \ref{sec:MesoGen} is imported into Abaqus/Standard \cite{Abaqus}. Each partition in the mesostructure is homogenised into an 8-node, quadrilateral Belytschko-Tsay shell element (S8R). The platelets and resin are assumed to be linear elastic, with properties shown in table \ref{tab:MechPropSim}. At one end a uniform uni-axial displacement is applied and the other end is fixed in all directions. For homogeneous geometrically scaled specimens, $g(\alpha)$ and $g'(\alpha)$ do not change with the structure size \cite{Bazant}. However, this is not generally true for DFCs since they have an inhomogeneous mesostructure. For this reason, $5-8$ specimens for every size and out time are simulated to verify if any size effects on $g(\alpha)$ and $g'(\alpha)$ are present.

\begin{figure}[htb!]
    \centering
    \includegraphics[width=0.8\textwidth]{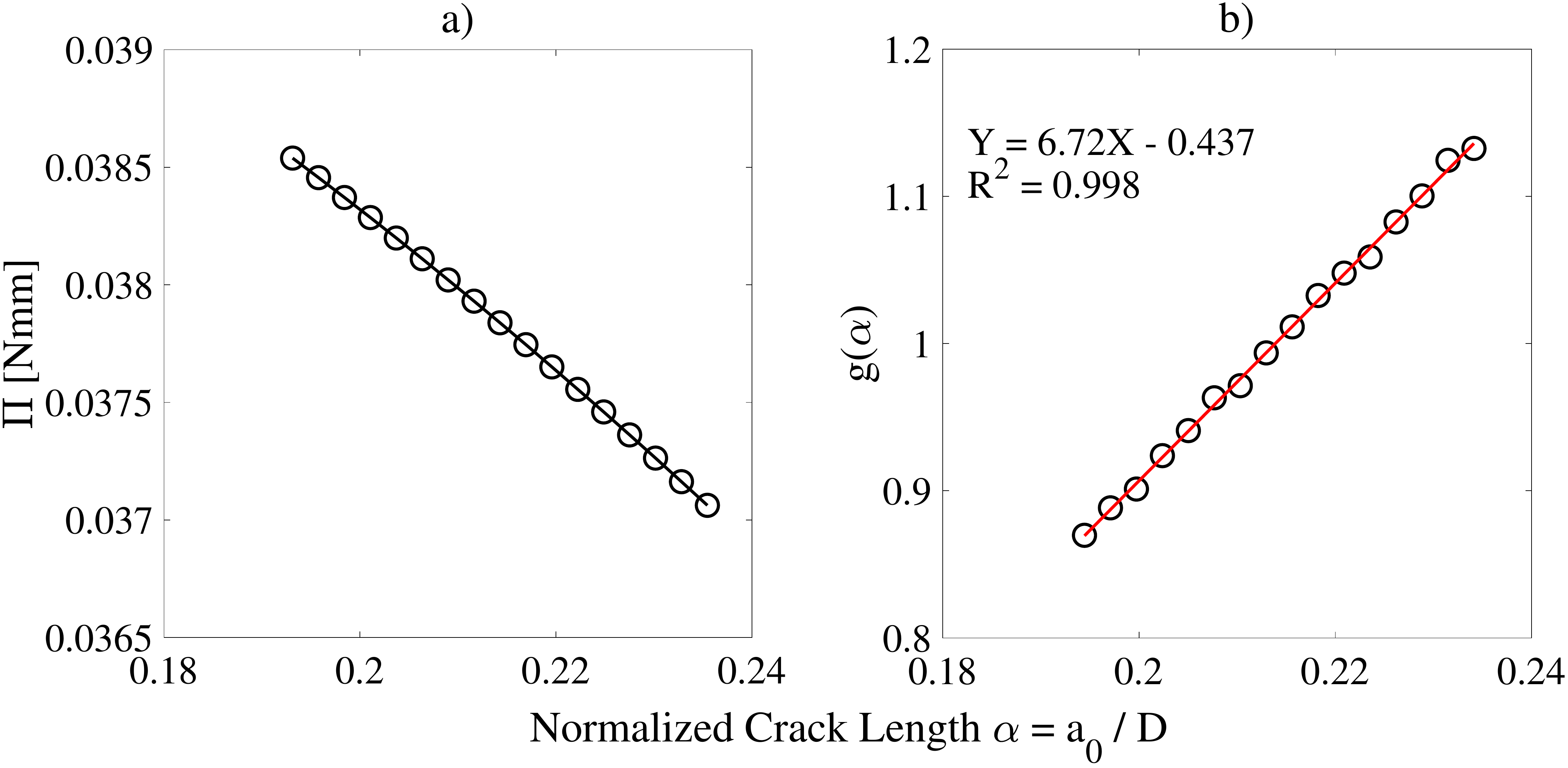}
    \caption{a) Strain energy as a function of the normalized crack length obtained from a typical SENT simulation. b) The calculated dimensionless energy release rate $g(\alpha_0)$ and its derivative $g'(\alpha_0$).}
    \label{fig:SE-Strain-Energy-g}
\end{figure}

Generally, a method to obtain the energy release rate is by using the J-integral \cite{RiceJint, SalviatoTextile}. This can not be done with DFCs because of their inhomogeneous mesostructure. To calculate $G(\alpha)$, its definition is used \cite{Bazant}:
\begin{equation}
    G(u,a) = -\frac{1}{b}\left[\frac{\partial\Pi(u,a)}{\partial{a}}\right]_u
\end{equation}
with $u$ being the applied displacement, $a$ being the crack length, $b$ the thickness, and $\Pi$ being the potential energy of the whole specimen. The subscript $u$ denotes that the potential energy is taken for a constant applied displacement. Figure \ref{fig:SE-Strain-Energy-g}a shows the potential energy of a typical DFC SENT specimen. To approximate $G(u,a)$ the central finite difference method is used to get the partial derivative of the potential energy as a function of the normalized crack length $\alpha = a/D$. Then, the dimensionless energy release rate, $g(\alpha)$, is found using eq. \ref{eq:G} and $g'(\alpha)$ can be found through linear regression as can be seen in fig. \ref{fig:SE-Strain-Energy-g}.

\begin{figure}[htb!]
    \centering
    \includegraphics[width=0.8\textwidth]{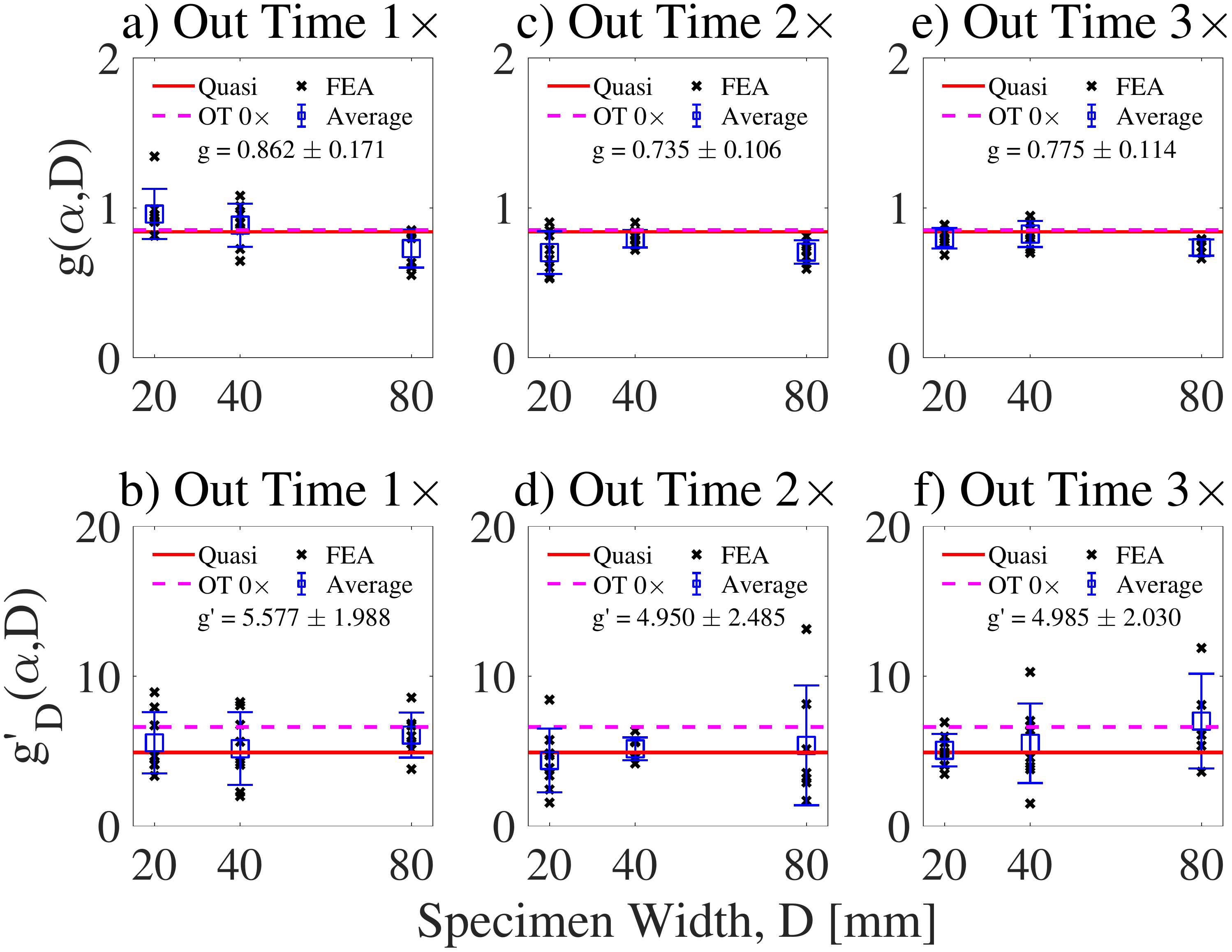}
    \caption{Comparison of dimensionless energy release rate and its derivative for the out times of a) and b) $1\times (28$ days), c) and d) $2\times (56$ days), and e) and f) $3\times (84$ days). Quasi-isotropic and non-aged DFC results are obtained from Ko et al. \cite{SeungPlateletSE}.}
    \label{fig:g and g'}
\end{figure}

Figure \ref{fig:g and g'} and table \ref{tab:Gf cf g g'} show the dimensionless energy release rate and its derivative at various out times for various sizes. It can be seen that $g$ and $g'$ do not change significantly with age or with structure size. In fact, the values of $g$ and $g'$ are about the same as the values for a quasi-isotropic laminate and for the non-aged DFC specimen. Since there is little change with respect to structure size, the average value of $g$ and $g'$ using results from all sizes are used for the fracture energy and characteristic length calculations. It is worth noting though that the average is not taken across out times since the mesostructure was not consistent through the various out times.

\subsection{Fracture energy and characteristic length}
With $g(\alpha_0)$ and $g'(\alpha_0)$ obtained from the previous section, the fracture energy $G_f$ and the FPZ length, $c_f$ can be calculated. By utilizing eq. \ref{eq:SEL} and \ref{eq:linReg} we get that:
\begin{equation}
    G_f = \frac{g(\alpha_0)}{E^*A}, \quad c_f = \frac{E^*G_fC}{g'(\alpha_0)}
\end{equation}
where $E^*$ is the equivalent elastic tensile modulus for the DFC specimen at the specific out time, and $A$ and $C$ are the slope and intercept found from the linear regression analysis. 

\begin{figure}[htb!]
    \centering
    \includegraphics[width=0.9\textwidth]{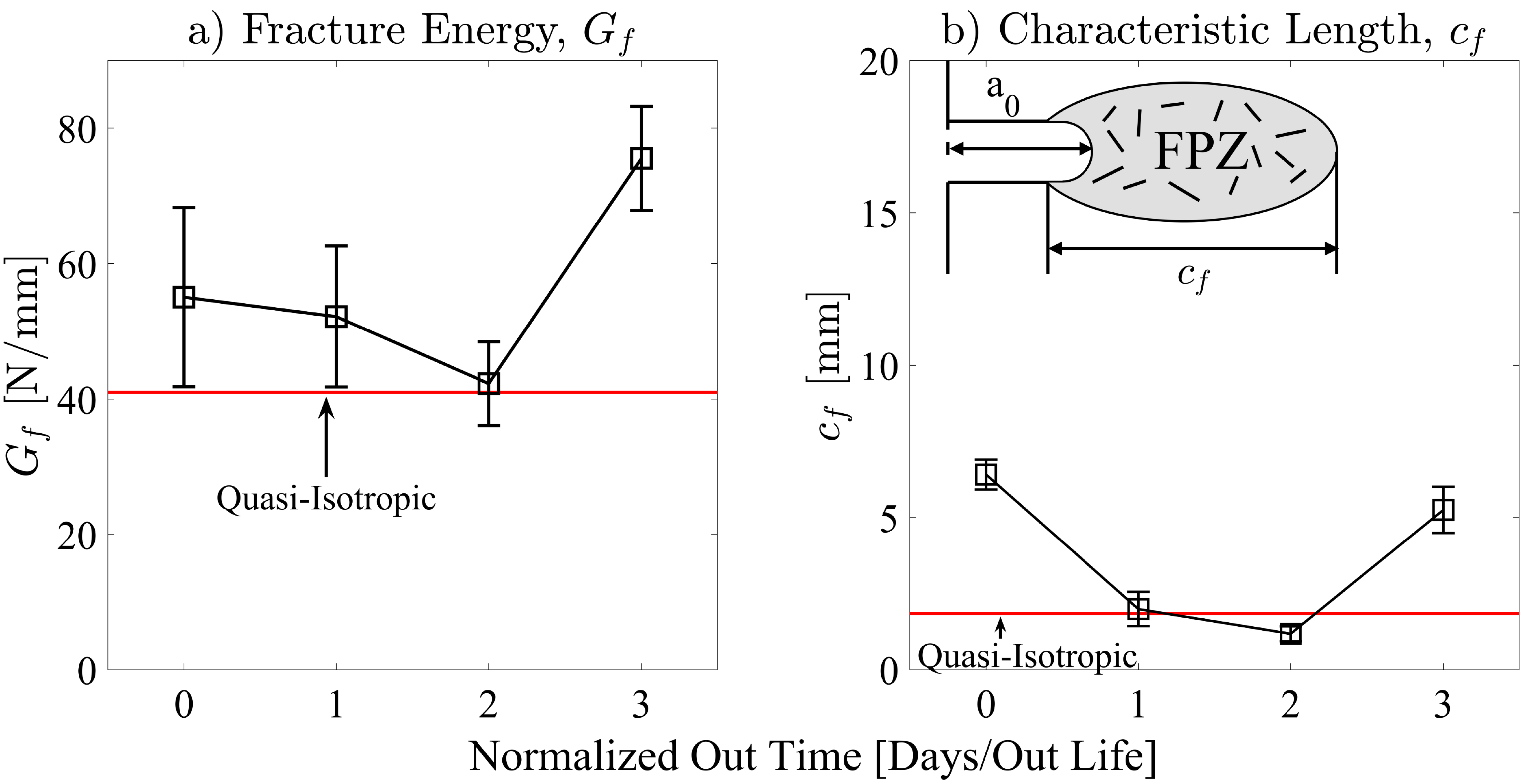}
    \caption{a) Fracture energy and b) characteristic length as a function of the prepreg out time compared to a pristine quasi-isotropic values. Quasi-isotropic and non-aged DFC results are taken from Ko et al. \cite{SeungPlateletSE}.}
    \label{fig:Gf-cf}
\end{figure}

Figure \ref{fig:Gf-cf} and table \ref{tab:Gf cf g g'} show the fracture properties. It can be seen that there is little change in the fracture energy when going from out time $0\times$ to $1\times$. When increasing the out time to $2\times$ there is a decrease in the fracture energy, and when increasing to $3\times$ the out life, there is a large increase in the fracture energy. The fracture energy of out time $2\times$ becomes close to the fracture energy of a quasi-isotropic layup, but then increases by $1.7\times$ when going to out time $3\times$. A similar trend found in the fracture energy can be seen for the characteristic length in out times $2\times$ and $3\times$. In addition to these trends, the characteristic length also decreases when going from out time $0\times$ to out time $1\times$. The characteristic length of out time $1\times$ and $2\times$ becomes about the same as a quasi-isotropic laminate.

\section{Discussion}
The present study investigated the effect of the age of repurposed prepregs on the mechanical performance of chopped fiber composites. As described in the previous sections, the study included the analysis of the remaining crosslinking degree via Differential Scanning Calorimetry (DSC) for different age times along with the investigation of the mechanical performance of the DFCs in tension, compression, shear, and nominal mode I fracture. To the best of the authors' knowledge, this investigation on aging covers the most extensive set of properties ever reported for DFCs.

In regard to the DSC results, the initial linear decrease of enthalpy with age reported in this work has been shown in many previous studies \cite{Blass, NASADSC, NuttDSC, NuttKimEnthalpy, ColeDSC, ScolaDSCMech}. However, the horizontal asymptote for large age times as the one shown in fig. \ref{fig:DSCResults} has only been reported in the study done by Blass et al. \cite{Blass}. One of the two prepreg tested by Blass et al. showed this plateau at $4\times$ the out life, while the other did not reach this plateau even after $12\times$ the out life. Both of these prepregs were aged to $120$ days which is less than half the aging times investigated in the present work. Thanks to the extensive duration, probably among the largest ever reported in the literature, the tests performed in this work clearly show a saturation of crosslinking occurring at ambient temperature for sufficient out times. In the context of repurposing the material to make chopped fiber composites, this observation is of utmost importance. In fact, tracking the out time of various material streams might not be practical or even possible. However, the results clearly suggest that while crosslinking occurs continuously during the aging of the material, it would take a significant amount of time to reach a state in which further curing is not possible. Simple DSC sampling of the repurposed material every e.g. $4-5$ out times might be enough to make sure the material can still be used. Of course, the extensive usability over large aging times makes the use of scrap prepregs from other composite manufacturing processes very attractive and convenient provided that the mechanical properties are not significantly affected by aging.   

From the morphological point of view, an increase in Discontinuous Fiber Composite (DFC) plate thickness with increasing age times was noted. After thorough mesostructure studies, it is hypothesized that the increase in thickness could be from an increase in viscosity of the prepreg caused by aging. It has been found in previous studies that with the increase in prepreg out time, there is an increase in the viscosity of epoxy \cite{NuttKimEnthalpy, ScolaDSCMech, NuttKimViscosityVoid}. This increase in viscosity would decrease the platelet flow during the curing process leading to less platelet reorientation during curing and causing the number of platelets through the thickness to increase for the same hot press temperature and pressure. This decrease in flow is also supported by the fact that there is an increase in void area fraction with age as shown in fig. \ref{fig:fiber-void-area-frac}. If there was flow during curing, platelets and resin can fill the volume where there are less platelets, which would lead to a lower void area fraction. This can be seen in the lower void area fraction in the non-aged DFCs compared to the aged ones. 

Another important morphological observation obtained in this work is that the fiber area fraction of the DFC specimens is all the same as the quasi-isotropic specimens, about $60\%$. This is different from typical short fiber composites who has a fiber area fraction of about $25-40\%$ \cite{ShortFiberVf}. When looking at the application of recycling, this is a huge improvement. Typically when recycling prepreg, the resin is removed from the fiber and the fiber is milled into powder, chopped into short fiber, or made into mats. The option most comparable to DFCs would be short fibers, which would have a much lower fiber area fraction than DFCs. If DFCs are used for recycling there would be no decrease in fiber area fraction. The void area fraction increases with age for DFCs. This is also seen in previous studies \cite{NuttKimViscosityVoid}. Even with the increase in void area fraction, the fiber volume fraction remains still very high. At the high end, void content was about $3\%$ for most of the samples investigated in this work, and at the low end it was comparable to the void content in the non-aged DFC and quasi-isotropic laminate. With changes in the curing process to account for the increase in viscosity with age and with extra care to evenly distribute platelets, it may be possible to decrease the void content to be comparable with the non-aged specimens. However, decreasing void content may not be necessary. Voids have less impact on the structural behavior of DFCs than with continuous fiber laminates and, different from continuous fiber composites, void content of $3\%$ may be acceptable for DFCs. Also, it was seen that even with the increase in void content, the modulus of aged specimen did not change and the strength increased. This shows the impact of voids on the mechanical properties of aged DFCs to be minimal.

From the mechanical point of view, it can be seen that the out time of the prepreg does not effect the various moduli of DFC specimens, even when aged to three times the out life. Figure \ref{fig:Norm-Prop} shows the DFCs' modulus normalized to their respective quasi-isotropic values. The figure shows how the modulus does not change with age and that the modulus of DFCs are about the same as a quasi-isotropic layup. The lowest moduli found was for the shear modulus at out time $3\times$ and is $81\%$ that of the quasi-isotropic shear modulus. For other test, the moduli are between $82-97\%$ that of their respective quasi-isotropic values and this does not change with age. This consistent modulus with age has also been seen in continuous fiber composites \cite{Blass, ScolaDSCMech, JonesAgedILSS, ColeMech, SchmidtFlex, AkayMech} and by Nilakantan et al. \cite{NuttAging} in DFC tension specimens. However, unlike with the modulus, an increase in strength with the aged specimens can be seen. This is different to what has been seen in previous studies done on continuous fiber composites \cite{Blass, ScolaDSCMech, JonesAgedILSS, ColeMech, SchmidtFlex, AkayMech}. In fact, in these studies it was shown that if the mechanical property had more dependence on the matrix, such as $90^{\circ}$ tension, then there would be a decrease in strength, and if it had more dependence on the fibers, such as $0^{\circ}$ tension, then the strength would not change with age. Figure \ref{fig:Norm-Prop} shows the strength of all tests normalized to their respective quasi-isotropic values and a clear aging effect can be seen. Non-aged specimen can be as low as $39\%$ that of the quasi-isotropic value. Then with age, the strength increases by $15\%$ on average, and has a maximum percent increase of $27\%$ for size $3$ out time $2\times$ specimen. It is worth mentioning that an increase in tensile strength with prepreg age for DFCs was also seen by Nilakantan et al. \cite{NuttAging}. However, to the best of the authors' knowledge, the present work is the first to ever investigate the effect of aging also on the compression and shear strengths and to confirm that such strengths increase with age. This is a very important result since recycled DFC components can be subject to a large variety of loading conditions, not only uniaxial tension. Thanks to the results presented in this work, it is possible to conclude with confidence that the structural capacity of a recycled DFC component is not affected by age. On the contrary, the structural capacity might slightly increase.

By looking at work done on aging of cured epoxy and composites we can gain insight on what could be the reason for the strength increase. Work by Zhou et al. \cite{ZhouMoisture} has shown that there are two types of bonding that can occur when cured epoxy is aged in distilled water. The first type of bond corresponds to one hydrogen bond between the water molecule and the resin network. This breaks the initial interchain Van der Waals forces resulting in increased mobility of the chains and plasticization of the epoxy. The second type of bonding happens when the water molecules form multiple hydrogen bonds and act as a pseudo-crosslink. This plasticization effect can be seen in aging of a cured laminate and aging of prepregs. Asp \cite{AspMoisture} found that when composites were aged in high humidity and temperature, mode I fracture toughness increased when testing at room temperature and at $100^{\circ}$C. Sharp \cite{SharpMoisture} allowed the prepreg used for the center plies of the laminate to absorb $0.5\%$ of its weight in water before curing. This resulted in an increase in mode I and mode II fracture toughness, similar to what was found to happen when cured laminates absorbed water. Blass et al. \cite{Blass} aged two prepregs in a standard ambient temperature and humidity and performed Double Cantilever Beam (DCB) tests to find mode I fracture toughness. Here one of the prepregs tested showed no changed in fracture toughness after $120$ days ($12\times$ out life), while the other prepreg tested had an increase in fracture toughness. The increases in fracture toughness seen in previous works can be linked to the plasticization of the epoxy. In DFCs, the failure is due to a combination of fiber and matrix damage mechanisms, so the plasticization allows the epoxy to absorb more energy before failure and the fibers can carry more load, resulting in a higher strength which is not seen in continuous fiber composites.


The present study also provided the first ever investigation on the effects of aging on the nominal mode I fracture energy of DFCs. This is an important property measuring the capability of the material of dissipating energy upon fracture. From the size effect analysis performed in this study it can be seen that the characteristic size of the Fracture Process Zone (FPZ) decreases when going from out time $0\times$ to out time $1\times$ and $2\times$, but it increases when going from out time $2\times$ to out time $3\times$. On the other hand, the fracture energy has little change for the first out time, decreases for the second, and increases for the final out time. 

There are two phenomena with contrasting effects at play that are causing the decrease and increase in these properties. The first phenomenon is the increase in thickness of the specimen with increasing age. Ko et al. \cite{SeungThicknessSE} reported that the fracture energy and characteristic length are strongly effected by the thickness. It was seen that the fracture properties will increase up to a certain thickness and then hit a plateau. At the plateau it is possible that the increase in thickness will cause a decrease in the fracture properties. Ko et al. saw that when the thickness increased from $2.2$ mm to $3.3$ mm the fracture energy increased, but when the thickness increased from $3.3$ mm to $4.1$ mm the fracture energy decreased. This phenomenon is exactly what is happening for out time $1\times$ and $2\times$. 

The second phenomenon affecting the fracture properties is the plasticization of the matrix with aging which leads to an increase of the fracture energy. The effect of plasticization can be seen in the fact that for even though there is a thickness increase from out time $0\times$ to out time $1\times$ there is no change in the fracture energy. It was seen by Blass et al. \cite{Blass}, that with increase in prepreg out time, the mode I fracture toughness would increase or remain constant. It is believed that this retention or increase is caused by the plasticization of the matrix that happens with aging. The thickness change and plasticization are fighting each other to change the fracture properties. This can also been seen in the fact that the rate at which the fracture properties change when going from out time $0\times$ to $1\times$ and out time $1\times$ to $2\times$ changes. The rate of decrease in properties becomes lower which indicates the plasticization taking effect and then when out time $3\times$ is reached, the plasticization is developed enough to increase the fracture properties.

\section{Conclusions}
\begin{figure}[htb!]
    \centering
    \includegraphics[width=\textwidth]{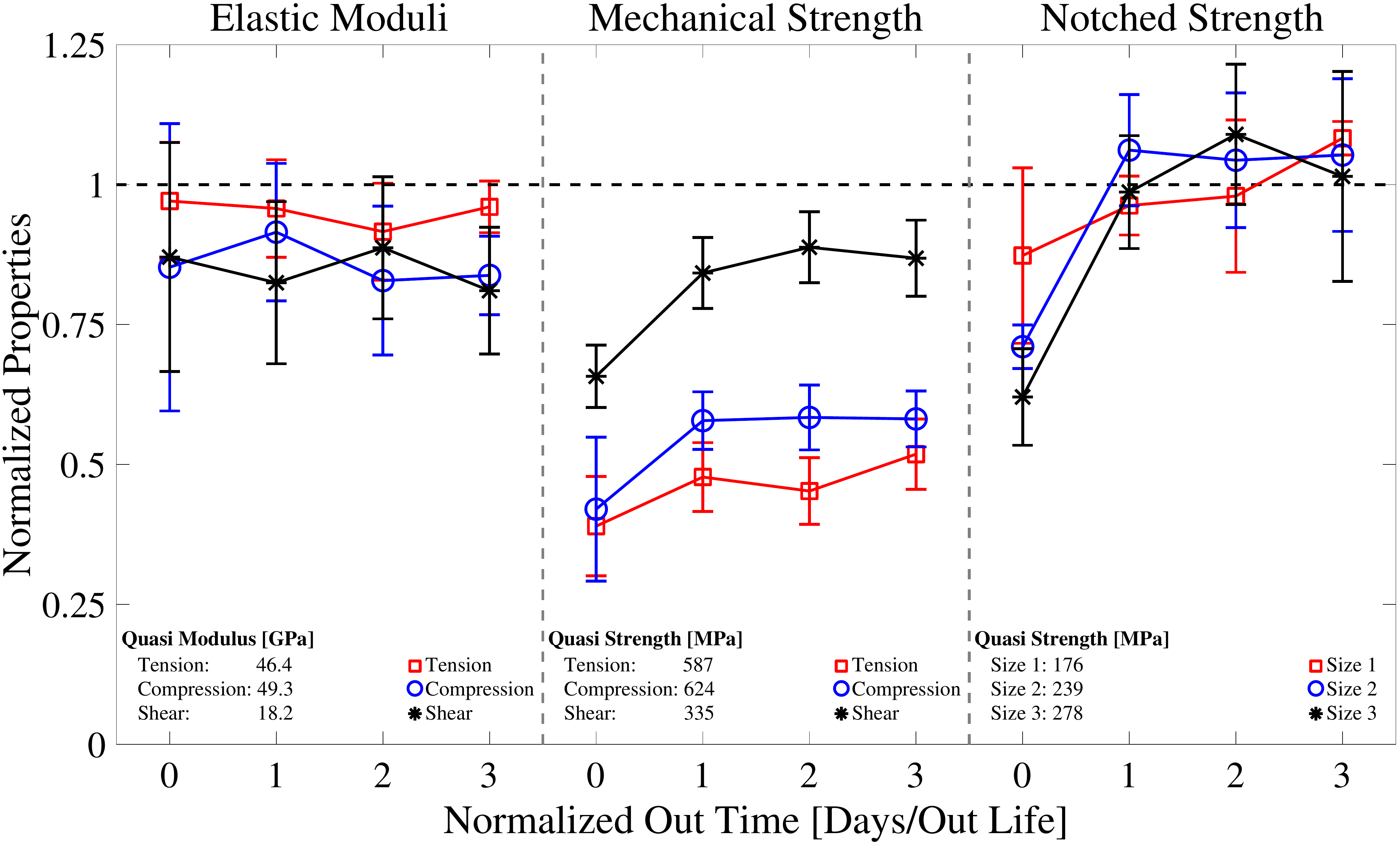}
    \caption{Normalized modulus and strength found from mechanical and size effect tests normalized to their respective quasi-isotropic values. Quasi-isotropic and non-aged DFC results for tension and SENT specimen are obtained from Ko et al. \cite{SeungPlateletSE}.}
    \label{fig:Norm-Prop}
\end{figure}

This work investigated the effects of out time on the mechanical and fracture properties of Discontinuous Fiber Composites (DFCs) for the use in recycling of thermoset prepreg material. Prepreg was aged to $1\times$, $2\times$, and $3\times$ the out life of the prepreg ($28$, $56$, and $84$ days) and compared with non-aged DFC and quasi-isotropic specimens. Based on this study the following conclusions can be drawn:

\begin{enumerate}
    
    \item As the out time of the prepreg increases, the prepreg is ambient curing and the degree of cure increases up to a an asymptotic value at around $9\times$ the out life of the prepreg ($252$ days). The effects of this ambient curing can be seen in the change in the mesostructure of the DFC specimen. With an increase in out time, there was an increase in the number of platelets through the thickness of the laminate and an increase in the laminate thickness. This is caused by the decrease in resin flow from the viscosity increase of the resin caused by aging. At the same time there was no change in the fiber content, but there was an increase in the void content of the aged specimens. However, voids do not impact the mechanical properties of DFCs when at the level void content seen here the same way it would for continuous fiber composites.
    
    \item The experimental results on tensile, compression, and shear specimens showed for the first time that prepreg out time has no effect on the moduli of DFCs, even when the out time reached $3\times$ the out life of the prepreg. Figure \ref{fig:Norm-Prop} shows the moduli found normalized to their respective quasi-isotropic values. Here it can be seen that the moduli of DFC specimens remain at $81-97\%$ that of the quasi-isotropic values, and with age this does not change. This is similar to what happens with continuous fiber composites.
    
    \item The experimental results on tensile, compression, shear, and geometrically-scaled Single Edge Notch Tension (SENT) specimens showed for the first time that prepreg out time had a strengthening effect on DFCs. Figure \ref{fig:Norm-Prop} shows the mechanical strength and notched strength found normalized to their respective quasi-isotropic values. It can be seen that the non-aged strength could drop to be as low as $39\%$ that of the quasi-isotropic strength. However, with age the strength increased, and for the notched strength, could be larger than the quasi-isotropic strength. This is contrary to what happens with continuous fiber composites, where aging has no effect on fiber dominated strength (i.e. $0^\circ$ tension), but decreases matrix-dominated properties (i.e. $90^\circ$ tension). The strengthening is suspected to be caused by plasticization of the matrix which allows for more energy absorption before failure. Since DFCs' properties depend on both fiber and matrix, this leads to an increase in strength which is not seen with continuous fiber composites.
    
    \item The experimental results on the geometrically-scaled SENT specimen for the various out times showed a significant size effect on the nominal strength. As the size of the specimen increases, the nominal strength approached Linear Elastic Fracture Mechanics (LEFM). For smaller specimen, the fracture can exhibit a more pseudo-ductile behavior and deviate from LEFM.
    
    \item At a given out time the fracture behavior of geometrically-scaled SENT specimens changes to be quasi-ductile to brittle and back to quasi-ductile. Out time $0\times$ specimens exhibit a much more quasi-ductile or quasi-brittle behavior, depending on the specimen size. For out time $1\times$ and $2\times$, the behavior was much closer to the LEFM regime, similar to the quasi-isotropic specimen, being much more brittle at the structure sizes tested when compared to out time $0\times$. However for out times $3\times$ the behavior became similar to that of out time $0\times$, exhibiting much more quasi-ductile behavior.
    
    \item The transition from quasi-ductile to brittle behavior with an increase in specimen size can be attributed to the development of the Fracture Process Zone (FPZ). For out times $0\times$ and $3\times$, the FPZ dimensions were found to be comparable to the platelet size. In the FPZ, significant non-linear deformations due to sub-critical damage mechanisms, such as platelet delamination, matrix microcracking, and platelet splitting/fracture, promote strain redistribution and mitigate the intensity of the stress field induced by the crack/notch.The size of a fully-developed FPZ is typically a material property and thus its influence on the structural behavior becomes increasingly significant as the structure size is reduced. For sufficiently large structures, the size of the FPZ becomes negligible compared to the structure’s characteristic size in agreement with the inherent assumption of the LEFM that non-linear effects are negligible during the fracturing process. For out times $1\times$ and $2\times$, the size of the FPZ becomes sufficiently small for the sizes tested, unlike the other two out times, and causes little deviation from LEFM for the smallest structure size.
    
    \item The transition from quasi-ductile to brittle and back to quasi-ductile fracture with an increase in out time can be attributed to two factors, the first being a thickness effect, and the second being the plasticization caused by aging. With more platelets through the thickness the behavior of the structure can become more brittle. However, with aging and plasticization of the matrix, the behavior can become more quasi-ductile. For out time $1\times$ the thickness increase and plasticization cause a decrease in the characteristic length $c_f$, but no change in the fracture energy, $G_f$. However, for out time $2\times$ the thickness increase has more weight causing both $c_f$ and $G_f$ to decrease. Finally, for out time $3\times$ the plasticization takes over causing both $c_f$ and $G_f$ to increase.
    
    \item  To investigate the effect of the prepreg out time on the fracture energy, $G_f$ , and the effective length of the FPZ, $c_f$ , the approach combining equivalent fracture mechanics and stochastic finite element modeling proposed in \cite{SeungPlateletSE} was used. This model accounts for the effects of the complex random mesostructure of the material by modeling the platelets explicitly. This theoretical framework was able to describe the scaling of structural strength and enabled the characterization of the mode I fracture energy of DFCs.
    
    \item An important takeaway from the results presented in this work is that utilizing repurposed prepregs to manufacture DFCs is a great way to mitigate the impact of scrap thermoset materials on the environment. In fact, in contrast to other recycling technologies which might damage the fibers, the mechanical properties of DFCs made from old prepregs compared to DFCs made from virgin materials are retained or improved. The recycling process is also safe, easy, and environmentally friendly. More work can be done to investigate ways to prevent the change in the void content and laminate thickness with aging, however with the current study it can already be seen that DFCs represent a great form of recycling thermoset composite materials largely produced by the aerospace and automotive industries.
    
\end{enumerate}




\section*{Acknowledgments}
This work was supported by the Joint Center for Aerospace Technology Innovation (JCATI) through the project ``Computationally-Assisted Modeling of the Manufacturing of Recycled Discontinuous Fiber Composites". We also wish to acknowledge the financial support by the Federal Aviation Administration (FAA)-funded Center of Excellence for Advanced Materials in Transport Aircraft Structures (AMTAS) and the Boeing Company.  

\clearpage
\bibliographystyle{elsarticle-num}
\bibliography{ref.bib}

\begin{thebibliography}{10}
\expandafter\ifx\csname url\endcsname\relax
  \def\url#1{\texttt{#1}}\fi
\expandafter\ifx\csname urlprefix\endcsname\relax\def\urlprefix{URL }\fi
\expandafter\ifx\csname href\endcsname\relax
  \def\href#1#2{#2} \def\path#1{#1}\fi

\bibitem{das2019preparation}
T.~K. Das, P.~Ghosh, N.~C. Das, Preparation, development, outcomes, and
  application versatility of carbon fiber-based polymer composites: a review,
  Advanced Composites and Hybrid Materials (2019) 1--20.

\bibitem{harris2002design}
C.~E. Harris, J.~H. Starnes~Jr, M.~J. Shuart, Design and manufacturing of
  aerospace composite structures, state-of-the-art assessment, Journal of
  aircraft 39~(4) (2002) 545--560.

\bibitem{othman2018application}
R.~Othman, N.~I. Ismail, M.~Pahmi, M.~H.~M. Basri, H.~Sharudin, A.~R. Hemdi,
  Application of carbon fiber reinforced plastics in automotive industry: A
  review, J. Mech. Manuf 1 (2018) 144--154.

\bibitem{Mazumdar2020}
S.~Mazumdar, M.~Pichler, W.~Benevento, R.~Seneviratine, W.~E., State of the
  industry report, Composites Manufacturing (2020).

\bibitem{ACMAsite}
\href{http://www.acma.net}{{A}merican {C}omposites {M}anufacturers
  {A}ssociation ({ACMA}) website} ((visited: December 2021)).
\newline\urlprefix\url{http://www.acma.net}

\bibitem{NuttAging}
G.~Nilakantan, S.~R. Nutt, Reuse and upcycling of thermoset prepreg scrap: Case
  study with out-of-autoclave carbon fiber/epoxy prepreg, J Compos Mater 52
  (2018) 341--60.

\bibitem{naqvi2018critical}
S.~Naqvi, H.~M. Prabhakara, E.~Bramer, W.~Dierkes, R.~Akkerman, G.~Brem, A
  critical review on recycling of end-of-life carbon fibre/glass fibre
  reinforced composites waste using pyrolysis towards a circular economy,
  Resources, conservation and recycling 136 (2018) 118--129.

\bibitem{ma2018chemical}
Y.~Ma, S.~Nutt, Chemical treatment for recycling of amine/epoxy composites at
  atmospheric pressure, Polymer degradation and stability 153 (2018) 307--317.

\bibitem{kuang2018recycling}
X.~Kuang, Y.~Zhou, Q.~Shi, T.~Wang, H.~J. Qi, Recycling of epoxy thermoset and
  composites via good solvent assisted and small molecules participated
  exchange reactions, ACS Sustainable Chemistry \& Engineering 6~(7) (2018)
  9189--9197.

\bibitem{pickering2006recycling}
S.~J. Pickering, Recycling technologies for thermoset composite
  materials—current status, Composites Part A: applied science and
  manufacturing 37~(8) (2006) 1206--1215.

\bibitem{goodship2009management}
V.~Goodship, Management, recycling and reuse of waste composites, CRC Press,
  Boca Raton, FL, USA, 2009.

\bibitem{witik2013carbon}
R.~A. Witik, R.~Teuscher, V.~Michaud, C.~Ludwig, J.-A.~E. M{\aa}nson, Carbon
  fibre reinforced composite waste: an environmental assessment of recycling,
  energy recovery and landfilling, Composites Part A: Applied Science and
  Manufacturing 49 (2013) 89--99.

\bibitem{CRTC}
\href{https://compositerecycling.org/}{{C}omposite {R}ecycling {T}echnology
  {C}enter, ({CRTC})} (2021).
\newline\urlprefix\url{https://compositerecycling.org/}

\bibitem{SeungPlateletSE}
S.~Ko, J.~Yang, M.~E. Tuttle, M.~Salviato, Effect of the platelet size on the
  fracturing behavior and size effect of discontinuous fiber composite
  structures, Compos Struct 227 (2019) 111245.

\bibitem{SeungThicknessSE}
S.~Ko, J.~Davey, S.~Douglass, J.~Yang, M.~E. Tuttle, M.~Salviato, Effect of the
  thickness on the fracturing behavior of discontinuous fiber composite
  structures, Compos Part A 125 (2019) 105520.

\bibitem{ASTMD4332}
{ASTM D}4332-14. {S}tandard practice for conditioning containers, packages, or
  packaging components for testing, ASTM International (2015).

\bibitem{SeungSAMPE}
S.~Ko, K.~Chan, R.~Hawkins, R.~Jayaram, C.~Lynch, R.~E. Mamoune, et~al,
  Characterization and computational modeling of the fracture behavior in
  discontinuous fiber composite structures, SAMPE Conference (2018).

\bibitem{ASTMD3418}
{ASTM D}3418-15. {S}tandard practice for conditioning containers, packages, or
  packaging components for testing, ASTM International (2015).

\bibitem{ASTME1356}
{ASTM E}1356-08. {S}tandard test method for assignment of the glass transition
  temperatures by differential scanning calorimetry, West Conshohocken, PA
  (2014).

\bibitem{matlab}
The MathWorks, Inc., “Matlab 2019b,” The MathWorks, Inc., Natick, MA, 2019.

\bibitem{gom}
GOM. Braunschweig, Germany. https://www.gom.com.

\bibitem{ASTMD3039}
{ASTM D}1356-17. {S}tandard test method for tensile properties of polymer
  matrix composite materials, West Conshohocken, PA (2017).

\bibitem{ASTMD3410}
{ASTM D}3410-16. {S}tandard test method for compressive properties of polymer
  matrix composite materials with unsupported gage section by shear loading,
  West Conshohocken, PA (2016).

\bibitem{ASTMD5379}
{ASTM D}5379-12. {S}tandard test method for shear properties of composite
  materials by the {v}-notch beam method, West Conshohocken, PA (2012).

\bibitem{SeleznevaTenCompShear}
M.~Selezneva, L.~Lessard, Characterization of mechanical properties of randomly
  oriented strand thermoplastic composites, J Compos Mater 50 (2016) 2833–51.

\bibitem{Feraboli1}
P.~Feraboli, E.~Peitso, F.~Deleo, T.~Cleveland, Characterization of
  prepreg-based discontinuous carbon fiber/epoxy systems, J Reinf Plast Comp 28
  (2009) 1191–214.

\bibitem{Feraboli2}
P.~Feraboli, E.~Peitso, T.~Cleveland, P.~Stickler, Modulus measurement for
  prepreg based discontinuous carbon fiber/epoxy systems, J Compos Mater 43
  (2009) 1947–65.

\bibitem{Takahashi1}
Y.~Wan, J.~Takahashi, Tensile and compressive properties of chopped carbon
  fiber tapes reinforced thermoplastics with different fiber lengths and
  molding pressures, Compos Part A 87 (2016) 271–81.

\bibitem{Takahashi2}
S.~Yamashita, K.~Hashimoto, H.~Suganuma, J.~Takahashi, Experimental
  characterization of the tensile failure mode of ultra-thin chopped carbon
  fiber tape-reinforced thermoplastics, J Reinf Plast Comp 35 (2016) 1342–52.

\bibitem{FeraboliNotch}
P.~Feraboli, E.~Peitso, T.~Cleveland, P.~Stickler, J.~Halpin, Notched behavior
  of prepreg-based discontinuous carbon fiber/epoxy systems, Compos Part A 40
  (2009) 289–99.

\bibitem{QianNotch}
C.~Qian, L.~Harper, T.~Turner, N.~Warrior, Notched behaviour of discontinuous
  carbon fibre composites: comparison with quasi-isotropic non-crimp fabric,
  Compos Part A 42 (2011) 293–302.

\bibitem{SalviatoTextile}
M.~Salviato, K.~Kirane, Z.~Bažant, G.~Cusatis, Experimental and numerical
  investigation of intra-laminar size effect in textile composites, Comps Sci
  Technol 135 (2016) 67–75.

\bibitem{Salbook}
Z.~P. Ba{\v{z}}ant, J.-L. Le, M.~Salviato, Quasibrittle Fracture Mechanics and
  Size Effect: A First Course, Oxford University Press, 2021.

\bibitem{Baz04}
Z.~P. Ba{\v{z}}ant, Scaling theory for quasibrittle structural failure,
  Proceedings of the National Academy of Sciences 101~(37) (2004) 13400--13407.

\bibitem{LiQia21}
W.~Li, Y.~Qiao, J.~Fenner, K.~Warren, M.~Salviato, Z.~P. Ba{\v{z}}ant,
  G.~Cusatis, Elastic and fracture behavior of three-dimensional ply-to-ply
  angle interlock woven composites: Through-thickness, size effect, and
  multiaxial tests, Composites Part C: Open Access 4 (2021) 100098.

\bibitem{SalKir19}
M.~Salviato, K.~Kirane, Z.~P. Ba{\v{z}}ant, G.~Cusatis, Mode {I} and {II}
  interlaminar fracture in laminated composites: a size effect study, Journal
  of Applied Mechanics 86~(9) (2019).

\bibitem{SalEsn16}
M.~Salviato, S.~E. Ashari, G.~Cusatis, Spectral stiffness microplane model for
  damage and fracture of textile composites, Composite Structures 137 (2016)
  170--184.

\bibitem{QiaSal19}
Y.~Qiao, M.~Salviato, Strength and cohesive behavior of thermoset polymers at
  the microscale: A size-effect study, Engineering Fracture Mechanics 213
  (2019) 100--117.

\bibitem{MefQia17}
C.~H. Mefford, Y.~Qiao, M.~Salviato, Failure behavior and scaling of graphene
  nanocomposites, Composite Structures 176 (2017) 961--972.

\bibitem{QiaSal19b}
Y.~Qiao, M.~Salviato, Study of the fracturing behavior of thermoset polymer
  nanocomposites via cohesive zone modeling, Composite Structures 220 (2019)
  127--147.

\bibitem{BazPfe87}
Z.~P. Bazant, P.~A. Pfeiffer, et~al., Determination of fracture energy from
  size effect and brittleness number, ACI Materials Journal 84~(6) (1987)
  463--480.

\bibitem{CarCus19}
C.~Carloni, G.~Cusatis, M.~Salviato, J.-L. Le, C.~G. Hoover, Z.~P.
  Ba{\v{z}}ant, Critical comparison of the boundary effect model with cohesive
  crack model and size effect law, Engineering Fracture Mechanics 215 (2019)
  193--210.

\bibitem{NguDon21}
H.~T. Nguyen, A.~A. D{\"o}nmez, Z.~P. Ba{\v{z}}ant, Structural strength scaling
  law for fracture of plastic-hardening metals and testing of fracture
  properties, Extreme Mechanics Letters 43 (2021) 101141.

\bibitem{Irwin}
G.~Irwin, Fracture Handbuch der Physik vol. 6, Springer, 1958.

\bibitem{Bazant}
J.~Planas, Z.~P. Bazant, Fracture and size effect in concrete and other
  quasibrittle materials, CRC Press, 1998.

\bibitem{SEL}
Z.~P. Ba{\v{z}}ant, Size effect in blunt fracture: concrete, rock, metal,
  Journal of engineering mechanics 110~(4) (1984) 518--535.

\bibitem{MesoGen1}
M.~Selezneva, S.~Roy, S.~Meldrum, L.~Lessard, A.~Yousefpour, Modelling of
  mechanical properties of randomly oriented strand thermoplastic composites, J
  Compos Mater 51 (2017) 831--845.

\bibitem{MesoGen2}
P.~Feraboli, T.~Cleveland, P.~Stickler, J.~Halpin, Stochastic laminate analogy
  for simulating the variability in modulus of discontinuous composite
  materials, Compos Part A 41 (2010) 557–70.

\bibitem{MesoGen3}
K.~Harban, M.~Tuttle, Reducing certification costs of discontinuous fiber
  composite structures via stochastic modeling, US Dept. of Transportation FAA
  (2017).

\bibitem{KoSampe20}
S.~Ko, J.~Yang, M.~E. Tuttle, M.~Salviato, Stochastic computational modeling of
  the fracturing behavior in discontinuous fiber composite structures, in:
  Proc. of the SAMPE 2020 conference, SAMPE 2020 Virtual Series, 2020, pp.
  1--12.

\bibitem{KoASC21}
S.~Ko, T.~Nakagawa, Z.~Chen, J.~Davey, T.~Abdullah, L.~Kuklenski, E.~J. Adams,
  M.~R. Soja, C.~Y. Park, W.~B. Avery, et~al., Experimental and numerical
  investigations of stochastic thickness effects in discontinuous fiber
  composites, in: Proceedings of the American Society for
  Composites—Thirty-Sixth Technical Conference on Composite Materials, 2021.

\bibitem{Abaqus}
Dassault Systemes ABAQUS. ABAQUS Documentation. Providence, RI; 2021.

\bibitem{RiceJint}
J.~Rice, A path independent integral and the approximate analysis of strain
  concentrations by notches and cracks, J Appl Mech ASME 35 (1969) 379–86.

\bibitem{Blass}
D.~Blass, S.~Kreling, K.~Dilger, The impact of prepreg aging on its
  processability and the postcure mechanical properties of epoxy-based
  carbon-fiber reinforced plastics, P I MECH ENG L-J MAT 231 (2017) 62--72.

\bibitem{NASADSC}
S.~Miller, J.~Sutter, D.~Scheiman, M.~Maryanski, M.~Schlea, Study of out-time
  on the processing and properties if im7/977-3 composites, SAMPE conference
  (2010).

\bibitem{NuttDSC}
L.~Grunenfelder, S.~Nutt, Prepreg age monitoring via differential scanning
  calorimetry, J Reinf Plast Comp 31 (2012) 295--302.

\bibitem{NuttKimEnthalpy}
D.~Kim, T.~Centea, S.~Nutt, Modelling and monitoring of out-time and moisture
  absorption effects on cure kinetics and viscosity for an out-of-autoclave
  ({O}o{A}) prepreg, Compos Sci Technol 138 (2017) 200--208.

\bibitem{ColeDSC}
K.~Cole, D.~No{\"{e}}l, J.~Hechler, Room temperature aging of narmco 5208
  carbon-epoxy prepreg. part i: physicochemical characterization, Polym
  Composite 10 (1989) 150--161.

\bibitem{ScolaDSCMech}
D.~Scola, J.~Vontell, Effects of ambient aging of 5245{C}/graphite prepreg on
  composition and mechanical properties of fabricated composites, Polym
  Composite 8 (1987) 244--252.

\bibitem{NuttKimViscosityVoid}
D.~Kim, T.~Centea, S.~Nutt, Effect of room-temperature out-time on tow
  impregnation in an out-of-autoclave prepreg, Compos Part A 45 (2013)
  119--216.

\bibitem{ShortFiberVf}
Y.~Pan, L.~Iorga, A.~Pelegri, Numerical generation of a random chopped fiber
  composite rve and its elastic properties, Compos Sci and Tech 68 (2008)
  2792--2798.

\bibitem{JonesAgedILSS}
R.~Jones, Y.~Ng, J.~McClelland, Monitoring ambient-temperature aging of a
  carbon-fiber/epoxy composite prepreg with photoacoustic spectroscopy, Compos
  Part A 29 (2008) 965--971.

\bibitem{ColeMech}
K.~Cole, D.~No\"{e}l, J.~Hechler, P.~Cielo, J.~Krapez, Polym Composite 12
  (1991) 203--212.

\bibitem{SchmidtFlex}
C.~Schmidt, P.~Weber, T.~Hocke, B.~Denkena, Influence of prepreg material
  quality on carbon fiber reinforced plastic laminates processed by automated
  fiber placement, 11th CIRP Conference (2017).

\bibitem{AkayMech}
M.~Akay, Effects of prepreg ageing and post-cure hygrothermal conditioning on
  the mechanical behaviour of carbon fibre/epoxy laminates, Compos Sci Technol
  38 (1990) 359--370.

\bibitem{ZhouMoisture}
J.~Zhou, J.~Lucas, Hygrothermal effects of epoxy resin. part {I}: the nature of
  water in epoxy, Polymer 40 (1999) 5505–5512.

\bibitem{AspMoisture}
L.~Asp, The effects of moisture and temperature on the interlaminar
  delamination toughness of a carbon/epoxy composite, Compos Sci and Tech 58
  (1998) 967–977.

\bibitem{SharpMoisture}
N.~Sharp, Effects of moisture on the properties of epoxies and carbon-epoxy
  composite laminates, Ph.D. thesis, Purdue University, (Publication No.
  3734368).

\end{thebibliography}


\newpage
\section*{Figures and Tables}
\begin{table}[htb!]
    \caption{Summary of the mesostructure analysis performed}
    \centering
    \begin{tabularx}{\textwidth}{Y Z Z Z Z Z Z}
        \hline\hline
        & Quasi & Out Time $0\times$ & Out Time $1\times$ & Out Time $2\times$ & Out Time $3\times$ \\
        \hline
        Laminate Thickness [mm] & 3.53 $\pm$ 0.05 & 3.01 $\pm$ 0.04 & 3.70 $\pm$ 0.06 & 3.99 $\pm$ 0.14 & 3.87 $\pm$ 0.10 \\
        Platelet Thickness [mm] & 0.145 $\pm$ 0.01 & 0.137 $\pm$ 0.05 & 0.133 $\pm$ 0.04 & 0.132 $\pm$ 0.04 & 0.131 $\pm$ 0.04 \\
        Number of Platelets & 24 $\pm$ 0 & 20.74 $\pm$ 4.47 & 24.06 $\pm$ 7.26 & 27.67 $\pm$ 7.85 & 28.36 $\pm$ 4.88 \\
        Fiber Area Fraction [\%] & 58.19 $\pm$ 0.90 & 59.19 $\pm$ 7.73 & 60.21 $\pm$ 7.77 & 57.43 $\pm$ 10.10 & 61.48 $\pm$ 6.61 \\
        Void Area Fraction [\%] & 0.29 $\pm$ 0.23 & 0.24 $\pm$ 0.18 & 0.98 $\pm$ 1.24 & 0.90 $\pm$ 1.17 & 1.10 $\pm$ 1.34 \\
        \hline\hline
    \end{tabularx}
    \label{tab:Meso}
\end{table}

\begin{table}[htb!]
    \caption{Tensile modulus and strength of non-aged quasi-isotropic and aged DFC. Quasi-isotropic and non-aged DFC results are taken from Ko et al. \cite{SeungSAMPE}.}
    \centering
    \begin{tabular}{c c c c}
        \hline\hline
        Type & Specimen Tested & Modulus [GPa] (\%) & Strength [MPa] (\%) \\
        \hline
        Quasi        & 11 & 46.40 $\pm$ 1.11 (2.39)  & 587.10 $\pm$ 27.86 (4.75) \\
        OT 0$\times$ & 11 & 45.04 $\pm$ 4.87 (10.81) & 228.90 $\pm$ 52.20 (22.80) \\
        OT 1$\times$ & 10 & 44.42 $\pm$ 4.04 (9.09)  & 280.41 $\pm$ 36.10 (12.87) \\
        OT 2$\times$ & 10 & 42.50 $\pm$ 4.02 (9.45)  & 265.73 $\pm$ 34.92 (13.14) \\
        OT 3$\times$ & 10 & 44.56 $\pm$ 2.15 (4.82)  & 304.43 $\pm$ 36.82 (12.09) \\
        \hline\hline
    \end{tabular}
    \label{tab:Ten}
\end{table}

\begin{table}[htb!]
    \caption{Compression modulus and strength of non-aged quasi-isotropic and aged DFC.}
    \centering
    \begin{tabular}{c c c c}
        \hline\hline
        Type & Specimen Tested & Modulus [GPa] (\%) & Strength [MPa] (\%) \\
        \hline
        Quasi        & 12 & 49.28 $\pm$ 6.55 (6.21)  & 623.81 $\pm$ 38.74 (13.29) \\
        OT 0$\times$ & 12 & 42.01 $\pm$ 12.65 (30.11)& 262.16 $\pm$ 80.14 (30.57) \\
        OT 1$\times$ & 8  & 45.10 $\pm$ 6.05 (13.42) & 360.79 $\pm$ 32.16 (8.91) \\
        OT 2$\times$ & 8  & 40.83 $\pm$ 6.55 (16.05) & 364.39 $\pm$ 36.18 (9.93) \\
        OT 3$\times$ & 8  & 41.29 $\pm$ 3.46 (8.38)  & 362.73 $\pm$ 31.17 (8.59) \\
        \hline\hline
    \end{tabular}
    \label{tab:Comp}
\end{table}

\begin{table}[htb!]
    \caption{Shear modulus and strength of non-aged, quasi-isotropic, and aged DFC.}
    \centering
    \begin{tabular}{c c c c}
        \hline\hline
        Type & Specimen Tested & Modulus [GPa] (\%) & Strength [MPa] (\%) \\
        \hline
        Quasi        & 12 & 18.19 $\pm$ 3.42 (18.82) & 334.67 $\pm$ 29.18 (8.72) \\
        OT 0$\times$ & 12 & 15.83 $\pm$ 3.72 (23.49) & 220.09 $\pm$ 18.71 (8.71) \\
        OT 1$\times$ & 12 & 15.00 $\pm$ 2.63 (17.54) & 281.84 $\pm$ 21.17 (7.51) \\
        OT 2$\times$ & 12 & 16.14 $\pm$ 2.31 (14.31) & 297.23 $\pm$ 21.22 (7.14) \\
        OT 3$\times$ & 12 & 14.74 $\pm$ 2.06 (13.96) & 290.67 $\pm$ 22.78 (7.84) \\
        \hline\hline
    \end{tabular}
    \label{tab:Shear}
\end{table}

\begin{table}[htb!]
    \caption{Failure strength of non-aged quasi-isotropic, and aged DFC SENT specimen and the number of specimen tested. Quasi-isotropic and non-aged DFC results are taken from Ko et al. \cite{SeungPlateletSE}.}
    \centering
    \begin{tabular}{c c c c}
        \hline\hline
        & \multicolumn{3}{c}{Average failure strength [MPa] (\%), Number of specimens tested} \\
        \cline{2-4}
        Type & Size 1 & Size 2 & Size 3 \\
        \hline
        Quasi        & 176.1 $\pm$ 9.73 (5.53), 3   & 238.6 $\pm$ 23.20 (9.72), 8 & 277.9 $\pm$ 31.54 (11.35), 6 \\
        OT 0$\times$ & 153.8 $\pm$ 27.61 (17.95), 3 & 198.2 $\pm$ 14.90 (5.49), 5 & 214.2 $\pm$ 14.58 (13.90), 4 \\
        OT 1$\times$ & 169.5 $\pm$ 9.27  (5.47), 4  & 253.3 $\pm$ 23.73 (9.37), 8 & 274.2 $\pm$ 28.03 (10.22), 8 \\
        OT 2$\times$ & 172.5 $\pm$ 23.97 (13.90), 2 & 249.0 $\pm$ 28.67 (11.52), 8 & 302.9 $\pm$ 34.85 (11.51), 8 \\
        OT 3$\times$ & 190.7 $\pm$ 5.28  (2.77), 3  & 251.2 $\pm$ 32.54 (12.95), 8 & 282.0 $\pm$ 52.11 (18.48), 8 \\
        \hline\hline
    \end{tabular}
    \label{tab:SizeEffect}
\end{table}

\begin{table}[htb!]
    \caption{Fracture properties obtained from size effect experiments and the stochastic finite element analysis. Quasi-isotropic and non-aged DFC results are taken from Ko et al. \cite{SeungPlateletSE}.}
    \centering
    \begin{tabular}{c c c c c}
        \hline\hline
        Type & $g(\alpha_\circ)$ & $g'(\alpha_\circ)$ & $G_f$ [N/mm] & $c_f$ [mm] \\
        \hline
        Quasi        & 0.84              & 4.92              & 41.01 $\pm$ 11.28 & 1.85 $\pm$ 0.31 \\
        OT 0$\times$ & 0.852 $\pm$ 0.096 & 6.607 $\pm$ 2.355 & 55.05 $\pm$ 13.23 & 6.41 $\pm$ 0.49 \\
        OT 1$\times$ & 0.862 $\pm$ 0.171 & 5.577 $\pm$ 1.988 & 52.19 $\pm$ 10.40 & 1.99 $\pm$ 0.56 \\
        OT 2$\times$ & 0.748 $\pm$ 0.116 & 4.759 $\pm$ 1.594 & 42.29 $\pm$ 6.18  & 1.18 $\pm$ 0.24 \\
        OT 3$\times$ & 0.811 $\pm$ 0.077 & 5.292 $\pm$ 1.977 & 75.53 $\pm$ 7.70  & 5.25 $\pm$ 0.76 \\
        \hline\hline
    \end{tabular}
    \label{tab:Gf cf g g'}
\end{table}

\begin{table}[htb!]
    \caption{Elastic properties of platelets and resin used in the stochastic finite element model.}
    \centering
    \begin{tabular}{ccc}
        \hline\hline
        Description & T700G & Resin \\
        \hline
        Platelet thickness, t [mm] & varies & varies \\
        In-plane longitudinal modulus, $E_1$ [GPa] & 135 & 3 \\
        In-plane transverse modulus, $E_2$ [GPa] & 10 & 3 \\
        In-plane shear modulus, $G_{12}$ [GPa] & 5 & 1.1 \\
        In-plane Poisson ratio, $\nu_{12}, \nu_{31}$ & 0.3 & 0.35 \\
        \hline\hline
    \end{tabular}
    \label{tab:MechPropSim}
\end{table}

\end{document}